\documentclass[10pt,twocolumn]{article}
\usepackage[numbers,sort&compress]{natbib}
\usepackage{authblk}
\usepackage[a4paper,margin=1.8cm,columnsep=0.7cm]{geometry}
\usepackage{longtable}
\usepackage{amsmath,amssymb}
\usepackage{bm}
\usepackage{tabularx}
\usepackage{booktabs}
\usepackage{braket}
\usepackage{graphicx}
\usepackage{hyperref}
\usepackage{xcolor}
\usepackage{comment} 
\usepackage{cleveref}
\newcommand{\BigO}[1]{\mathcal{O}\!\left(#1\right)}

\newcommand{\Tr}{\operatorname{Tr}}

\begin{document}

\title{A fermionic framework for quantum machine learning benchmarks}

\author[1,3]{Jérémie Gince}

\author[1,3]{Ayana Sarkar}

\author[1,2,3,4]{Stefanos Kourtis}

\affil[1]{Département de physique, Université de Sherbrooke, Sherbrooke, Québec J1K 2R1, Canada}
\affil[2]{Département d'informatique, Université de Sherbrooke, Sherbrooke, Québec J1K 2R1, Canada}
\affil[3]{Institut quantique, Université de Sherbrooke, Sherbrooke, Québec J1K 2R1, Canada}
\affil[4]{Mila -- Quebec AI Institute, Montréal, Québec, H2S 3H1, Canada}

\date{\today}

\twocolumn[
\begin{@twocolumnfalse}
\maketitle
\begin{abstract}
We present a general framework for the construction and benchmarking of scalable machine learning (ML) models based on the simulation of free-fermion systems.
We demonstrate that our fermionic ML (FermiML) framework is (i) versatile enough to encompass a wide range of ML techniques, including kernel methods and neural networks, (ii) applicable to common ML tasks, including classification and regression, and (iii) scalable to real-world datasets. FermiML models are quantum circuits built out of the same building blocks as most quantum ML (QML) models, namely, (parameterized) qubit rotations and entangling gates. FermiML thus allows us to build fermionic surrogates that mimic the architecture -- and hence the quantum resources and inductive bias -- of any given QML model, while remaining efficiently simulable classically. This, in turn, enables us to perform equal-footing benchmarks of QML models against close fermionic counterparts on any dataset and to ascertain whether any purported performance gain in QML is truly due to classically inaccessible resources or simply an accidental side-effect. To demonstrate this, we comprehensively benchmark quantum kitchen sinks (QKS),instantaneous quantum polynomial (IQP) and generic parameterized quantum circuit (PQC)-based QML models against fermionic surrogates across five classification datasets comprising 31 configurations. We find that fermionic-\textsc{QKS} (f\textsc{QKS}) significantly outperforms \textsc{QKS} and remains statistically indistinguishable from \textsc{IQP} and \textsc{PQC}, while fermionic-\textsc{PQC} (f\textsc{PQC}) outperforms \textsc{IQP} and \textsc{QKS} and remains statistically indistinguishable from \textsc{PQC}, tying it in 28 configurations. Moving beyond classification, we introduce fermionic fidelity and Pfaffian covariance kernels for continuous-target regression and establish baselines for QML in this harder task. Finally, we introduce a fermionic energy-based classifier scalable to large text (IMDb, AG News) and image (CIFAR2, Fashion) datasets, and find that it outperforms prominent QML models on image data, while beating or closely trailing them on text data and remaining close to classical ML baselines on all datasets tested. Overall, our results show that while performance depends sensitively on model architecture, the FermiML framework can be used to build fermionic surrogates that challenge or even outperform their unrestricted QML counterparts and design fermionic QML models that match the scale of classical ML baselines and reach real-world datasets.
\end{abstract}

\end{@twocolumnfalse}]

\section{Introduction}
\label{sec:introduction}

Machine learning (ML) tasks are promising near-term use cases of scientific and commercial relevance for quantum computation~\cite{Biamonte2017QuantumMachineLearning, Perdomo-Ortiz_2018, Wang2024ComprehensiveQML, Gujju2024NearTermQML, Lamata2023QMLImplementations, zhao2026exponentialquantumadvantageprocessing}. Quantum machine learning (QML) is motivated by the idea that quantum resources---superposition,
interference, and entanglement---may be exploited to reveal patterns in data that are difficult
or impossible to resolve using classical computation. This possibility is supported by
provable quantum-classical separations for carefully designed learning
problems and datasets~\cite{kubler2021inductivebiasquantumkernels,
QKL-ECfree-Huang2021,Huang2022,Liu2021}.

Whether QML confers any benefit in conventional ML tasks on real-world data remains
unresolved, partly because systematic benchmarking at scale is difficult.
Generic quantum circuits are expensive to simulate, while current quantum computing hardware
constraints severely limit the training, inference, and hyperparameter exploration required by proposed QML methods. Studies are
therefore often restricted to modest circuit sizes and small and/or synthetic datasets, where
conclusions can depend strongly on data encoding, model selection, and the
choice of baseline. Moreover, even when promising QML candidate models seem to perform
well at small scale, the observed accuracy gains are often an accidental side-effect of the aforementioned factors
and unrelated to quantum computational resources inaccessible to classical computation~\cite{bowles2024betterclassicalsubtleart}.

The subtlety of the classical-quantum separation for computational resources is crisply illustrated by fermionic quantum computation (FQC)~\cite{valiant2001quantum, knill2001fermioniclinearopticsmatchgates, bravyi2002fermionic, terhal2002classical, josza2008matchgates, bravyi2005lagrangian, FGS_matchgates_2022_surace}. FQC encompasses only quantum operations equivalent to evolution under Hamiltonians that describe systems of noninteracting (or free) fermions. Free-fermion evolution generates Gaussian transformations that can be decomposed as quantum circuits of two-qubit gates acting on adjacent qubits on a line. Free-fermion physics dictates that these gates satisfy algebraic constraints tied to the theory of perfect matchings in graphs and are hence called matchgates. Circuits composed exclusively of matchgates can produce superposition, interference, and entanglement, and are in fact equivalent to polynomial-time and logarithmic-space bounded universal unitary quantum computation~\cite{Jozsa_2009}, yet remain classically simulable in polynomial time.

Another characteristic of fermionic circuits is that, unlike other restricted classes of quantum circuits such as the Clifford hierarchy, their gates can be continuously parametrized. Moreover, in many settings, the gradient of expectation values with respect to gate parameters is provably finite for fermionic circuits~\cite{Ragone2024_LieAlgebraicBP,
diaz2023showcasingbarrenplateautheory,khatri2025trainability}. These properties make fermionic circuits appealing for building QML models that are intrinsically free of barren plateaus~\cite{mcclean2018barren,diaz2023showcasingbarrenplateautheory,
Larocca2025_BP_review}, thus avoiding a major roadblock in QML~\cite{wan2023matchgate,mele2025efficient,bittel2025pac,
strobl2026unflattening,xu2026spectral,
jäger2026provablescalablequantumgaussian,matos2023characterization,argyle2026measurement,bako2025fermionic}. In conjunction with classical simulability, they also allow for ML benchmarks at scales beyond the reach of today's quantum computers or their classical simulation~\cite{gince2024fermionic,kerenidis2026scalable}.

An important question for the practical relevance of QML in the near term is whether the performance of a model built using an unrestricted gate set can be matched by a comparable fermionic model. More precisely, if the fermionic model mimics the patterns of superposition, interference, and entanglement that its unrestricted counterpart generates and yet achieves the same performance, then the quantum resources used by the latter cannot confer a significant quantum advantage. Fermionic circuits thus provide an invaluable baseline against which proposed QML methods can be empirically benchmarked on equal footing.

In this work, we develop the above insights into a general fermionic ML (FermiML) framework and use it to build QML models that we then comprehensively test. FermiML is at the same time versatile enough to formulate a wide range of QML methodologies, including quantum kernel methods (QKMs), quantum neural networks (QNNs), and Hamiltonian-based learning~\cite{verdon2019quantum, tiblias2025efficient} -- all of which we demonstrate within FermiML, and scalable to large datasets and qubit counts.

Our evaluation of FermiML models is twofold. First, we build fermionic counterparts of selected QML models that preserve the original data preprocessing, encoding strategy, readout, classical estimator, and hyperparameter selection budget as closely as possible. This enables us to perform controlled comparisons between original and fermionic counterparts, in which the principal change is the restriction of circuits to the free-fermionic manifold. As a concrete illustration, we evaluate quantum kitchen sinks (QKS), instantaneous quantum polynomial (IQP) and a generic parameterized quantum circuit-based (PQC) QML models across five classification datasets (3 synthetic, Bars-and-stripes, and downscaled MNIST) comprising 31 configurations against their fermionic (f) surrogates, fQKS, fIQP, fPQC respectively. We find that across the 31 dataset configurations, f\textsc{IQP} significantly underperforms all three unrestricted baselines. In contrast, f\textsc{QKS}
significantly outperforms \textsc{QKS} and remains statistically indistinguishable from \textsc{IQP} and \textsc{PQC}, while f\textsc{PQC} outperforms \textsc{IQP} and \textsc{QKS} and remains statistically indistinguishable from \textsc{PQC}, tying it in 28 configurations. To facilitate further benchmarking of QML models, we have developed \textit{MajoranaKaraoke}, a PennyLane plugin that automatically builds a fermionic surrogate for a given quantum circuit and simulate it efficiently using the MatchCake library
~\cite{Gince_MatchCake_A_Python_2026,gince_torchpfaffian}. The translation is exact for circuits expressible in terms of nearest-neighbor free-fermionic gates and approximate otherwise.

Second, we build FermiML models that enable us to establish baselines for tasks and/or scales that have remained largely inaccessible to QML models thus far. Specifically, we introduce fermionic fidelity and Pfaffian covariance kernels for continuous-target regression using either a support vector machine (SVM) or a Gaussian process (GP) and establish QML baselines for three datasets (1 synthetic, Friedman \#1, and California housing) in this harder supervised ML task. Furthermore, we introduce NIF-Ising, a trainable fermionic energy-based classifier scalable to large datasets. We find that NIF-Ising outperforms prominent QML models on image data (CIFAR2, Fashion), while beating or closely trailing them on text data (IMDb, AG News) and remaining close to classical ML baselines on all datasets tested.

Our results establish the free-fermion limit as a rigorous reference point for QML benchmarking. Our conclusion is that any putative performance gains of QML observed empirically in a concrete task should be attributed to classically intractable quantum phenomena only if the evaluated model can outperform closely matched fermionic counterparts.

\section{Methodology}
\label{sec:methodology}




\subsection{Matchgate circuits and covariance matrices}
\label{subsec:fermionic-representation}

We consider a system of $n$ fermionic modes, represented after a Jordan-Wigner transformation by $n$ qubits. Since each fermionic mode contributes two Majorana operators, the Hilbert space of the fermion system is spanned by the vector $\bm{c}=(c_1,\ldots,c_{2n})^T$,
whose components are Hermitian ($c_j^\dagger=c_j$) and satisfy the canonical anti-commutation relations $\{c_j,c_k\}=2\delta_{jk}$.
We use the Jordan--Wigner transformation
\begin{equation}
\begin{aligned}
c_{2q-1}
&=
\left(\prod_{\ell=1}^{q-1}Z_\ell\right)X_q,
&
c_{2q}
&=
\left(\prod_{\ell=1}^{q-1}Z_\ell\right)Y_q,
\,q=1,\ldots,n.
\end{aligned}
\label{eq:JW_transformation}
\end{equation}
For an arbitrary state $\rho$ of $n$ fermionic modes, we define the
$2n\times2n$ Majorana covariance matrix by
\begin{equation}
    \Gamma_{jk}(\rho)
    =
    \frac{i}{2}\,
    \Tr\!\left(
        \rho\,[c_j,c_k]
    \right),
    \quad
    [c_j,c_k]=c_jc_k-c_kc_j \,.
    \label{eq:majorana-covariance}
\end{equation}
The matrix $\Gamma(\rho)$ is real and antisymmetric. For $j\neq k$, the
Majorana anti-commutation relations give
\begin{equation}
    \Gamma_{jk}(\rho)
    =
    i\,\Tr(\rho c_jc_k).
\end{equation}
The covariance matrix is defined for any fermionic state, but in general it
contains only its two-point correlations. Parity-preserving fermionic
Gaussian states are special because Wick's theorem determines all their
higher-order correlations from these two-point functions
~\cite{bravyi2005lagrangian}. Such a Gaussian state is therefore completely
specified by its covariance matrix. Pure fermionic Gaussian states can,
equivalently, be prepared from Fock-basis states using unitaries generated by quadratic fermionic Hamiltonians.
Any real antisymmetric covariance matrix can be brought to the canonical form
\begin{equation}
    \Gamma
    =
    O
    \left[
        \bigoplus_{q=1}^{n}
        \begin{pmatrix}
            0 & \nu_q \\
            -\nu_q & 0
        \end{pmatrix}
    \right]
    O^{T},\,
    O\in\mathrm{O}(2n), \,
    0\leq\nu_q\leq1.
    \label{eq:williamson-form}
\end{equation}
The quantities $\nu_q$ are the fermionic Williamson eigenvalues; equivalently, the eigenvalues of $\Gamma$ occur in pairs $\pm i\nu_q$. They are invariant under fermionic Gaussian transformations and characterize the mixedness of
the independent fermionic modes in the canonical basis. A Gaussian state is
pure precisely when
\begin{equation}
    \Gamma^2=-I_{2n},
    \label{eq:pure-gaussian-covariance}
\end{equation}
or, equivalently, when $\nu_q=1$ for every $q$
~\cite{bravyi2005lagrangian}.

More explicitly, let
$S=\{s_1<\cdots<s_{2p}\}\subseteq\{1,\ldots,2n\}$ and let
$\Gamma|_S$ denote the corresponding principal sub-matrix of $\Gamma$. Wick's theorem
gives
\begin{equation}
    \Tr\!\left[
        \rho\,c_{s_1}\cdots c_{s_{2p}}
    \right]
    =
    (-i)^p
    \operatorname{Pf}\!\left(\Gamma|_S\right).
    \label{eq:majorana-wick-pfaffian}
\end{equation}
Thus, every even-order Majorana correlator of a parity-preserving Gaussian state is obtained from a Pfaffian of its covariance matrix. Its odd-order Majorana correlators vanish. More generally, odd-order correlators vanish for any state with definite fermionic parity. 

\subsubsection{Free-fermion dynamics and matchgate circuits}

Free-fermion dynamics are generated by Hamiltonians that are quadratic in Majorana operators. Up to an additive constant, such a Hamiltonian is written as
\begin{equation}
    H
    =
    \frac{i}{4}
    \sum_{j,k=1}^{2n}
    \mathcal{A}_{jk}c_jc_k,
    \quad
    \mathcal{A}^T=-\mathcal{A},
    \quad
    \mathcal{A}\in\mathbb{R}^{2n\times 2n}.
    \label{eq:quadratic-majorana-hamiltonian}
\end{equation}
The corresponding unitary evolution $U=\exp(-iH)$ preserves Gaussianity and
acts linearly on the Majorana operators:
\begin{equation}
    U^\dagger c_j U
    =
    \sum_{k=1}^{2n}R_{jk}c_k,
    \qquad
    R\in\mathrm{SO}(2n).
    \label{eq:majorana-orthogonal-action}
\end{equation}
The matrix $R$ is the single-particle representation of the many-body
unitary $U$, also called its single-particle transition matrix (SPTM). Since $R$ is orthogonal, its inverse is $R^T$, and
\begin{equation}
    U c_j U^\dagger
    =
    \sum_{k=1}^{2n}R_{kj}c_k.
    \label{eq:inverse-majorana-action}
\end{equation}
A fermionic circuit is a sequence of $L$ Gaussian unitaries, $U=U_L\cdots U_1$. Applying Eq.~\eqref{eq:majorana-orthogonal-action} gate by gate shows that the circuit is represented by the product of the corresponding SPTMs,
\begin{equation}
    R=R_L\cdots R_1\in\mathrm{SO}(2n).
    \label{eq:orthogonal-circuit-composition}
\end{equation}
Under the Jordan-Wigner transformation of Eq.~\eqref{eq:JW_transformation}, such circuits are implemented by nearest-neighbor matchgates. A two-qubit matchgate preserves parity and acts independently on the even- and odd-parity subspaces. In the computational basis, it has the form
\begin{equation}
    G(A,B)
    =
    \begin{pmatrix}
        A_{11} & 0      & 0      & A_{12} \\
        0      & B_{11} & B_{12} & 0      \\
        0      & B_{21} & B_{22} & 0      \\
        A_{21} & 0      & 0      & A_{22}
    \end{pmatrix},
    \,\,
    \det A=\det B,
    \label{eq:matchgate-matrix}
\end{equation}
where $A$ and $B$ specify the action on the even- and odd-parity subspaces, respectively. Equivalently, after extracting a common global phase, both blocks may be chosen in $\mathrm{SU}(2)$. A matchgate acting on qubits $q$ and $q+1$ is generated by a quadratic Hamiltonian involving only $c_{2q-1},\ldots,c_{2q+2}$, so its SPTM differs from the identity only on the corresponding $4\times4$ block.

Starting from a Gaussian reference state $\rho_0$ with covariance matrix $\Gamma_0$, the circuit prepares the state $\rho=U\rho_0U^\dagger$. Combining Eqs.~\eqref{eq:majorana-covariance} and~\eqref{eq:majorana-orthogonal-action}, its covariance matrix is
\begin{equation}
    \Gamma=R\,\Gamma_0R^T.
    \label{eq:covariance-propagation}
\end{equation}
Since the output state remains Gaussian, $\Gamma$ specifies it completely. Simulating a matchgate circuit therefore amounts to building the SPTM of each gate and propagating the initial covariance matrix using Eq.~\eqref{eq:covariance-propagation}. A covariance matrix contains $\BigO{n^2}$ independent real entries, compared to the $2^n$ complex amplitudes of a generic $n$-qubit state. Dense covariance transformations and Gaussian fidelity calculations thus require only polynomial-time linear algebra. An individual fixed-order correlator is also obtained efficiently from a Pfaffian through Eq.~\eqref{eq:majorana-wick-pfaffian}. If all order-$m$ correlators are retained as features, their number is $\binom{2n}{m}=\BigO{n^m}$ for fixed $m$, which remains polynomial in $n$. This compact description is the source of the classical simulability of fermionic quantum circuits.

Fermionic circuits can, moreover, be continuously parameterized. Each gate or layer $U_\ell(\bm{\theta}_\ell)$ depends on a vector of real \emph{gate angles} $\bm{\theta}_\ell$, which determine the entries of the corresponding matchgate and, hence, of its SPTM $R_\ell(\bm{\theta}_\ell)$. For instance, the blocks $A$ and $B$ in Eq.~\eqref{eq:matchgate-matrix} can be chosen as single-qubit rotations such as $R_X(\theta)$ or $R_Z(\theta)$. More generally, the gate angles may directly set the entries of the generator $\mathcal{A}$ in Eq.~\eqref{eq:quadratic-majorana-hamiltonian}. The complete circuit and its SPTM are then
\begin{equation}
    U(\bm{\theta})=U_L(\bm{\theta}_L)\cdots U_1(\bm{\theta}_1),
    \quad
    R(\bm{\theta})=R_L(\bm{\theta}_L)\cdots R_1(\bm{\theta}_1).
    \label{eq:parameterized-circuit}
\end{equation}
Changing the gate angles only changes the entries of the SPTMs and leaves the simulation cost unchanged.

In a learning model, the gate angles are not directly the quantities being learned. Instead, they depend on an \emph{input sample}, i.e., a feature vector $\bm{x}\in\mathbb{R}^{d}$ drawn from a dataset of the form described in Eq.~\eqref{eq:supervised-dataset}, and on a set of model parameters $\bm{\Theta}$, which are either trained or kept fixed. This dependence is specified by \emph{encoding functions} that compute the gate angles of each layer,
\begin{equation}
    \bm{\theta}_\ell=f_\ell(\bm{x};\bm{\Theta}).
    \label{eq:encoding-functions}
\end{equation}
Different layers may use different encoding functions and may therefore encode different components or transformations of the same input sample. Writing $U(\bm{x};\bm{\Theta})\equiv U(\bm{\theta}(\bm{x};\bm{\Theta}))$ and $R(\bm{x};\bm{\Theta})\equiv R(\bm{\theta}(\bm{x};\bm{\Theta}))$, the encoded state and its covariance matrix are
\begin{equation}
    \rho(\bm{x};\bm{\Theta})=U(\bm{x};\bm{\Theta})\,\rho_0\,U^\dagger(\bm{x};\bm{\Theta}),
    \label{eq:encoded-gaussian-state}
\end{equation}
\begin{equation}
    \Gamma(\bm{x};\bm{\Theta})=R(\bm{x};\bm{\Theta})\,\Gamma_0\,R^T(\bm{x};\bm{\Theta}).
    \label{eq:encoded-gaussian-cov}
\end{equation}
This formulation covers all the models considered in this work. In angle embeddings, the encoding functions depend only on $\bm{x}$ and contain no model parameters. In all cases, the many-body circuit is simulated through the $2n\times2n$ matrix $R$, without ever constructing the $2^n$-dimensional state vector.

\subsection{Feature map families}
\label{subsec:fermionic-feature-maps}

Below we consider two fermionic feature maps, each leading to distinct kernel. The fidelity kernel (Sec.~\ref{subsec:fermionic-fidelity-kernel}) compares the complete
states through their squared overlap, whereas the order-$m$ covariance kernel (Sec.~\ref{subsec:pfaffian-covariance-kernel})
compares only their order-$m$ Majorana correlations. These quantum kernels are then used in standard classical kernel learning algorithms. The resulting model families differ in their data encoding, circuit architecture, and readout.

\subsubsection{Fermionic fidelity kernel}
\label{subsec:fermionic-fidelity-kernel}

Let $\rho_0=\ket{\psi_0}\!\bra{\psi_0}$ be a pure Gaussian reference state and let $U(\bm{x})$ be a data-dependent matchgate circuit.
Here we drop the $\boldsymbol\Theta$-dependence for simplicity since we only consider fixed kernels, but parameterized kernels are straightforwardly accommodated in our framework.
The corresponding
encoded state is
\begin{equation}
    \rho(\bm{x})
    =
    U(\bm{x})\rho_0U^\dagger(\bm{x})
    =
    \ket{\psi(\bm{x})}\!\bra{\psi(\bm{x})}.
\end{equation}
The fermionic fidelity kernel is defined as
\begin{equation}
    k^{\mathrm F}(\bm{x},\bm{x}')
    =
    \Tr\!\left[
        \rho(\bm{x})\rho(\bm{x}')
    \right]
    =
    \left|
        \braket{\psi(\bm{x})|\psi(\bm{x}')}
    \right|^2.
    \label{eq:fermionic_fidelity_kernel_original}
\end{equation}
For Gaussian states, this overlap can be evaluated directly from their covariance matrices. Let $\rho$ and $\rho'$ be fermionic Gaussian states with covariance matrices
$\Gamma$ and $\Gamma'$, respectively. Wick's theorem expresses even-order Majorana correlators as
Pfaffians of principal covariance sub-matrices
~\cite{bravyi2005lagrangian}. Their Hilbert-Schmidt overlap can therefore be
written as
\begin{equation}
    \Tr(\rho\rho')
    =
    2^{-n}
    \sum_{\substack{
        S\subseteq\{1,\ldots,2n\}\\
        |S|\ {\rm even}
    }}
    \operatorname{Pf}\!\left(\Gamma|_S\right)
    \operatorname{Pf}\!\left(\Gamma'|_S\right),
    \label{eq:gaussian-overlap-pfaffian-sum}
\end{equation}
where $\Gamma|_S$ denotes the principal sub-matrix indexed by $S$, and the
Pfaffian of the empty sub-matrix is defined as one. The sum has closed
form
\begin{equation}
    \Tr(\rho\rho')
    =
    2^{-n}
    \sqrt{
        \det\!\left(
            I_{2n}-\Gamma\Gamma'
        \right)
    },
    \label{eq:gaussian-overlap-covariance}
\end{equation}
where the non-negative square root is taken
~\cite{bravyi2005lagrangian,FGS_matchgates_2022_surace}. We consider a supervised-learning dataset containing $N_s$ labeled examples,
\begin{equation}
    \mathcal{D}
    =
    \left\{
        \bigl(\bm{x}_i,y_i\bigr)
    \right\}_{i=1}^{N_s},
    \qquad
    \bm{x}_i\in\mathbb{R}^{d},
    \label{eq:supervised-dataset}
\end{equation}
where $\bm{x}_i$ is a $d$-dimensional input vector whose components are called \textit{features} and $y_i$ is its
associated target. For classification, $y_i\in\{1,\ldots,K\}$ denotes one of
$K$ classes, while for regression, $y_i\in\mathbb{R}$ is a continuous
target. The resulting Gram matrix $\mathcal{K}^{\mathrm F}\in\mathbb{R}^{N_{s}\times N_{s}}$ has entries
\begin{equation}
    \mathcal{K}^{\mathrm F}_{ij}
    =
    k^{\mathrm F}(\bm{x}_i,\bm{x}_j).
    \label{eq:fermionic-fidelity-gram}
\end{equation}
This feature map can now be combined with different classical learning algorithms. In the classification experiments, this is supplied to a SVC, whereas the regression experiments use SVR or GPR.

\subsubsection{Order-$m$ Pfaffian covariance kernel}
\label{subsec:pfaffian-covariance-kernel}

For an even order $m$, we define the Pfaffian feature
vector by
\begin{equation}
    \bigl[v_m(\bm{x})\bigr]_S
    =
    \operatorname{Pf}\!\left(
        \Gamma(\bm{x})\big|_S
    \right),
    \qquad
    \dim v_m=\binom{2n}{m},
    \label{eq:orderm-pfaffian-features}
\end{equation}
where $\Gamma(\bm{x})|_S$ is the principal submatrix indexed by $S$, The definition of subset $S$ is given above \ref{eq:majorana-wick-pfaffian}. By
Wick's theorem, these features are proportional to the corresponding
$m$-point Majorana correlators.

The order-$m$ covariance kernel is then
\begin{equation}
    \kappa_m(\bm{x},\bm{x}')
    =
    a_{n,m}\,
    v_m(\bm{x})^T v_m(\bm{x}'),
    \qquad
    a_{n,m}>0,
    \label{eq:orderm-pfaffian-covariance-kernel}
\end{equation}
where $a_{n,m}$ is an input-independent normalization. Because \eqref{eq:orderm-pfaffian-covariance-kernel} is an ordinary Euclidean inner-product, it defines a positive semi-definite kernel. Individual
kernel entries can nevertheless be negative because the
Pfaffian features are signed.

\subsection{Classification models}
\label{subsec:classification-models}

\subsubsection{Kernel and random-feature models}
\label{subsec:classification-fixed-models}

The classification benchmark comprises six circuit-based models arranged
into three matched pairs: \textsc{IQP}-f\textsc{IQP},
\textsc{QKS}-f\textsc{QKS}, and
\textsc{PQC}-f\textsc{PQC}.

\paragraph{Non-matchgate quantum baselines.}

We consider two well established QML circuit ans\"{a}tze: an instantaneous-quantum-polynomial (\textsc{IQP}) fidelity kernel and quantum kitchen sinks (\textsc{QKS})~\cite{bowles2024betterclassicalsubtleart,wilson2018qks,lloyd2020quantumembeddingsmachinelearning_IQP, shen2026characterizingtrainabilityinstantaneousquantum_IQP, Bremner_2010_IQC, jcdk-q3xc_IQP}. These models provide the unrestricted quantum baseline against which we assess the effect of imposing a fermionic Gaussian structure.

The \textsc{IQP} model is based on the quantum feature map introduced by Havl\'{i}\v{c}ek \emph{et al.}~\cite{havlicek2019supervised}. Since then, IQP circuits have been widely employed across a range of QML tasks~\cite{lerch2026iqpbornmachinesdatadependent,bowles2024betterclassicalsubtleart,raj2026trainclassicaldeployquantum}. It acts on $n=d$ qubits and alternates layers of Hadamard gates with diagonal operators containing data-dependent single-qubit and two-qubit phases. Schematically, the diagonal encoding layer has the form
\begin{equation}
D_{\Phi}(\bm{x})
=\exp\!\left[i\sum_{j=1}^{n}\phi_j(\bm{x})Z_j+
i\sum_{j<k}\phi_{jk}(\bm{x})Z_jZ_k
\right],
\label{IQPmodel_equation}
\end{equation}
where the functions $\phi_j$ and $\phi_{jk}$ specify the single-feature and pairwise-feature encodings used in the implementation. The complete encoding circuit, denoted by $U_{\mathrm{IQP}}^{(r)}(\bm{x})$, contains $r$ repetitions of the prescribed mixing and diagonal-encoding layers. The fidelity kernel is then given by
\begin{equation}
k_{\mathrm{IQP}}(\bm{x},\bm{x}')
=
\left|
\bra{0}^{\otimes n}
U_{\mathrm{IQP}}^{(r)\dagger}(\bm{x}')
U_{\mathrm{IQP}}^{(r)}(\bm{x})
\ket{0}^{\otimes n}
\right|^2.
\label{eq:IQP_kernel}
\end{equation}
We evaluate the IQP kernel via exact state-vector simulation and are thus restricted to small system $n$.

The \textsc{QKS} model follows the random-feature construction of Wilson \emph{et al.}~\cite{wilson2018qks}. For each episode $e=1,\ldots,E$, it samples an independent random affine map
\begin{equation}
\bm{\theta}_e(\bm{x})=
\Omega_e\bm{x}+\bm{\beta}_e,
\label{QKS_wilson_construction}
\end{equation}
whose components determine the angles of single-qubit $R_X$ rotations. These rotations are followed by the CNOT entangling pattern prescribed by the original implementation in \cite{wilson2018qks}. Measuring the circuit in the computational basis produces a bit string $\bm{z}_e(\bm{x})\in \{0,1\}^{\otimes n}$. The outcomes of all episodes are then concatenated to form the explicit random-feature vector
\begin{equation}
\bm{z}_{\mathrm{QKS}}(\bm{x})
=
\bm{z}_1(\bm{x})
\oplus\cdots\oplus
\bm{z}_E(\bm{x})
\in \{0,1\}^{\otimes E_{n}},
\label{QKS_random_feature_vector}
\end{equation}
which is used as the input to logistic regression. Unlike \textsc{IQP}, the \textsc{QKS} is therefore an explicit random-feature model. We consider the full encoding with $n=d$ and the reduced encoding with
$n=\lceil d/2\rceil$.

The \textsc{PQC} baseline uses the YZ-CX architecture of
Haug \emph{et al.}~\cite{haug2023quantum}. Each input feature $x_j$ is
encoded exactly once through the angle
\begin{equation}
    \theta_j(\bm{x})
    =
    \frac{\pi}{2}
    \left(
        \theta_{r,j}+x_j
    \right),
    \label{eq:pqc-feature-encoding}
\end{equation}
where the offsets $\theta_{r,j}$ are sampled independently and uniformly
from $[0,1]$, and are then held fixed throughout training and evaluation.
These angles parameterize layers of $R_YR_Z$ rotations followed by CNOT
entanglers acting on alternating pairs of wires.

We vary the circuit width by choosing the number of qubits to be the full,
half, quarter, or eighth fraction of the input dimension. When fewer qubits
are used, the features are distributed over additional encoding layers so
that every feature is still loaded once. These four width settings therefore
correspond to approximately one, two, four, and eight encoding layers,
respectively. The resulting \textsc{PQC} circuits are evaluated using exact state-vector
simulation.

\paragraph{Matchgate-constrained counterparts.}

To assess whether the performance of the aforementioned quantum baselines 
is indeed due to quantum resources inaccessible to classical simulation, we construct matchgate counterparts of the quantum
baselines. These counterparts are designed manually to preserve the original
architectures as closely as possible; more generally, this translation can
be automated using our \href{https://github.com/MatchCake/MajoranaKaraoke}{\textit{MajoranaKaraoke} tool}, which produces an inspectable matchgate surrogate of a given circuit.

The f\textsc{IQP} ansatz preserves the repeated structure of \textsc{IQP} while replacing each layer by a matchgate. Within each repetition, pairs of single-feature rotations $R_Z(x_i)$ and $R_Z(x_j)$ are replaced by the two-qubit matchgates $G\!\left(R_Z(x_i),R_Z(x_j)\right).$
Here, $G(A,B)$ denotes a matchgate given in Eq.~\eqref{eq:matchgate-matrix}  and is not equivalent to the tensor product $A\otimes B$. The pairwise IQP interaction $\exp\!\left(-\frac{i}{2}x_i x_j Z_iZ_j\right)$
is replaced by
$G\!\left(
        R_Z(x_i x_j),
        R_Z(x_i x_j)
    \right).$
Thus, the product $x_i x_j$ determines the angle of the
$G(R_Z,R_Z)$ gate. Fixed $G(Z,X)$ gates, corresponding to fSWAPs, are inserted to route the features so that the required pairwise terms can be implemented
using the circuit connectivity. The mixing layer is similarly replaced by $G(H,H)$.
Because the resulting circuit is Gaussian, its action is represented by $R_{\mathrm{fIQP}}^{(r)}(\bm{x})\in$ SO$(2n)$, and its output is evaluated using the fermionic fidelity kernel
\begin{equation}
k_{\mathrm{fIQP}}(\bm{x},\bm{x}')=2^{-n}
\sqrt{\det \!\left[I_{2n}-
\Gamma_{\mathrm{fIQP}}(\bm{x})
\Gamma_{\mathrm{fIQP}}(\bm{x}')
\right]
}.
\label{fermionic_fidelity_kernel}
\end{equation}
The repetition values $r$ are matched to those used for \textsc{IQP}.

Similarly, f\textsc{QKS} uses the same independently sampled affine maps
$\bm{\theta}_e(\bm{x})=\Omega_e\bm{x}+\bm{\beta}_{e}$ as \textsc{QKS}, but replaces the single-qubit $R_X$ layer by $G(R_X, R_X)$ matchgates and the CNOT entangling pattern by two nearest-neighbor brickwork layers of fSWAPs. Instead of sampling computational-basis bit strings, f\textsc{QKS} constructs its explicit features deterministically from SPTMs generated by each episode. Concatenating the features from the $E$ episodes gives
\begin{equation}
\bm{z}_{\mathrm{fQKS}}(\bm{x})
=
\bm{z}^{\mathrm f}_1(\bm{x})
\oplus\cdots\oplus
\bm{z}^{\mathrm f}_E(\bm{x}),
\label{fQKS_circuit_construction}
\end{equation}
which is supplied to the same logistic-regression procedure as the \textsc{QKS} features. 

The f\textsc{PQC} model is the matchgate counterpart of the
\textsc{PQC} baseline and uses the same circuit architecture introduced in
our earlier work on fermionic machine learning~\cite{gince2024fermionic}. It preserves the feature angles of
\cref{eq:pqc-feature-encoding}, the number of qubits, the encoding depth,
and the alternating-pair circuit layout. The $R_YR_Z$ blocks are replaced
by matchgate-compatible rotations, while the CNOT entanglers are replaced
by $U(Z,X)$ matchgate fSWAP gates acting on the same wire pairs.

The comparisons \textsc{IQP}-f\textsc{IQP}, \textsc{QKS}-f\textsc{QKS} and \textsc{PQC}-f\textsc{PQC} are intended as controlled architectural comparisons. They test how the restriction to classically simulable fermionic Gaussian dynamics changes predictive performance while preserving the overall encoding strategy and the corresponding classical learning method.

\paragraph{Native free-fermionic kernels.}
\label{free_fermionic_kernels_classifications}
Compared with f\textsc{IQP}, f\textsc{QKS}, and f\textsc{PQC}, the native
free-fermionic models are not obtained by converting a particular
unrestricted circuit into matchgates. They instead use the
\emph{Vanilla encoder}, which maps
$\bm{x}\in\mathbb{R}^{d}$ directly to a fermionic Gaussian
transformation. The affine map
\begin{equation}
    \bm{a}(\bm{x})=W\bm{x}+\bm{b},
    \qquad
    W\in\mathbb{R}^{\binom{2n}{2}\times d},
    \quad
    \bm{b}\in\mathbb{R}^{\binom{2n}{2}},
    \label{eq:vanilla-affine-map}
\end{equation}
generates the independent entries of a real antisymmetric matrix $h(\bm{x})\in\mathbb{R}^{2n\times2n}$. The coefficients $W$ and $\bm{b}$ are sampled from random distributions once for each model realization and are not optimized
during classifier training. These parameters are initialised with the PyTorch default for linear layers~\cite{Paszke2019}, i.e drawn independently from $\mathcal{U}(-1/\sqrt{d}, 1/\sqrt{d})$ which corresponds to the Kaiming-uniform scheme~\cite{He2015} with negative slope $a=\sqrt{5}$. The resulting Majorana transformation is
\begin{equation}
    R_{\mathrm{Vanilla}}(\bm{x})
    =
    \exp\!\left[h(\bm{x})\right]
    \in\mathrm{SO}(2n).
    \label{R_matrix_for_vanillaM_models}
\end{equation}
For the reference state $\ket{0}^{\otimes n}$ with covariance matrix
$\Gamma_0$,
\begin{equation}
    \Gamma_{\mathrm{Vanilla}}(\bm{x})
    =
    R_{\mathrm{Vanilla}}(\bm{x})
    \Gamma_0
    R_{\mathrm{Vanilla}}^{T}(\bm{x}).
    \label{covariance_matrix_for_vanilla}
\end{equation}
This encoder differs from the construction introduced in our earlier work~\cite{gince2024fermionic}. The earlier model loads the input features gate-by-gate into a layered nearest-neighbor matchgate circuit. This encoder on the other hand, constructs the full single-particle transformation directly, without prescribing a layered gate architecture. It therefore provides a simple native free-fermionic encoding that can be combined with various readouts. 

Combining the Vanilla encoder with the fermionic fidelity kernel defines the
\textsc{Vanilla} model. Replacing the fidelity readout by the order-$m$ Pfaffian covariance kernel defines \textsc{VanillaMpf}. These two models therefore
use the same encoder and the same reference state
$\ket{0}^{\otimes n}$, but extract different similarity measures from the
resulting Gaussian states. \textsc{AngleMpf} uses the same Pfaffian covariance readout as \textsc{VanillaMpf}, but replaces the random affine Vanilla encoder with a direct angle encoding of the input features. It therefore provides a comparison that separates the effect of the Pfaffian readout from that of the random affine encoding in
\cref{eq:vanilla-affine-map}. The reference-state variant
\textsc{VanillaMpf+}, discussed next, instead retains the Vanilla encoder
while replacing $\ket{0}^{\otimes n}$ with
$\ket{+}^{\otimes n}$. All four native free-fermionic models are evaluated
using the complete twenty-split benchmark.

Note here that although, \textsc{AngleMpf} and f\textsc{PQC} encode classical features
through matchgate rotation angles, they are distinct models.
f\textsc{PQC} is constructed as the fermionic counterpart of unrestricted
\textsc{PQC} baseline. It preserves the shifted feature angles, circuit
width, encoding depth, feature-to-wire assignment, and alternating-pair
layout of \textsc{PQC}, while replacing its rotations and CNOT gates with matchgate-compatible unitaries. Its readout is the fermionic fidelity
kernel, which compares the complete encoded Gaussian states.

In contrast, \textsc{AngleMpf} is a natively free-fermionic model. The input features directly determine the angles of successive matchgate layers, and the
resulting state is evaluated through the order-$m$ Pfaffian covariance
kernel. Therefore, it compares only the selected order-$m$ Majorana
correlations. Thus although the two models share the general principle of angle encoding, they differ in their data-to-angle map, circuit architecture, and
readout.

\paragraph{Reference-state variants.}

The \textsc{VanillaMpf+} model investigates how the native
fermionic construction depends on the choice of reference state. It replaces
the fermionic vacuum $\ket{0}^{\otimes n}$ used by
\textsc{VanillaMpf} with the product state $\ket{+}^{\otimes n}$, while
leaving the Vanilla data encoder unchanged. Because
$\ket{+}^{\otimes n}$ does not have definite fermion parity, this model
requires the generalized product-state simulation supported by
MatchCake~\cite{Gince_MatchCake_A_Python_2026}, rather than the standard
fixed-parity covariance construction.

The readout of \textsc{VanillaMpf+} is the generalized
Pfaffian feature map provided by MatchCake. The resulting kernel remains
positive semidefinite because it is constructed as an inner product between
explicit feature vectors. It should nevertheless be distinguished from the
standard covariance features of \textsc{VanillaMpf}, which are obtained
from the fixed-parity reference state $\ket{0}^{\otimes n}$.The comparison between \textsc{VanillaMpf} and
\textsc{VanillaMpf+} is performed using the twenty-split protocol described
in Sec.~\ref{subsec:classification-benchmark}, and the corresponding results
are presented in Sec.~\ref{subsubsec:classification_kernel_results}.

Table~\ref{tab:classification-model-summary} specifies how each model
represents the input data, including its encoder, reference state, and
readout. Table~\ref{tab:classification-hyperparameter-summary}
summarizes the scale and outcome of the validation search. Hyperparameters
are selected independently for each dataset configuration using the mean
validation accuracy, and test results are accessed only after this
selection.
\begin{table*}[t]
    \centering
    \small
    \setlength{\tabcolsep}{3pt}
    \begin{tabular}{@{}lllll@{}}
        \toprule
        \textbf{Family}
        & \textbf{Model}
        & \textbf{Encoder}
        & \textbf{Reference state(s)}
        & \textbf{Readout} \\
        \midrule

        Unrestricted quantum
        & \textsc{IQP}
        & IQP circuit
        & $\ket{0}^{\otimes n}$
        & Fidelity \\

        Unrestricted quantum
        & \textsc{QKS}
        & Random affine $+$ $R_X$/CNOT
        & $\ket{0}^{\otimes n}$
        & Sampled bit strings \\

        Unrestricted quantum
        & \textsc{PQC}
        & YZ--CX circuit
        & $\ket{0}^{\otimes n}$
        & Fidelity \\
        \midrule

        Matchgate counterpart
        & f\textsc{IQP}
        & Matchgate IQP
        & $\ket{0}^{\otimes n}$
        & Fermionic fidelity \\

        Matchgate counterpart
        & f\textsc{QKS}
        & Random affine $+$ matchgates
        & $\ket{0}^{\otimes n}$
        & Single-particle features \\

        Matchgate counterpart
        & f\textsc{PQC}
        & Matchgate YZ--CX
        & $\ket{0}^{\otimes n}$
        & Fermionic fidelity \\
        \midrule

        Native free-fermionic
        & \textsc{Vanilla}
        & Vanilla
        & $\ket{0}^{\otimes n}$
        & Fermionic fidelity \\

        Native free-fermionic
        & \textsc{VanillaMpf}
        & Vanilla
        & $\ket{0}^{\otimes n}$
        & Order-$m$ Pfaffian covariance \\

        Reference-state variant
        & \textsc{VanillaMpf+}
        & Vanilla
        & $\ket{+}^{\otimes n}$
        & Generalized Pfaffian features \\

        Native free-fermionic
        & \textsc{AngleMpf}
        & Angle
        & $\ket{0}^{\otimes n}$
        & Order-$m$ Pfaffian covariance \\
        \bottomrule
    \end{tabular}

    \caption{\textbf{Models retained for classification benchmark.}
    The table summarizes the encoder, reference state, and readout of each
    model evaluated using the complete twenty-split protocol. The
    \textsc{IQP}--f\textsc{IQP},
    \textsc{QKS}--f\textsc{QKS}, and
    \textsc{PQC}--f\textsc{PQC} pairs compare unrestricted circuits with
    matchgate-constrained counterparts based on the same overall encoding
    strategy. The native free-fermionic models use either a fermionic
    fidelity readout or Pfaffian covariance features. In particular,
    \textsc{VanillaMpf} combines the Vanilla encoder with the order-$m$
    Pfaffian covariance kernel, while \textsc{VanillaMpf+} uses the same
    encoder with the $\ket{+}^{\otimes n}$ reference state and generalized
    Pfaffian features.}
    \label{tab:classification-model-summary}
\end{table*}
\begin{table*}[t]
    \centering
    \begin{tabular}{@{}lrrrlr@{}}
        \toprule
  Model & Points & Runs & Ever best & Most often best & $W$ \\
  \midrule
  IQP & 12 & 7\,440 & 3 & $r=1$, $C=1$ & 16 \\
  fIQP & 12 & 7\,440 & 7 & $r=1$, $C=1$ & 7 \\
  QKS & 8 & 4\,960 & 2 & $E=2000$, $qf=\mathrm{half}$ & 20 \\
  fQKS & 8 & 4\,960 & 5 & $E=2000$, $qf=\mathrm{full}$ & 9 \\
  PQC & 16 & 8\,720 & 3 & $n=\mathrm{full}$, $C=1$ & 16 \\
  fPQC & 20 & 11\,200 & 5 & $n=\mathrm{double}$, $C=1$ & 16 \\
  Vanilla & 8 & 4\,960 & 4 & $n=d$, $C=1$ & 17 \\
  VanillaMpf & 16 & 9\,760 & 8 & $n=d$, $m=2$, $C=0.1$ & 8 \\
  VanillaMpf+ & 16 & 9\,920 & 8 & $n=32$, $m=2$, $C=0.1$ & 6 \\
  AngleMpf & 4 & 2\,480 & 4 & $n=d$, $m=2$, $C=100$ & 15 \\
  \midrule
  Total & 120 & 71\,840 & & & \\
  \bottomrule
    \end{tabular}
    \caption{\textbf{Summary of the classification hyperparameter search.} For each model, ``Points'' is the number of hyperparameter configurations included in the search grid, and ``Runs'' is the total number of completed fits across all applicable dataset configurations and the twenty data splits. Some model--configuration combinations are unavailable under the considered qubit-allocation rules, so the number of runs need not equal $\mathrm{Points}\times31\times20$. ``Best Ever'' counts the hyperparameter points that attained the unique highest mean validation accuracy for at least one dataset configuration. ``Most often best'' identifies the hyperparameter point with the largest number of such unique validation wins, and $W$ reports that number. Here, $r$ is the number of IQP repetitions, $E$ is the number of QKS episodes, $qf$ is the QKS qubit-allocation rule, $n$ specifies the number or allocation of qubits, $m$ is the Majorana-sector order, and $C$ is the SVM regularization parameter. Complete point-level win, tie, loss, rank, and test-accuracy statistics are reported in the Supplementary Material.}
    \label{tab:classification-hyperparameter-summary}
\end{table*}
\subsubsection{Trainable Hamiltonian classifier}
\label{subsec:nif-ising}

Scalability in QML depends both on the cost of encoding each input and on the number of model evaluations required by the learning procedure. Kernel methods evaluate similarities between pairs of training examples, so constructing the corresponding Gram matrix for $\mathcal{D}$ examples requires $\BigO{\mathcal{D}^2}$ kernel evaluations. This quadratic dependence on the dataset size becomes a limitation to large-scale learning. Ref.~\cite{tiblias2025efficient} avoids this pairwise construction by encoding each input directly into a Hamiltonian and using its energy as the model output. Each example can then be processed independently, so the number of dataset-dependent evaluations scales as $\BigO{\mathcal{D}}$ for fixed model size and a fixed number of classes.This linear dependence on size of the dataset makes the Hamiltonian approach more suitable for larger datasets compared to methods that require the complete Gram matrix.

\begin{figure*}[h!]
    \centering
    \includegraphics[width=0.7\linewidth]{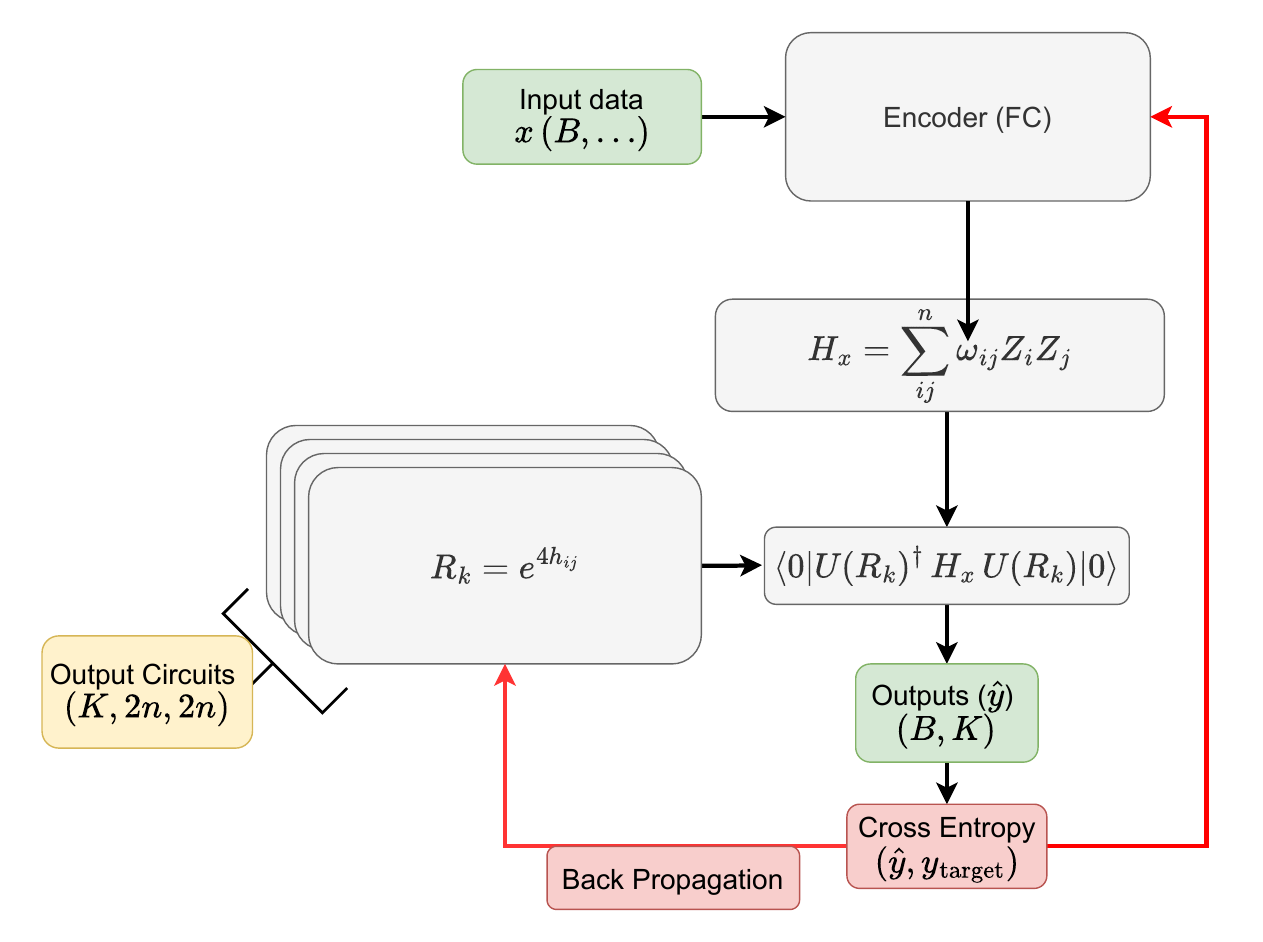}
    \caption{
    \textbf{The NIF-Ising model for classification}.
    First, the input data are encoded into an Ising Hamiltonian by predicting the weights $\omega_{ij}$ of the $ZZ$-interactions using an encoder, implemented here as a FC layer. Next, we construct $K$ free-fermion circuits by defining $K$ parameterized  SPTMs, $R_k \in \mathbb{R}^{2n \times 2n}$. The corresponding $K$ energies are then computed and mapped to probability densities via a softmax function. The model parameters are optimized using gradient descent, with gradients obtained through backpropagation. In this figure, the dimension $B$ denotes the batch size sampled from the dataset. The trainable parameters consist of those of the encoder and those of the SPTMs, amounting to $K\cdot \left(2n^2 - n\right)$ parameters associated with the Hermitian matrix $h$. The number of qubits $n$ is treated as a hyperparameter.}
    \label{fig:nif-ising-cls}
\end{figure*}

Following the same approach, we introduce the NIF-Ising classifier shown in Fig.~\ref{fig:nif-ising-cls}. For $n$ qubits, the model includes one coupling for each unordered pair $(i,j)$ with $i<j$. A classical fully connected (FC) encoder maps the input $\bm{x}\in\mathbb{R}^{d}$ to the coupling vector
\begin{equation}
    \bm{\omega}(\bm{x})
    =
    f_{\Theta_{\mathrm{enc}}}(\bm{x})
    =
    \varphi\!\left(
        W\widetilde{\bm{x}}+\bm{b}
    \right)
    \in\mathbb{R}^{M},
    \label{eq:nif-ising-encoder}
\end{equation}
where $\widetilde{\bm{x}}$ is the batch-normalized input,
$W\in\mathbb{R}^{M\times d}$, $\bm{b}\in\mathbb{R}^{M}$ are the
trainable weights and bias while $\varphi (\cdot)$ is the activation function which is either the hyperbolic-tangent or the identity. Here, $M=\binom{n}{2}$ denotes the number of such
couplings, where $n$ is the number of qubits. The encoder parameters are denoted collectively
by $\Theta_{\mathrm{enc}}=\{W,\bm{b}\}$. The entries of $\bm{\omega}(\bm{x})$ are indexed by qubit pairs:
\begin{equation}
    \bm{\omega}(\bm{x})
    =
    \bigl(
        \omega_{12}(\bm{x}),
        \omega_{13}(\bm{x}),
        \ldots,
        \omega_{n-1,n}(\bm{x})
    \bigr)^T.
    \label{eq:nif-ising-coupling-vector}
\end{equation}
Thus, $\omega_{ij}(\bm{x})$ is the component associated with the interaction between qubits $i$ and $j$. These couplings define the
input-dependent Ising Hamiltonian
\begin{equation}
    H(\bm{x})=
    \sum_{1\leq i<j\leq n}
    \omega_{ij}(\bm{x})Z_iZ_j.
    \label{eq:nif-ising-hamiltonian}
\end{equation}
For a classification problem with $K$ classes, the model contains $K$ trainable free-fermion circuits. The $k$-th circuit prepares a fermionic Gaussian state $\ket{\psi_k}$ associated with class $k$. The energy assigned to input $\bm{x}$ and class $k$
is 
\begin{equation}
    E_k(\bm{x})=\langle\psi_k|H(\bm{x})|\psi_k\rangle=
    \sum_{1\leq i<j\leq n}
    \omega_{ij}(\bm{x})
    \langle\psi_k|Z_iZ_j|\psi_k\rangle.
    \label{eq:nif-ising-class-energy}
\end{equation}
The class probabilities are obtained by treating the negative energies as softmax logits. The encoder and circuit parameters are optimized jointly so that the correct class is assigned a lower energy.

In NIF-Ising, the class-dependent circuits do not depend on $\bm{x}$, while their gate angles are trainable model parameters. Since $\Gamma$ (the covariance matrix defined in~\eqref{eq:encoded-gaussian-cov}) depends smoothly on these angles, gradients with respect to the model parameters follow from the chain rule through the encoding functions and the SPTMs. NIF-Ising can therefore be viewed as an energy-based classifier in a learned space of Ising couplings. The encoder maps each input to the couplings $\bm{\omega}(\bm{x})$, while the class-dependent SPTMs determine the fermionic correlation patterns against which these couplings are evaluated. Restricting the class states to fermionic Gaussian states limits the accessible correlation patterns, but allows the energies and their gradients to be evaluated using polynomial-size single-particle representations without constructing the full many-body state vector.

\subsection{Regression Models}
\label{subsec:regression-models}

The regression benchmark contains five fermionic models and two classical baselines, presented concisely in Table~\ref{tab:regression-model-summary}. The fermionic models are divided into two families according to the kernel used to compare the encoded Gaussian states. Four of the models use Pfaffian covariance kernels with GPR and one of them uses the fidelity kernel with SVR. For the classical baselines we use RBF kernels with the same two regression methods.
\subsubsection{Covariance-kernel GPR models}
\label{subsec:regression-covariance-models}

The covariance-kernel regression models use the Pfaffian feature maps defined
in Sec.~\ref{subsec:pfaffian-covariance-kernel}. Models initialized in a
definite-parity state retain only the even-order sectors $m\in\{2,4\}$,
because parity-preserving matchgate evolution gives vanishing expectation
values for all odd products of Majorana operators. The Plus models instead
use the parity-breaking reference state $\ket{+}^{\otimes n}$ together with
the generalized first-moment representation, allowing us to include the sector sets
\begin{equation}
    M_s\in
    \left\{
        \{1,2,3\},
        \{1,2,3,4\}
    \right\}.
    \label{eq:definitionM}
\end{equation}
We restrict the benchmark to sectors of order at most four, as higher sectors are increasingly expensive. Each resulting covariance kernel is used with Gaussian-process regression. The kernel parameters are held fixed, while the diagonal regularization parameter $\alpha\in\{10^{-6},10^{-4},10^{-3},10^{-2}\}$ is selected using validation data. Further details are provided in the SM.
\paragraph{Single-sector models.}
\textsc{AngleNIFGP} uses the parameter-free
\texttt{SptmAngleEmbedding}~\cite{Gince_MatchCake_A_Python_2026}, in which the transformed input features determine
the angles of successive $R_Y$ and $R_X$ matchgate layers. The covariance kernel retains one Majorana order, selected from $m\in\{2,4\}$. \textsc{VanillaNIFGP} replaces the direct angle embedding with the vanilla encoder. A linear map sends the input to a real antisymmetric generator $h(\bm{x})$, from which the single-particle transformation $R(\bm{x})=\exp[h(\bm{x})]\in\mathrm{SO}(2n)$ is constructed. The kernel again retains one order, selected from $m\in\{2,4\}$. 
\paragraph{Multi-sector models.}
AngleMpf \textsc{+} and \textsc{VanillaMpf+} use angle and vanilla encoders, along with the reference state
$\ket{+}^{\otimes n}$ respectively. Since this state is not a fermionic-parity eigenstate, the corresponding representation can contain both odd- and
even-order Majorana features. The retained sectors are selected as given in $M_s$ in Eq.~\eqref{eq:definitionM}. 
Further details of the odd-order Pfaffian construction are given in the SM.
\subsubsection{Fidelity-kernel SVR model}
\label{subsec:regression-fidelity-model}

\textsc{VanillaSVR} uses the vanilla encoder but compares the resulting Gaussian states through the fermionic fidelity kernel.

The corresponding Gram matrix is supplied to an $\varepsilon$-SVR using the
precomputed-kernel interface. The regularization parameter $C$ and the width
$\varepsilon$ of the insensitive-loss region are selected from
\begin{align}
\nonumber
    C &\in \{0.1,1,10,100\},\\ 
    \varepsilon &\in \{0.01,0.1,1\}.
\end{align}
\subsubsection{Classical regression baselines}
\label{subsec:classical-regression-baselines}

Two RBF-kernel models provide classical reference points.
\textsc{ClassicalGP} uses
\begin{equation}
    k_{\mathrm{classical}}(\bm{x},\bm{x}')
    =
    k_{\mathrm{RBF}}(\bm{x},\bm{x}';\ell)
    +
    k_{\mathrm{white}}(\bm{x},\bm{x}'),
\end{equation}
with the RBF length scale and white-noise level optimized by maximizing the log marginal likelihood. The initial length scale is selected from
$\ell_0\in\{0.1,1,10\}$. The \textsc{ClassicalSVR} uses an RBF kernel with the factor $\gamma$ set to ``scale". The $C$ and $\varepsilon$ values are selected from the same grids used for \textsc{VanillaSVR}, providing a matched regularization budget for the two SVR models.

As in the classification benchmark section, the model definitions and the hyperparameter search serve separate purposes. Table~\ref{tab:regression-model-summary} specifies how each regression model represents the input data, while Table~\ref{tab:regression_hyperparameter_search_summary} summarizes the hyperparameter search summary for these regression models.

\begin{table*}[t]
    \centering
    \small
    \begin{tabular}{llllll}
        \toprule
        \textbf{Family}
        & \textbf{Model}
        & \textbf{Encoder}
        & \textbf{Reference state}
        & \textbf{Kernel}
        & \textbf{Regressor} \\
        \midrule

        Fermionic covariance
        & \textsc{AngleNIFGP}
        & Angle
        & $|0\rangle^{\otimes n}$
        & Single sector, $m\in\{2,4\}$
        & GPR \\

        Fermionic covariance
        & \textsc{VanillaNIFGP}
        & Vanilla
        & $|0\rangle^{\otimes n}$
        & Single sector, $m\in\{2,4\}$
        & GPR \\

        Fermionic covariance
        & \textsc{AngleMpf+}
        & Angle
        & $\ket{+}^{\otimes n}$
        & $M_s \in
          \{\{1,2,3\},\{1,2,3,4\}\}$
        & GPR \\

        Fermionic covariance
        & \textsc{VanillaMpf+}
        & Vanilla
        & $\ket{+}^{\otimes n}$
        & $M_s\in
          \{\{1,2,3\},\{1,2,3,4\}\}$
        & GPR \\

        Fermionic fidelity
        & \textsc{VanillaSVR}
        & Vanilla
        & $|0\rangle^{\otimes n}$
        & Fidelity
        & SVR \\

        \midrule

        Classical
        & \textsc{ClassicalGP}
        & Direct input
        & --
        & RBF $+$ white noise
        & GPR \\

        Classical
        & \textsc{ClassicalSVR}
        & Direct input
        & --
        & RBF
        & SVR \\

        \bottomrule
    \end{tabular}
    \caption{
        \textbf{Regression models included in the benchmark}. The fermionic covariance
        models differ in their encoder, reference state, and retained
        Majorana-correlation sectors. The fidelity and classical RBF kernels
        are evaluated using SVR, while the covariance kernels and classical
        RBF-plus-white-noise kernel are used with GPR.
    }
    \label{tab:regression-model-summary}
\end{table*}
\begin{table*}
\centering
\begin{tabular}{@{}lrrrlr@{}}
  \toprule
  Model & Points & Runs & Best Ever & Most often best & $W$ \\
  \midrule
  ClassicalGP & 15 & 4\,800 & 1 & $\ell=1$, $s=\mathrm{standard}$ & 4 \\
  ClassicalSVR & 60 & 19\,200 & 3 & $C=100$, $\epsilon=0.01$, $s=\mathrm{standard}$ & 2 \\
  VanillaSVR & 60 & 19\,200 & 8 & $C=0.1$, $\epsilon=0.1$, $s=\mathrm{quantile}$ & 2 \\
  AngleNIFGP & 40 & 12\,800 & 5 & $m=4$, $\alpha=0.001$, $s=\mathrm{none}$ & 4 \\
  VanillaNIFGP & 40 & 12\,800 & 9 & $m=4$, $\alpha=0.001$, $s=\mathrm{minmax01}$ & 4 \\
  AngleMpf$+$ & 40 & 12\,800 & 10 & $M_s=\{1,2,3\}$, $\alpha=10^{-6}$, $s=\mathrm{minmax01}$ & 4 \\
  VanillaMpf$+$ & 40 & 12\,800 & 10 & $M_s=\{1,2,3,4\}$, $\alpha=0.001$, $s=\mathrm{minmax01}$ & 2 \\
  \midrule
  Total & 295 & 94\,400 & & & \\
  \bottomrule
\end{tabular}
\caption{\textbf{Regression hyperparameter-search summary.}
For each model, \emph{Points} gives the number of hyperparameter configurations in the search grid, and \emph{Runs} gives the total number of model fits across all dataset configurations and train-validation splits. \emph{Best ever} counts the grid points that achieved the highest mean validation $R^2$ for at least one dataset configuration. \emph{Most often best} reports the configuration selected most frequently, and $N_{W}$ is the number of dataset configurations on which that point achieved the best validation score. Here, $s$ denotes the input transformation, $\ell$ the Gaussian-process length scale, $m$ the Majorana order, $M_s$ the set of retained Majorana orders, and $\alpha$ the diagonal regularization parameter.}
\label{tab:regression_hyperparameter_search_summary}
\end{table*}
\section{Benchmark design}
\label{sec:benchmarks}

We evaluate FermiML models in three settings. First, the classification benchmark compares fermionic models with unrestricted quantum and classical
models across several synthetic datasets. Second, the
NIF-Ising benchmark evaluates a trainable fermionic Hamiltonian classifier on larger text and image classification datasets. Third, the regression benchmark compares fermionic fidelity and covariance kernels with classical regression models. Because these experiments use different datasets, training
procedures, and statistical comparisons, their evaluation protocols are described separately below.
\subsection{Classification benchmark}
\label{subsec:classification-benchmark}
\subsubsection{Datasets and configurations}
\label{subsec:classification-datasets}

We use five binary-classification datasets from Ref.~\cite{bowles2024betterclassicalsubtleart}, loaded through
PennyLane~\cite{bergholm2022pennylaneautomaticdifferentiationhybrid}.
Table~\ref{tab:classification-data} summarizes the \(31\) configurations,
which form the basis of the paired comparisons in Sec.~\ref{sec:results}. i) \textbf{Linearly separable~:}
The inputs are sampled from a $d_I$-dimensional hypercube and divided into two classes by a hyperplane. This family tests whether the models can recover a simple linear decision boundary. ii) \textbf{Hidden manifold~:}
The inputs are generated from a low-dimensional latent manifold and embedded in a $d_a$-dimensional space, while the labels are assigned by a randomly initialized neural network acting on the latent coordinates. The classification rule therefore depends on the hidden structure of the data. iii) \textbf{Hidden manifold-diff~:}
This family uses the same generator but varies the dimension $m_{l}$ of the latent manifold while keeping the ambient dimension ($d_a$) fixed. It tests how the model performance changes as the intrinsic dimension of the data increases. iv) \textbf{Bars-and-stripes~:}
This family contains noisy $4\times4$ images formed from horizontal bars or
vertical stripes. The label is determined by the overall spatial pattern of the image. We use one configuration from this family. v) \textbf{Downscaled MNIST~:} MNIST images are reduced to $d_m=2,\ldots,20$ principal components via principal component analysis (PCA) producing nineteen configurations. For each split, $250$ training and $250$ test examples are sub-sampled using the split-specific random seed, so different splits may contain different examples. The complete set of configurations used in this work is summarized in
Table~\ref{tab:classification-data}.

\begin{table*}[t]
    \centering
    \caption{\textbf{Classification benchmark configurations for kernel methods.}
    This table discusses the datasets and parameter settings used in the classification benchmark. Each setting defines one dataset configuration,
    giving \(31\) configurations in total. Paired comparisons are performed
    across this common set of configurations.}
    \label{tab:classification-data}
    \begin{tabular}{@{}lll@{}}
        \toprule
        \textbf{Dataset family} & \textbf{Swept variable} & \textbf{Settings} \\
        \midrule
        Linearly separable
        & Input dimension \(d_I\)
        & \(10,15,20\) \\

        Hidden manifold
        & Ambient dimension \(d_a\)
        & \(2,6,10,20\) \\

        Hidden manifold-diff
        & Latent dimension $m_l$
        & \(4,6,10,14\) \\

        Bars-and-stripes
        & Image size
        & \(4\times4\) \\

        Downscaled MNIST
        & Number of PCA components \(d_m\)
        & \(2,3,\ldots,20\) \\
        \bottomrule
    \end{tabular}
\end{table*}
\begin{table*}[t]
    \centering
    \caption{\textbf{Regression benchmark configurations.}
    This table discusses datasets and parameter settings used in the regression benchmark.
    Each combination of settings defines one configuration, giving \(16\)
    configurations in total. All regression models are evaluated on this
    common set of configurations.}
    \label{tab:regression-data}
    \begin{tabular}{@{}lll@{}}
        \toprule
        \textbf{Dataset family}
        & \textbf{Swept variables}
        & \textbf{Settings} \\
        \midrule

        Friedman~\#1
        & Sample size \(N_s\); noise \(\sigma\)
        & \(N_s\in\{200,500\}\);
          \(\sigma\in\{0,0.1\}\) \\

        California Housing
        & Training-set size \(N_{\mathrm{tr}}\)
        & \(N_{\mathrm{tr}}\in
          \{500,2000,8000,16512\}\) \\

        Magnetization
        & Modes \(n_{m}\); sample size \(N_s\); noise \(\sigma\)
        & \(n_{m}\in\{4,8\}\);
          \(N_s\in\{200,500\}\);
          \(\sigma\in\{0,0.1\}\) \\

        \bottomrule
    \end{tabular}
\end{table*}
\subsubsection{Preprocessing and model selection}
\label{subsubsec:classification-selection}

All models are evaluated on the same training and test data for each dataset configuration. We use 20 reproducible stratified shuffle splits, where stratification preserves approximately the same class proportions in the training and validation subsets as in the original dataset. For each split, the pre-processing transformation is fitted exclusively on the training subset. The fitted parameters are kept fixed when transforming the validation and test subsets. This prevents information from these subsets from entering model fitting or selection.

For the kernel and random-feature classifiers, model selection is performed by exhaustive grid search. Every pre-defined hyperparameter configuration is
fitted and evaluated on each of the twenty splits. Hyperparameters are ranked by their mean validation accuracy across the splits. If several configurations
have the same mean validation accuracy, a fixed ordering of the hyperparameters is used to select one of them. This makes the selection
independent of the order in which the rows are stored or processed. Table~\ref{tab:classification-hyperparameter-summary}
summarizes the size of each model's search space and the configurations most
frequently selected. The complete results for every hyperparameter point are
reported in the SM.

\subsubsection{Evaluation and statistical comparisons}
\label{subsubsec:classification-statistics}

Classification performance is quantified via the test accuracy
\begin{equation}
    \operatorname{Acc}
    =\frac{\text{number of correctly classified test examples}}
         {\text{total number of test examples}}.
\end{equation}
This is the quantity displayed in the figures of the results section in the main text.
For each model and dataset configuration, we retain the test accuracies
obtained over the 20 splits after selecting the hyperparameters using
validation accuracy.
Training accuracies are also reported in the SM to show the generalization performance of these models.

We use two complementary statistical comparisons. In each case, the reported $p$-value is calculated from the corresponding two-sided statistical test. It is the probability, assuming that there is no systematic performance difference between the two models, of obtaining a difference at least as extreme as the one observed. A small $p$-value therefore indicates that the
observed difference would be unlikely under this null hypothesis and it does not give the probability that either model is better.

The \emph{pooled test} compares two models on a given dataset. Their retained test accuracies are pooled over all configurations and splits
belonging to that dataset and compared using a two-sided Mann-Whitney
test~\cite{mann1947test}. The test evaluates whether the scores of one model
tend to be larger than those of the other without pairing individual runs.
Since the pooled samples can combine configurations of different difficulty
and overlapping data partitions, the resulting $p$-values are interpreted
as descriptive summaries.

The \emph{paired test} compares two models across the entire benchmark.
For each model, the test accuracies are first averaged over the 20 splits
within each of the 31 dataset configurations. For each dataset configuration \(q\), we compare a fermionic model $F$ with a non-fermionic baseline $B$ using their mean test accuracies over
the 20 splits. We define the paired difference as
\begin{equation}
    \Delta_q
    =
    \overline{\operatorname{Acc}}_{B,q}
    -
    \overline{\operatorname{Acc}}_{F,q},
    \qquad q=1,\ldots,31,
    \label{eq:classification-paired-difference}
\end{equation}
where \(\overline{\operatorname{Acc}}_{B,q}\) and
\(\overline{\operatorname{Acc}}_{F,q}\) denote the mean test accuracies of
the baseline and fermionic models, respectively, on configuration \(q\).
Under this convention, \(\Delta_q<0\) favours the fermionic model, whereas
\(\Delta_q>0\) favours the non-fermionic baseline. The plot in Fig.~\ref{fig:all_plots_forest_classification} reports the mean paired difference
\begin{equation}
    \overline{\Delta}=
    \frac{1}{31}\sum_{q=1}^{31}\Delta_q.
    \label{eq:classification-mean-paired-difference}
\end{equation}
We summarize the $31$ configuration-level differences by their mean
$\overline{\Delta}$, with a $95\%$ confidence interval obtained from
$10^4$ percentile-bootstrap resamples~\cite{efron1994bootstrap}. A
two-sided Wilcoxon signed-rank test is applied to the same differences.
Because several configurations belong to the same dataset family, these
statistics describe the consistency of the observed performance difference
across the benchmark rather than variation across independent datasets.

The win, tie, and loss counts $W$, $T$, and $L$ denote the numbers of
configurations for which the non-fermionic baseline performs better than, ties with, or performs worse than the fermionic model, respectively. Thus, a win corresponds to $\Delta_q>0$, and $W+T+L=31$ when results are available for every configuration. Statistical significance is denoted by $^{*}$ for $p<0.05$, $^{**}$ for $p<0.01$, and $^{***}$ for $p<0.001$. ``ns'' denotes $p\geq0.05$ or an undefined test. 
\subsection{Trainable NIF-Ising benchmark}
\label{subsec:nifising-benchmark}

\subsubsection{Datasets}
\label{subsubsec:nifising-datasets}

The NIF-Ising classifier is evaluated separately from the kernel-based classification benchmarks on the IMDb sentiment dataset~\cite{maas2011learning}, the AG News topic-classification dataset~\cite{zhang2015character}, CIFAR2, a binary subset of CIFAR-10~\cite{krizhevsky2009learning}, and Fashion-MNIST~\cite{xiao2017fashion}. These datasets allow our trainable fermionic model to be compared with the classical and quantum models reported in Ref.~\cite{tiblias2025efficient}.

\subsubsection{Training and hyperparameter selection}
\label{subsubsec:nifising-selection}

NIF-Ising is computationally more expensive to train compared to the kernel models because its encoder and class-dependent free-fermion transformations are optimized by gradient descent. We therefore use a staged hyperparameter-selection procedure rather than an exhaustive full-budget grid search. This procedure uses Bayesian hyperparameter optimization (BO) from the ax platform~\cite{olson2025ax}.

Candidate hyperparameters are first evaluated using BO for a fixed number of trials (20) using a reduced training budget (128 epochs). The configuration with the highest validation accuracy is then trained again using the full epoch budget (1024 epochs). This procedure allows a broader range of configurations to be examined without training every candidate to the maximum number of epochs. The selection of hyperparameters uses only the training and validation data; the test set is evaluated after the configuration has been selected. The selected hyperparameters for every split are reported in the SM in Tables~\ref{tab:hparams-agnews}-\ref{tab:hparams-imdb}.
\subsubsection{Evaluation}
\label{subsubsec:nifising-evaluation}

For each dataset, we report the mean and standard deviation of the training, validation, and test accuracies across the NIF-Ising data splits. These
results are compared with the classical and quantum models reported in
Ref.~\cite{tiblias2025efficient}. Because the reference results and the NIF-Ising results were obtained using separate experimental pipelines, these comparisons are descriptive in nature.
\subsection{Regression benchmark}
\label{subsec:regression-benchmark}

\subsubsection{Datasets and configurations}
\label{subsubsec:regression-datasets}

We evaluate the regression models on three datasets: Friedman~\#1, California Housing, and a fermionic magnetization dataset. The magnetization dataset is designed to test the covariance kernel when the target is directly represented by its fermionic features. These datasets are summarized in Table~\ref{tab:regression-data}. 
\paragraph{Friedman~\#1.}

Friedman~\#1 is a synthetic nonlinear regression benchmark with ten input features~\cite{friedman1991,breiman1996bagging_freidman}. Each feature $x_i$ is drawn independently from the uniform distribution in $[0,1]$. The noiseless target is
\begin{equation}
y(\bm{x})= 10\sin\!\left(\pi x_1x_2\right)+20\left(x_3-\frac{1}{2}\right)^2+
10x_4
+
5x_5 \,,
\label{eq:friedman-target}
\end{equation}
so only the first five features are informative; the remaining five act as irrelevant covariates. Noise $\eta$ is added to the target to obtain the noisy observations
\begin{equation}
y_{\mathrm{obs}}(\bm{x}) = y(\bm{x})+\eta,
\qquad
\eta\sim\mathcal{N}(0,\sigma^2).
\label{eq:friedman-noisy-target}
\end{equation}
We consider sample sizes $N_s\in \{200,500\}$ and noise levels $\sigma\in \{0,0.1\}$. This dataset tests the fermionic kernels on a nonlinear target that is not
constructed to match their feature representation.
\paragraph{California Housing.} The California Housing dataset is a real-world regression benchmark derived from demographic and geographic information of the California census districts~\cite{pace1997_Calhousing,pedregosa2018scikitlearnmachinelearningpython}. It contains $20{\small,}640$ samples described by eight real-valued features, including median income, housing occupancy, population, geographic coordinates, and statistics of the local housing stock. The target is the median house value of the corresponding district. We first separate the dataset into a training pool of $16{\small,}512$ samples and an independent test set of $4{\small,}128$ samples. To study the dependence on training-set size while keeping the test distribution fixed, we use training subsets containing
\begin{equation}
N_{\mathrm{tr}}\in\left\{500,\,2000,\,8000;\,16512\right\}
\label{eq:california-training-sizes}
\end{equation}
samples. The eight input variables have substantially different physical units and numerical ranges. California Housing therefore also tests the sensitivity of each model to the input transformation selected during validation.

\paragraph{Magnetization.}
The magnetization dataset is a quantum-native synthetic benchmark designed to test the fermionic covariance kernel under a controlled match between the target and the model's feature space. Each input is a $n$-dimensional vector
\begin{equation}
\bm{x} = (x_1,\ldots,x_n)^T,
\qquad
x_j\sim\mathcal{U}(0,\pi),
\label{eq:magnetization-inputs}
\end{equation}
whose components are used as the rotation angles of a one-layer $R_Y$-$R_X$ matchgate embedding. Denoting the resulting fermionic Gaussian state by $\rho(\bm{x})$, the noiseless target is the total Pauli-$Z$ magnetization
\begin{equation}
y(\bm{x})=\sum_{j=1}^{n}
\Tr\!\left[
\rho(\bm{x})Z_j
\right] = \Tr\!\left[
\rho(\bm{x})
\sum_{j=1}^{n}Z_j
\right].
\label{eq:magnetization-target}
\end{equation}
Each $Z_j$ is quadratic in the Majorana operators and is therefore a second-order Majorana observable. Consequently, the target in Eq.~\eqref{eq:magnetization-target} belongs to the feature space of the order-2 covariance kernel when the same angle embedding and number of modes are used. In the absence of added noise, deviations from perfect prediction can, therefore, be attributed to finite-sample estimation, regularization, or numerical effects. We consider $n_{m}\in\{4,8\}$ fermionic modes and sample sizes $N_s\in\{200,500\}$. Noisy targets are generated according to
\begin{equation}
y_{\mathrm{obs}}(\bm{x}) = y(\bm{x})+\eta,
\qquad
\eta\sim\mathcal{N}(0,\sigma^2),
\label{eq:noisy-magnetization-target}
\end{equation}
with $\sigma\in \{0,0.1\}$. The targets are calculated using the matchgate simulator and independently checked against exact state-vector calculations.

\subsubsection{Preprocessing and model selection}
\label{subsubsec:regression-selection}

The training pool is divided using $20$ reproducible \emph{ShuffleSplit} partitions, with $20\%$ of each partition reserved for validation. Hyperparameters are selected separately within each split. All seven regression models are evaluated using the same five input pre-processing choices: no transformation, min--max scaling to $[0,1]$,
min--max scaling to $[-1,1]$, standardization to zero mean and unit variance, and a quantile transformation. Each transformation is fitted using
only the fitting portion of the split and is then applied, without refitting,
to the corresponding validation and test data.

The regression targets are standardized in the same way, using statistics computed from the fitting portion only. Predictions are subsequently returned
to the original target scale and clipped to the range of target values observed in the fitting data.

For every kernel-based regression model, all configurations in the pre-defined
hyperparameter grid are fitted on the fitting portion and evaluated on the validation portion. The configuration with the highest validation coefficient of determination, $R^2$, is selected within that split. Its performance is then evaluated on the held-out test set. The test set is not
used to select the preprocessing transformation, kernel parameters, or regression hyperparameters.

For Gaussian-process regression, the kernel parameters are treated as model choices and are not optimized by maximizing the log-marginal-likelihood. A diagonal regularization parameter is included in the validation grid. Further
details concerning the Gaussian-process implementation and numerical
conditioning are provided in the SM.

\subsubsection{Evaluation and statistical comparisons}
\label{subsubsec:regression-statistics}

Regression performance is evaluated using $R^2$, mean absolute error (MAE), mean-square error (MSE), and root-mean-square error (RMSE). For the Gaussian-process models, we additionally evaluate predictive uncertainty using the negative log predictive density and the empirical coverage of the nominal $68\%$ and $95\%$ posterior intervals.

Statistical comparisons are performed separately for each dataset. If a
dataset contains multiple configurations, their test $R^2$ values are first averaged within each split. For a fermionic model $F$ and classical reference
$C$, we then define
\begin{equation}
    \Delta_s^{(R^2)}
    =
    \overline{R}^{\,2}_{C,s}
    -
    \overline{R}^{\,2}_{F,s},
    \qquad
    s=1,\ldots,20,
    \label{eq:regression-paired-difference}
\end{equation}
where the bar denotes the average over the configurations included in the comparison and $s$ indexes the splits. Negative values favor the fermionic model, while positive values favor the classical reference. We report the median difference, a $95\%$ confidence interval obtained from $10^4$ bootstrap resamples, a two-sided
Wilcoxon signed-rank $p$-value, and split-level win/tie/loss counts.

Since the 20 splits use different partitions of the same training pool and
the same held-out test set, these statistics measure sensitivity to the
train--validation partition rather than variation across independent
datasets. All other statistical conventions follow
those detailed in Sec.~\ref{subsubsec:classification-statistics}.

\begin{table*}[h!]
    \centering
    \begin{tabular}{c|ccc|ccc}
        \toprule
        & \multicolumn{3}{c}{\textbf{IMDb}}
        & \multicolumn{3}{c}{\textbf{AG News}} \\
        \textbf{Model} & \textbf{Train} & \textbf{Val} & \textbf{Test} & \textbf{Train} & \textbf{Val} & \textbf{Test} \\
        \midrule
        LOG   & 85.8 & -- & 85.5 & 90.1 & -- & 89.2 \\
        MLP   & 88.1 & -- & 85.8 & 90.2 & -- & 89.1 \\
        LSTM  & \textbf{99.3} & -- & \textbf{88.4}
              & $91.5$ & -- & \textbf{90.5} \\
        RNN   & 79.7 & -- & 78.1 & 88.8 & -- & 88.1 \\
        CIRC  & 86.1 & -- & 85.8 & 89.7 & -- & 89.2 \\
        QLSTM & 76.8 & -- & 67.9 & 87.2 & -- & 86.7 \\
        \midrule
        HAM   & \textit{91.0} & -- & \textit{88.1} & -- & -- & -- \\
        PEFF  & 84.0 & -- & 83.8 & -- & -- & -- \\
        SIM   & 86.0 & -- & 85.3 & 89.4 & -- & 88.6 \\
        \midrule
        NIF-Ising
              & $87.3 \pm 0.5$
              & $85.7 \pm 0.4$
              & $85.5 \pm 0.3$
              & $\mathbf{92.8 \pm 0.5}$
              & $90.5 \pm 0.3$
              & \textit{89.8} $\pm$ \textit{0.2}\\
        \bottomrule
    \end{tabular}
    \caption{\textbf{Text classification performance.} Train, validation, and test accuracies (\%) on IMDb and AG News. Results for LOG, MLP, LSTM, RNN, CIRC, QLSTM, HAM, PEFF, and SIM are taken from Ref.~\cite{tiblias2025efficient}. Note that the latter three are quantum models from ~\cite{tiblias2025efficient}. These results are averages over ten training runs, and validation scores were not reported. NIF-Ising results are reported as the mean $\pm$ standard deviation over twenty train-validation splits. Dashes indicate unreported results. The best performance among quantum and quantum-inspired models are italicized whereas the best results across all are marked in bold.}
    \label{tab:nifising-text-comparison}
\end{table*}

\begin{table*}[h!]
    \centering
    \begin{tabular}{cccc|ccc}
        \toprule
         & \multicolumn{3}{c}{\textbf{CIFAR2}} & \multicolumn{3}{c}{\textbf{Fashion}} \\
         \textbf{Model} & \textbf{Train} & \textbf{Val} & \textbf{Test} & \textbf{Train} & \textbf{Val} & \textbf{Test} \\
         \hline
         LOG & 73.9 & - & 72.9 & 85.5 & - & 83.5 \\
         MLP & 89.0 & - & 83.4 & 87.9 & - & 85.8 \\
         CNN & \textbf{96.3} & - & \textbf{94.1} & \textbf{95.4} & - & \textbf{90.7} \\ 
         CIRC & 85.1 & - & 84.6 & 89.2 & - & 86.5\\
         QCNN & 84.9 & - & 84.7 & 76.7 & - & 76.3 \\
         \hline
         HAM & 89.6 & - & 81.0 & - & - & - \\
         PEFF & 79.3 & - & 78.7 & - & - & - \\
         SIM & 68.9 & - & 68.2 & 87.6 & - & 84.4 \\
         \hline
         NIF-Ising & \textit{90} $\pm$ \textit{2} & $85.6 \pm 0.5$ & \textit{85.5} $\pm$ \textit{0.5} & \textit{94} $\pm$ \textit{2} & $88.9 \pm 0.8$ & \textit{87.8} $\pm$ \textit{0.8} \\
         \bottomrule
    \end{tabular}
    \caption{\textbf{Image classification performance.} Train, validation, and test accuracies (\%) on CIFAR2 and Fashion. Results for LOG, MLP, CNN, CIRC, QCNN, HAM, PEFF, and SIM are taken from Ref.~\cite{tiblias2025efficient}. These results are averages over ten training runs, and validation scores were not reported. NIF-Ising results are reported as the mean $\pm$ standard deviation over twenty train--validation splits. Dashes indicate unreported results. The best performance among quantum and quantum-inspired models are italicized whereas the best results across all are marked in bold.}
    \label{tab:nifising-image-comparison}
\end{table*}
\section{Results}
\label{sec:results}
The classification and regression benchmarks differ in their performance measures, hyperparameter-selection procedures, and units of statistical comparison. We therefore describe their evaluation protocols separately. In both cases, hyperparameters are selected exclusively from validation performance, and the held-out test data are used only for the final evaluation of the selected model. Also note that in the case of classification benchmarks, the violin plots shown below pool configurations from the same datasets with different dimensions. Their shape therefore reflects both intra-configuration differences and variation across splits. We use the dimension-resolved figures in the SM to distinguish these effects.
\subsection{Classification: QML models versus FermiML surrogates}
\label{subsubsec:classification_kernel_results}

\paragraph{Linearly separable~:}

\begin{figure*}
    \centering
    \includegraphics[width=0.92\linewidth]{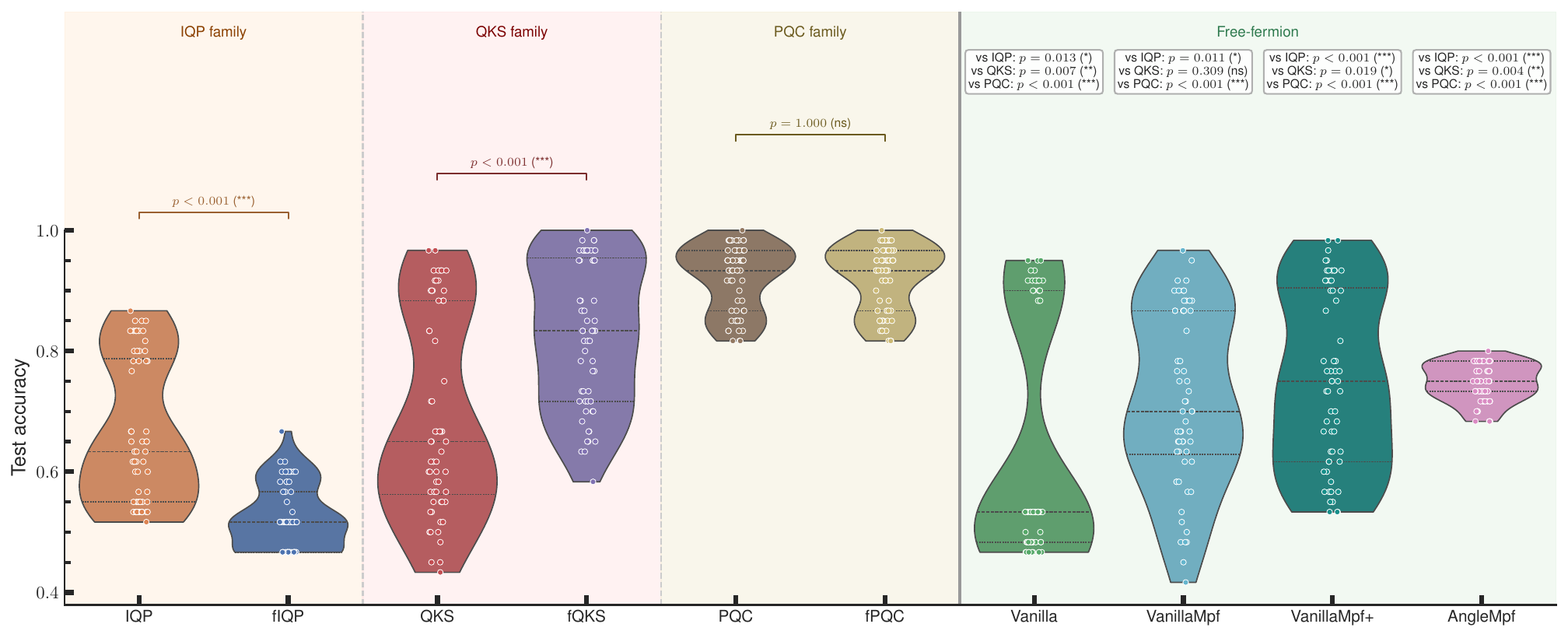}
   \caption{\textbf{Classification performance on the linearly separable datasets.} Each violin pools the test accuracies obtained after validation-based hyperparameter selection over the three input dimensions $d_I\in\{10,15,20\}$ and the twenty data splits. White circles show the individual test scores, while the horizontal lines indicate the median and inter-quartile range. Background shading separates the models into the \textsc{IQP}, \textsc{QKS}, and \textsc{PQC} families and the native free-fermionic models. Brackets compare each unrestricted model with its matchgate counterpart, while the annotations above the native free-fermionic models compare them separately with \textsc{IQP}, \textsc{QKS}, and \textsc{PQC}. All reported $p$-values are obtained from two-sided Mann-Whitney tests applied to the pooled scores. Because the violins combine different input dimensions, their multimodal structure reflects both dimension-dependent performance and variation between data splits. Therefore, they should not be interpreted as the distribution for a single dataset configuration. Significance levels are denoted by $^{*}$ for $p<0.05$, $^{**}$ for $p<0.01$, and $^{***}$ for $p<0.001$. $\mathrm{ns}$ denotes a nonsignificant difference.}
\label{fig:Classification_linearly_separable}
\end{figure*}

Figure~\ref{fig:Classification_linearly_separable} summarizes the classification results for the linearly separable datasets. Each violin combines the results for the input dimension $d_I=10$, $15$, and $20$, together with the twenty data splits at each dimension. The width and shape of each violin reflects two sources of variation: changes in performance with the input dimension and variation among data splits at fixed dimension. In particular, the multiple modes visible for several models are likely associated primarily with different values of $d_I$, rather than with distinct performance regimes for
a single dataset configuration. The pooled distributions provide an overall summary, but the dimension-resolved results in Figures~\ref{fig:classification_kernel_vs_dim}
and~\ref{fig:classification_kernel_diffvs_dim} of the SM are needed to determine how the differences evolve with $d_I$.

The three pairs of matched models exhibit different effects of imposing a matchgate structure. IQP significantly outperforms fIQP ($p<0.001$), whereas \emph{fQKS significantly outperforms QKS ($p<0.001$)}. PQC has the highest pooled accuracies among the three unrestricted models and significantly outperforms fPQC ($p<0.001$). For most dataset configurations, validation selects the full-width circuit, in which all features are loaded in a single rotation layer followed by a fixed entangling layer. Because the same final entangling operation is applied to both states entering the fidelity, its effect is canceled from their overlap.
The resulting fidelity kernel is therefore equivalent to a product of single-qubit kernels and does not depend on the entangling layer.Thus, the strong performance of PQC in this regime cannot be attributed directly to entangling quantum dynamics.

The native free-fermionic models show distinct performance profiles. Vanilla has a strongly multimodal distribution, containing both near-random and high-accuracy scores. Its pooled distribution differs significantly from IQP ($p=0.013$), QKS ($p=0.007$), and PQC ($p<0.001$). VanillaMpf is significantly different from IQP ($p=0.011$) and PQC ($p<0.001$) as well, but is statistically indistinguishable from QKS ($p=0.309$). VanillaMpf+ differs
significantly from all three unrestricted baselines, with
$p<0.001$ against IQP, $p=0.019$ against QKS, and $p<0.001$
against PQC. AngleMpf produces a narrower distribution concentrated at intermediate-to-high accuracies and also differs significantly from IQP ($p<0.001$), QKS ($p=0.004$), and PQC ($p<0.001$).

The figure does not report a direct statistical comparison between VanillaMpf and VanillaMpf+. Therefore, it does not, by itself, establish whether replacing the fixed-parity reference state with $\ket{+}^{\otimes n}$ provides an advantage. More generally, the pooled tests indicate differences across the full dimension sweep but do not show whether those differences are uniform in $d_I$ or driven by a subset of dimensions. This distinction is examined using dimension-resolved results in the SM.
\paragraph{Hidden manifold~:}

\begin{figure*}
    \centering
    \includegraphics[width=0.92\linewidth]{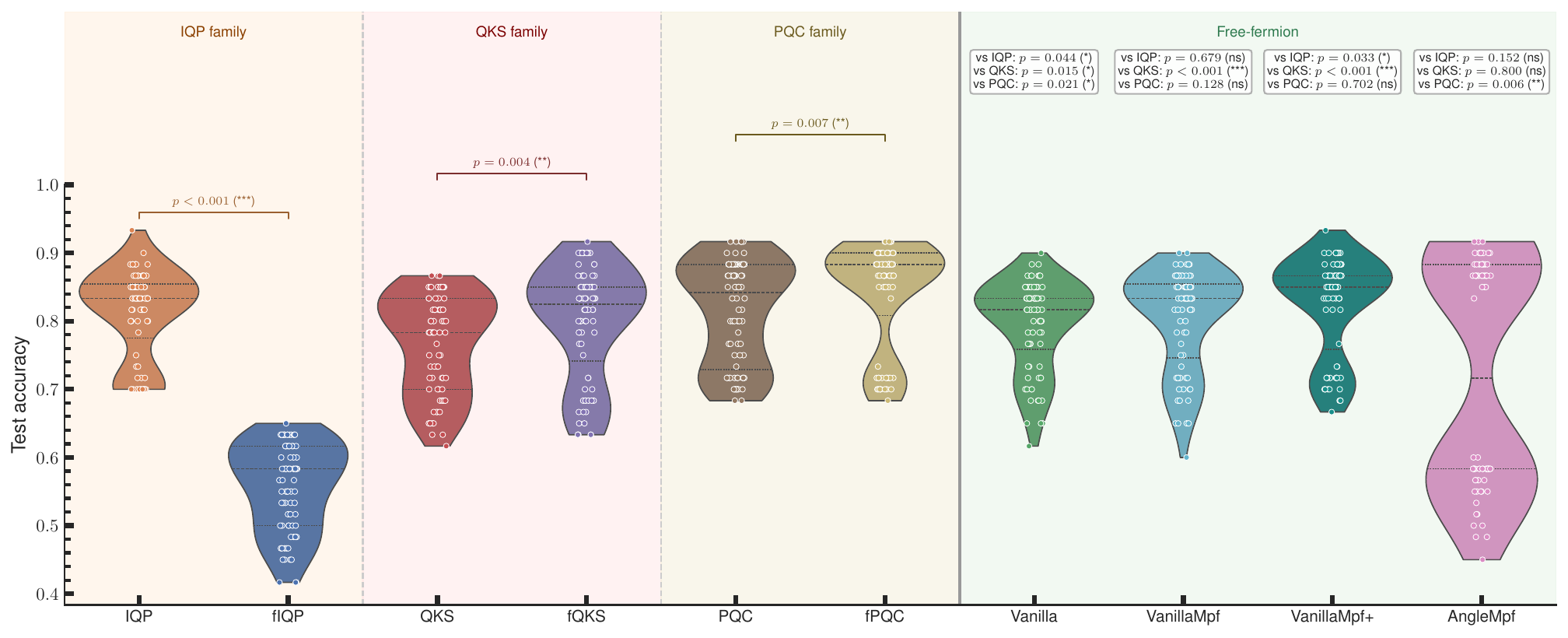}
   \caption{\textbf{Classification performance on the hidden manifold datasets.} Each violin pools the test accuracies obtained after validation-based hyperparameter selection over the ambient dimensions $d_a\in\{2,6,10,20\}$ and the twenty data splits. White circles show the individual test scores, while the horizontal lines indicate the median and inter-quartile range. Background shading groups the models into the \textsc{IQP}, \textsc{QKS}, and \textsc{PQC} families and the native free-fermionic models. Brackets compare each unrestricted model with its matchgate counterpart, while the annotations above the native free-fermionic models compare them separately with \textsc{IQP}, \textsc{QKS}, and \textsc{PQC}. All reported $p$-values are obtained from two-sided Mann-Whitney tests applied to the pooled scores. Because the violins combine different ambient dimensions, their shapes reflect both dimension-dependent performance and variation among data splits at fixed dimension. Significance levels are denoted by $^{*}$ for $p<0.05$, $^{**}$ for $p<0.01$, and $^{***}$ for $p<0.001$. $\mathrm{ns}$ denotes a non-significant difference.}
   \label{Classification_hidden_manifold}
\end{figure*}

Figure~\ref{Classification_hidden_manifold} summarizes the results for the hidden manifold datasets. In this family, the labels are determined by structure in a low-dimensional latent manifold that is subsequently embedded
in an ambient space of dimension $d_a$. The figure pools the results for $d_a=2$, $6$, $10$, and $20$, together with the twenty data splits at each dimension. The violin shapes therefore reflect both changes in performance
with the ambient dimension and variation among splits at fixed dimension. The multiple modes visible for several models are likely associated primarily with different values of $d_a$, rather than with distinct
performance regimes for a single dataset configuration.

Within the IQP family, IQP significantly outperforms fIQP
($p<0.001$). The IQP scores are concentrated mainly at high accuracies, whereas fIQP has a much larger concentration of results near random
classification. Imposing the matchgate restriction on this particular architecture therefore produces a clear reduction in performance on the
hidden manifold datasets. The opposite trend is observed for the QKS pair: fQKS significantly outperforms QKS ($p=0.004$), although their distributions overlap substantially. The PQC and fPQC distributions differ
visibly in shape, with fPQC displaying a particularly strong separation between low- and high-accuracy modes. Nevertheless, their pooled Mann-Whitney comparison is not statistically significant ($p=0.100$). The pooled results therefore provide no evidence of a systematic performance difference between PQC and fPQC across this dataset.

The native free-fermionic models show different relationships with the three unrestricted baselines. Vanilla differs significantly from IQP
($p=0.044$), QKS ($p=0.015$), and PQC ($p=0.021$).
VanillaMpf is statistically indistinguishable from IQP ($p=0.679$) and PQC ($p=0.128$), but differs significantly from QKS ($p<0.001$).
VanillaMpf+ differs significantly from IQP ($p=0.033$) and QKS
($p<0.001$), but not from PQC ($p=0.702$). Thus, on the pooled
hidden-manifold results, the two Vanilla-based Pfaffian models attain
performance comparable to at least one of the stronger unrestricted
baselines.

\begin{figure*}
    \centering
    \includegraphics[width=0.92\linewidth]{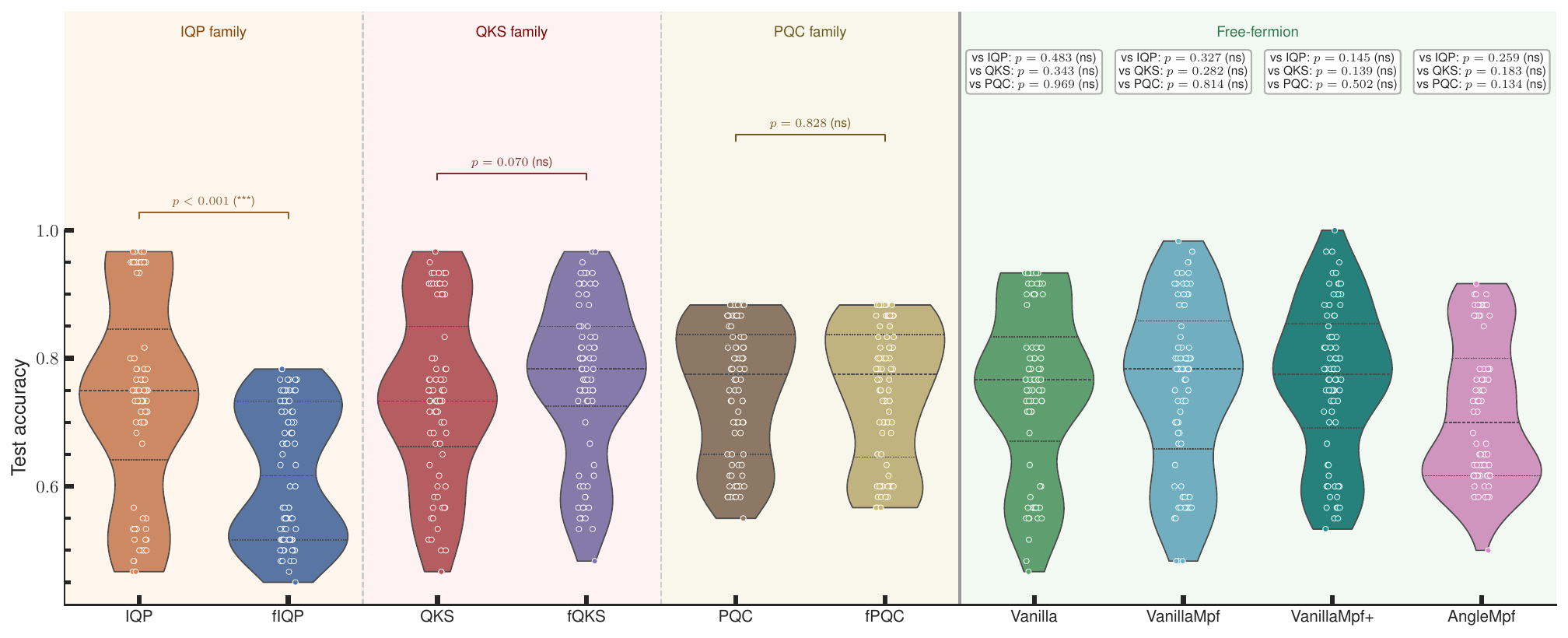}
   \caption{\textbf{Classification performance on the hidden manifold-diff datasets.} Each violin pools the test accuracies obtained after validation-based hyperparameter selection over the four latent-manifold dimensions $m_l\in\{4,6,10,14\}$ and the twenty data splits, with the ambient dimension fixed at $d_a=10$. White circles show the individual test scores, while the horizontal lines indicate the median and inter-quartile range. Background shading groups the models into the \textsc{IQP}, \textsc{QKS}, and \textsc{PQC} families and the native free-fermionic models. Brackets compare each unrestricted model with its matchgate counterpart, while the annotations above the native free-fermionic models compare them separately with \textsc{IQP}, \textsc{QKS}, and \textsc{PQC}. All reported $p$-values are obtained from two-sided Mann-Whitney tests applied to the pooled scores. Because the violins combine different latent dimensions, their shapes reflect both changes in dataset difficulty and variation among data splits at fixed $m_l$. Significance levels are denoted by $^{*}$ for $p<0.05$, $^{**}$ for $p<0.01$, and $^{***}$ for $p<0.001$. $\mathrm{ns}$ denotes a non-significant difference.}
\label{fig:Classification_Hidden_Manifold_diff}
\end{figure*}

AngleMpf displays the broadest bimodal structure among the native free-fermionic models, with one group of scores near random classification and another at high accuracy. Its pooled distribution is statistically
indistinguishable from IQP ($p=0.152$) and QKS ($p=0.800$), but differs
significantly from PQC ($p=0.006$). This broad distribution indicates that the effectiveness of direct angle encoding is strongly dependent on dataset. These pooled tests do not show whether a difference is present at every ambient dimension. A significant result may be driven primarily by a subset of values of $d_a$, while a nonsignificant pooled result may conceal opposing dimension-dependent trends. These effects are examined thoroughly, through the dimension-resolved results in the SM.
\paragraph{Hidden manifold-diff~:}
Figure~\ref{fig:Classification_Hidden_Manifold_diff} shows the results for the hidden manifold-diff datasets. The ambient dimension is fixed at $d_a=10$, while latent manifold dimension is varied over $m_l=4$, $6$, $10$, and $14$. Each violin therefore combines results from
four latent dimensions and the twenty data splits at each dimension. Thus its shape reflects both systematic changes in difficulty with $m_l$ and variation among splits at fixed $m_l$. In particular, the multiple modes
visible for several models may correspond to different latent dimensions rather than to separate performance regimes for a single dataset
configuration.

IQP significantly outperforms fIQP ($p<0.001$). Although both distributions span several accuracy regimes, fIQP contains a substantially larger concentration of scores near random classification. The matchgate restriction therefore produces a clear reduction in performance for the IQP architecture on this dataset family.
For the QKS pair, fQKS has a somewhat higher pooled median than QKS, but the difference is not statistically significant ($p=0.070$). Similarly, PQC and fPQC are not significantly different in the pooled comparison
($p=0.061$), despite visible differences in the shapes of their distributions.

\begin{figure*}
    \centering
    \includegraphics[width=0.92\linewidth]{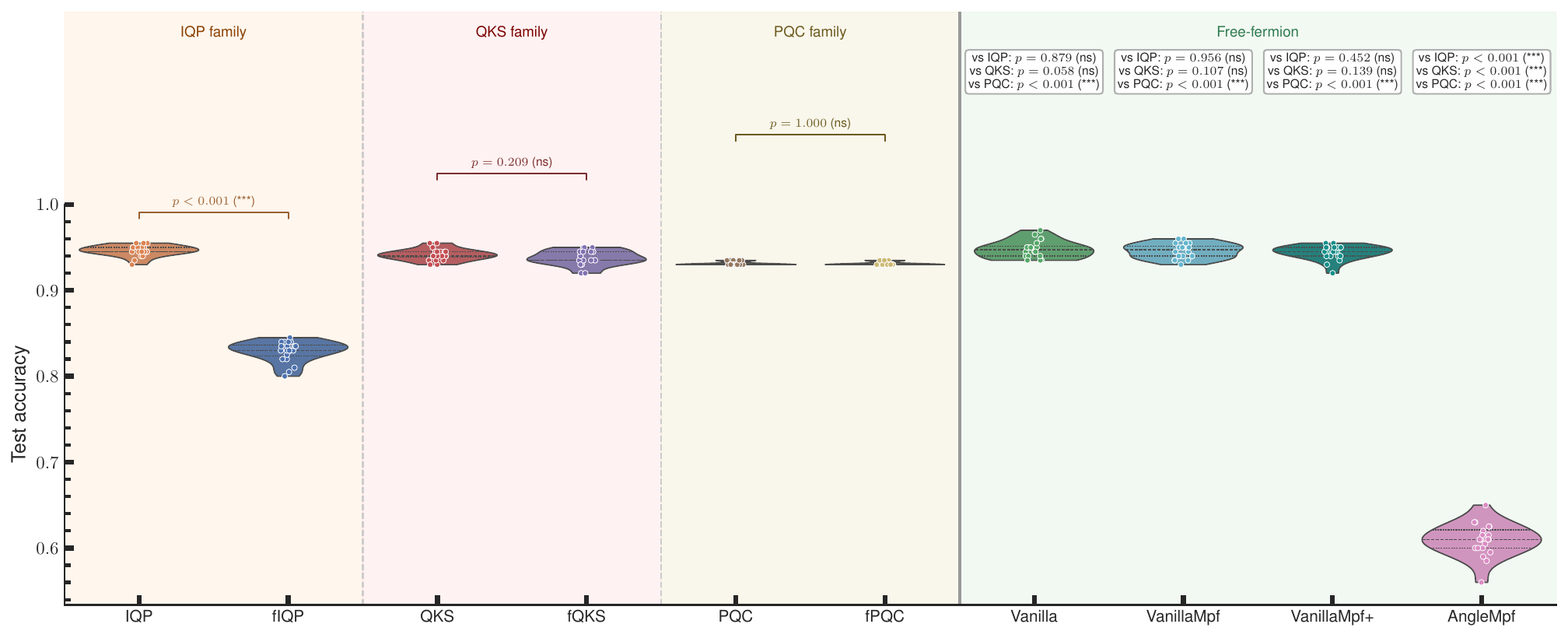}
    \caption{\textbf{Classification performance on the bars-and-stripes dataset.} Each violin shows the test accuracies obtained from the twenty data splits after validation-based hyperparameter selection. White circles represent the individual test scores, while the horizontal lines indicate the median and inter-quartile range. Background shading groups the models into the \textsc{IQP}, \textsc{QKS}, and \textsc{PQC} families and the native free-fermionic models. Brackets compare each unrestricted model with its matchgate counterpart, while the annotations above the native free-fermionic models compare them separately with \textsc{IQP}, \textsc{QKS}, and \textsc{PQC}. All reported $p$-values are obtained from two-sided Mann-Whitney tests. Significance levels are denoted by $^{*}$ for $p<0.05$, $^{**}$ for $p<0.01$, and $^{***}$ for $p<0.001$. $\mathrm{ns}$ denotes a nonsignificant difference.}
\label{fig:classification_bars_stripes}
\end{figure*}

The native free-fermionic models have broad distributions that overlap strongly with those of the unrestricted baselines. None of the comparisons of Vanilla, VanillaMpf, VanillaMpf+, or AngleMpf with IQP, QKS, or PQC is statistically significant. All corresponding tests give $p\geq0.134$. \emph{Thus, the pooled results provide no evidence that these native free-fermionic models perform systematically differently from the three unrestricted baselines across the complete latent-dimension sweep.}  The figure also does not include direct comparisons among the native free-fermionic models. Dimension-resolved results are provided in the SM and are used to determine whether performance changes systematically with $m_l$ or is driven only by a subset of latent dimensions.
\paragraph{Bars-and-Stripes~:}
Figure~\ref{fig:classification_bars_stripes} compares the test accuracies of ten classification models on the bars-and-stripes dataset~\cite{bowles2024betterclassicalsubtleart}. Unlike the preceding
figures, this comparison contains only one dataset configuration. Consequently, the spread of each violin represents variation among the twenty data splits.

IQP substantially outperforms fIQP ($p<0.001$) and shows scores tightly concentrated near $0.94$, whereas for the fIQP scores are centered near $0.83$, indicating a clear reduction in classification performance. By contrast,both QKS and fQKS attain accuracies close to $0.94$ and are statistically indistinguishable ($p=0.209$). The matchgate restriction therefore has no statistically detectable effect on the QKS architecture for this particular dataset.

Both members of the PQC family also have narrow distributions, \emph{but fPQC achieves consistently higher accuracies than PQC ($p<0.001$).} Thus, the three matched pairs display three different outcomes: the matchgate counterpart performs worse in the case of IQP, comparably for QKS, and better for PQC. Vanilla, VanillaMpf, and VanillaMpf+ all achieve test accuracies near $0.94$. Their distributions are statistically indistinguishable from that of IQP, with $p=0.879$, $p=0.956$, and $p=0.452$, respectively. They are also statistically indistinguishable from QKS, with $p=0.058$, $p=0.107$, and $p=0.139$, respectively. All three significantly outperform PQC ($p<0.001$ in each comparison). The figure provides no direct evidence on whether the Pfaffian readout or the $\ket{+}^{\otimes n}$ reference state systematically improves the Vanilla fidelity model for this dataset with the three Vanilla-based models
perform similarly near the upper end of the observed accuracy range.

AngleMpf behaves differently from the other native free-fermionic models.
Its scores are concentrated near $0.61$, and it performs significantly
worse than IQP, QKS, and PQC ($p<0.001$ in all three comparisons). Since
AngleMpf and VanillaMpf use the same type of Pfaffian readout but different
encoders, this result indicates that the random affine Vanilla encoder is
important for the strong performance of the Pfaffian model on
bars-and-stripes. Direct comparisons among the native free-fermionic models
are not included in the figure, so this interpretation is based on their
observed distributions rather than on an additional pairwise statistical
test.
\paragraph{Downscaled MNIST~:}

\begin{figure*}
    \centering
    \includegraphics[width=0.92\linewidth]{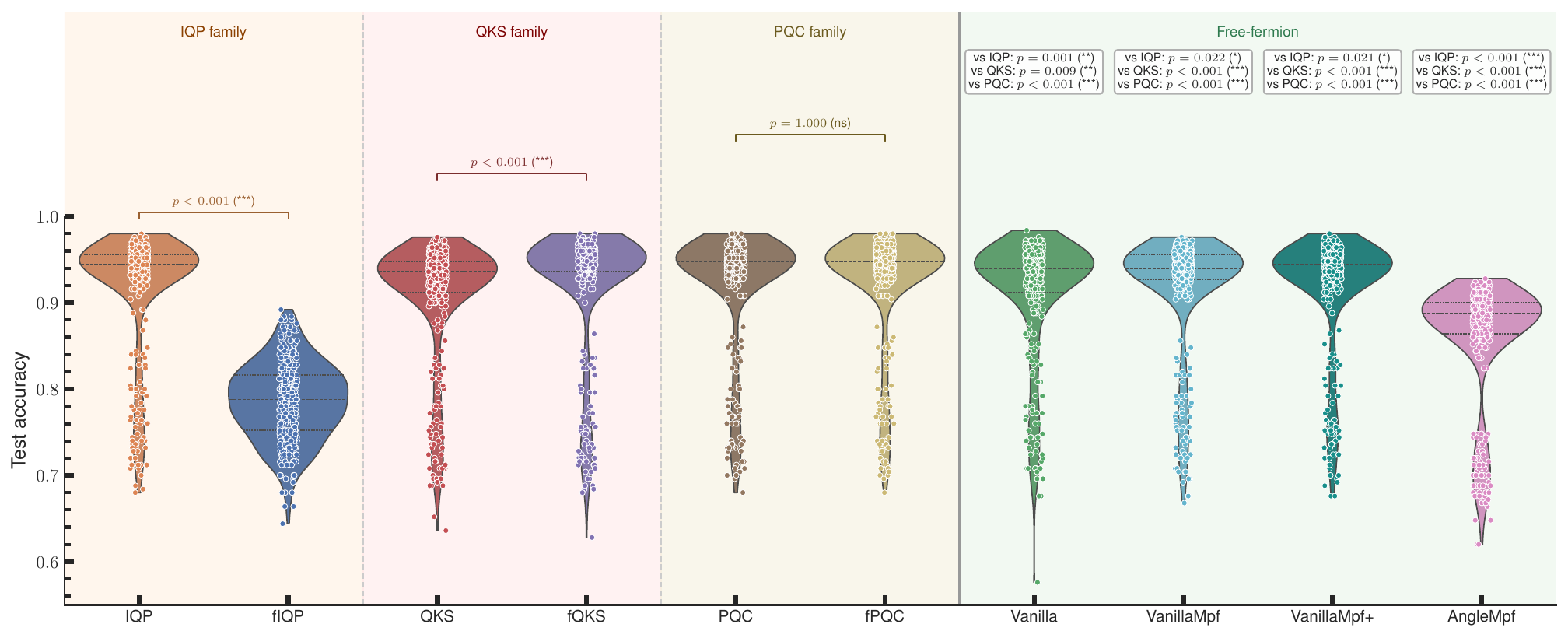}
    \caption{\textbf{Classification performance on the downscaled-MNIST datasets.} Each violin pools the test accuracies obtained after validation-based hyperparameter selection over the nineteen retained PCA dimensions $d_m=2,\ldots,20$ and the twenty data splits. White circles show the individual test scores, while the horizontal lines indicate the median and inter-quartile range. Background shading groups the models into the \textsc{IQP}, \textsc{QKS}, and \textsc{PQC} families and the native free-fermionic models. Brackets compare each unrestricted model with its matchgate counterpart, while the annotations above the native free-fermionic models compare them separately with \textsc{IQP}, \textsc{QKS}, and \textsc{PQC}. All reported $p$-values are obtained from two-sided Mann-Whitney tests applied to the pooled scores. Because the violins combine different PCA dimensions, their shapes reflect both dimension-dependent performance and variation among data splits at fixed $d_m$. Significance levels are denoted by $^{*}$ for $p<0.05$, $^{**}$ for $p<0.01$, and $^{***}$ for $p<0.001$. $\mathrm{ns}$ denotes a nonsignificant difference.}
    \label{fig:Classification_Downscaled_MNIST}
\end{figure*}

Figure~\ref{fig:Classification_Downscaled_MNIST} summarizes the results for
the downscaled-MNIST datasets. The figure combines all nineteen retained
PCA dimensions, $d_m=2,\ldots,20$, together with the twenty data splits at
each dimension. Different splits also contain different subsamples of the
$250$ training and $250$ test examples. The pooled distributions therefore
mix variation due to the PCA dimension, the sampled data, and the fitted
model. Their multiple modes and lower tails may consequently reflect
systematic changes with $d_m$, rather than instability at a fixed dataset
configuration. The dimension-resolved behavior is examined in
Figures~\ref{fig:classification_kernel_vs_dim}
and~\ref{fig:classification_kernel_diffvs_dim} of the SM.

Within the IQP family, IQP significantly outperforms fIQP
($p<0.001$). IQP has a large concentration of scores near $0.95$, whereas
fIQP is centred at substantially lower accuracies. Restricting the IQP
architecture to matchgates therefore produces a clear loss of performance on downscaled MNIST. Both QKS models perform strongly, \emph{but fQKS
significantly outperforms QKS ($p<0.001$), with a larger concentration of
scores in the highest-accuracy region.} The PQC family shows the opposite
behavior and significantly outperforms its matchgate-restricted counterpart fPQC ($p<0.001$). Thus, as in the
other datasets, the effect of imposing a matchgate structure depends
strongly on the underlying circuit architecture.

Vanilla, VanillaMpf, and VanillaMpf+ also have large concentrations of
scores near $0.93$--$0.96$, together with lower tails associated with the
PCA-dimension sweep. Vanilla differs significantly from IQP
($p=0.001$), QKS ($p=0.009$), and PQC ($p<0.001$).
VanillaMpf likewise differs from IQP ($p=0.022$), QKS ($p<0.001$),
and PQC ($p<0.001$). The corresponding results for VanillaMpf+ are
$p=0.021$, $p<0.001$, and $p<0.001$, respectively. \emph{These tests establish
that the pooled distributions are not identical, but they do not imply that
one model is uniformly better at every PCA dimension.} \emph{Visually, the three Vanilla-based models attain accuracies comparable to the strongest unrestricted baselines over a substantial part of the sweep.} AngleMpf on the other hand, performs less strongly and its main concentration lies near
$0.87$--$0.91$, with a lower tail extending to substantially smaller
accuracies. It differs significantly from IQP, QKS, and PQC
($p<0.001$ in all three comparisons). Since AngleMpf and VanillaMpf use the same type of Pfaffian readout but different encoders, their contrasting
distributions indicate that the random affine Vanilla encoder contributes substantially to the performance of VanillaMpf on downscaled MNIST.

\paragraph{Take-home points.}

\begin{figure*}
    \centering
    \includegraphics[width=1.02\linewidth]{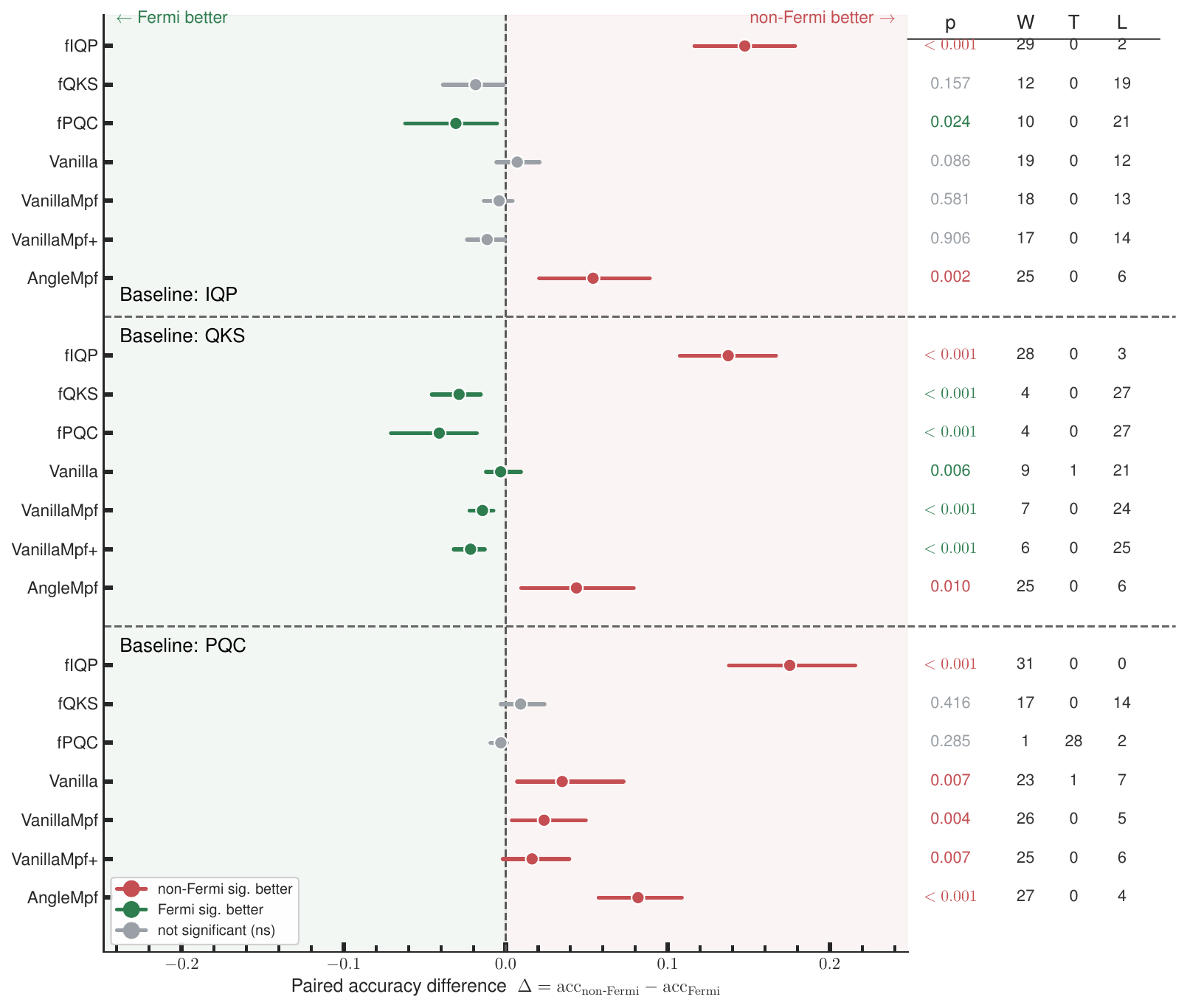}
    \caption{\textbf{Paired classification performance across all dataset configurations.} Each row compares the indicated fermionic model with one of the three non-fermionic baselines: \textsc{IQP}, \textsc{QKS}, or \textsc{PQC}. For each of the $31$ dataset configurations, the test accuracy is first averaged over the $20$ data splits, and the paired difference is defined as $\Delta_q =\overline{\operatorname{Acc}}_{\mathrm{non\text{-}Fermi},q}-\overline{\operatorname{Acc}}_{\mathrm{Fermi},q}$. Points show the mean difference across the $31$ configurations, and horizontal bars show the corresponding $95\%$ percentile-bootstrap confidence intervals obtained from $10^4$ resamples of the configuration-level differences. Negative values favor the fermionic model, whereas positive values favor the non-fermionic baseline. The reported $p$-values are obtained from two-sided Wilcoxon signed-rank tests applied to the paired differences. The columns $W$, $T$, and $L$ give the numbers of configurations for which the non-fermionic baseline performs better than, ties with, or performs worse than the fermionic model, respectively. Green and red markers denote statistically significant differences favoring the fermionic and non-fermionic models, respectively, while gray markers denote non-significant differences.} 
    \label{fig:all_plots_forest_classification}
\end{figure*}

Figure~\ref{fig:all_plots_forest_classification} summarizes the paired comparisons across all dataset configurations. Our results show that fermionic Gaussian dynamics can retain, and in some cases produce better classification performance, with their effectiveness depending strongly on the encoding architecture. f\textsc{IQP} performs significantly worse than \textsc{IQP}, \textsc{QKS}, and \textsc{PQC}. In contrast, \emph{f\textsc{QKS} significantly outperforms
\textsc{QKS} and remains statistically indistinguishable from \textsc{IQP} and \textsc{PQC}}. Similarly, \emph{f\textsc{PQC} significantly outperforms \textsc{IQP} and \textsc{QKS} and is statistically indistinguishable from
\textsc{PQC}, with the two models tying on $28$ of the $31$ dataset configurations.}

The \textsc{PQC}-f\textsc{PQC} comparison must be interpreted in light of the circuit depth selected during validation. For most configurations, validation selects the full-width, depth-one \textsc{PQC}. In this setting, all features are loaded into a single layer of local rotations followed by a fixed entangling layer. Because the same terminal entangling unitary is applied to both inputs, it cancels from the fidelity overlap. The resulting kernel is therefore determined entirely by the product-state encoding and does not depend on the final CNOT layer. Moreover, this product kernel admits an exact fermionic Gaussian realization on twice as many modes by encoding
$\ket{0}\mapsto\ket{00}$ and $\ket{1}\mapsto\ket{11}$. Within each encoded pair, the data-dependent single-qubit transformation can be implemented by a matchgate acting in the even-parity subspace. This kernel-level equivalence
accounts for the large number of ties between \textsc{PQC} and f\textsc{PQC}.

This equivalence concerns only the fidelity kernel. It does not imply that the two constructions produce identical output states, computational-basis samples, or expectation values of general observables, since the terminal
CNOT layer remains part of the original \textsc{PQC} state preparation. It also does not extend directly to deeper circuits in which additional data-dependent layers follow the entanglers. To isolate this genuinely
multilayer regime, we repeat the paired comparison using the best validation-selected \textsc{PQC} circuit of depth greater than one. The result, restricted to the $29$ configurations for which such a circuit is
available, is presented in Fig.~\ref{fig:classification-pqc-depth-forest} of the Supplementary Material.

The contrast between the \textsc{IQP}-f\textsc{IQP} and
\textsc{QKS}-f\textsc{QKS} pairs further demonstrates that the usefulness of the fermionic restriction depends on how the features are constructed. The \textsc{QKS} representation combines many independently randomized encoding episodes. Restricting the dynamics within each episode to matchgates still retains a diverse collection of random features, which can be combined by the classical classifier. The \textsc{IQP} construction is more sensitive to the matchgate restriction. Its data-dependent $ZZ$ phases encode products of input features through interacting many-body phases, and the fidelity kernel
converts the resulting interference structure into a similarity measure. These non-Gaussian phase interactions are absent from f\textsc{IQP}. Although its matchgate replacements retain a dependence on the same feature products, they fail to reproduce the action of the original $ZZ$ rotations. 

The native free-fermionic models provide the same general lesson. \textsc{Vanilla}, \textsc{VanillaMpf}, and \textsc{VanillaMpf+} frequently perform comparably to the strongest unrestricted baselines while remaining efficiently classically simulable. Conversely, the weaker performance of \textsc{AngleMpf} shows that a Pfaffian readout alone is not sufficient to
obtain a strong model but the encoder and the information supplied to the readout are equally important. Overall, neither non-Gaussianity nor an entangling circuit structure guarantees improved classification performance. Instead, good learning performance comes from a delicate balance between the inductive bias of the encoding, circuit architecture, readout, and dataset. Fermionic Gaussian models therefore provide a powerful, scalable and classically simulable benchmark, setting a meaningful performance threshold that general quantum models must surpass before their additional quantum resources can be
regarded as useful for practically relevant tasks.


\begin{figure*}
    \centering
    \includegraphics[width=1\linewidth]{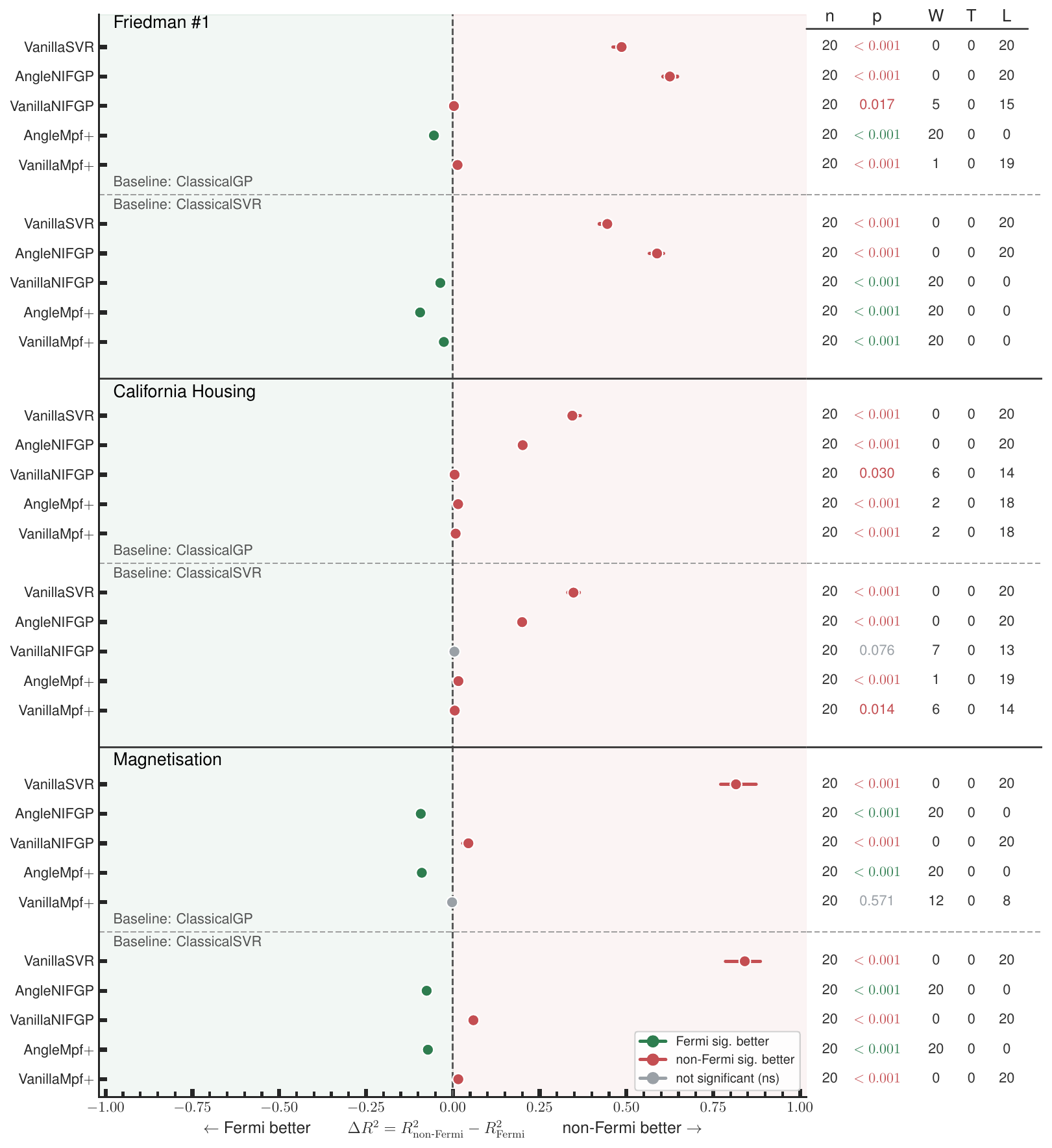}
    \caption{\textbf{Paired regression performance across datasets.} The FermiML regression models are compared with ClassicalGP and ClassicalSVR on Friedman~\#1, California Housing, and the magnetisation dataset. For each dataset, the upper and lower blocks use ClassicalGP and ClassicalSVR as the respective classical baselines. Points show the median paired difference $\Delta_s^{(R^2)}=R^2_{C,s}-R^2_{F,s}$ over twenty matched splits, and horizontal bars show the corresponding $95\%$ percentile-bootstrap confidence intervals. Negative values favor the fermionic model, while positive values favor the classical baseline. The reported $p$-values are obtained from two-sided Wilcoxon signed-rank tests. The columns $W$, $T$, and $L$ give the numbers of splits on which the fermionic model performs better than, ties with, or performs worse than the classical reference. Green and red markers indicate statistically significant differences in favor of the fermionic and classical models, respectively; grey markers indicate no significant difference.}
    \label{fig:regression_all_data}
\end{figure*}

\subsection{NIF-Ising classification}
\label{subsec:hamml-results}

Tables~\ref{tab:nifising-text-comparison} and
\ref{tab:nifising-image-comparison} compare NIF-Ising with the models reported in Ref.~\cite{tiblias2025efficient}. The comparison with HAM, PEFF, and SIM is particularly relevant because these models also formulate
classification in terms of data-dependent Hamiltonians. 

NIF-Ising gives the best reported result among the quantum and Hamiltonian-based models on AG News, CIFAR2, and Fashion. It reaches $89.8\%$ test accuracy on AG News, compared with $89.2\%$ for CIRC and $88.6\%$ for SIM. On CIFAR2, it reaches $85.5\%$, exceeding CIRC ($84.6\%$), QCNN ($84.7\%$), HAM ($81.0\%$), PEFF ($78.7\%$), and SIM
($68.2\%$). On Fashion, it reaches $87.8\%$, compared with $86.5\%$ for
CIRC, $84.4\%$ for SIM, and $76.3\%$ for QCNN. The exception is IMDb, where HAM obtains $88.1\%$ and NIF-Ising obtains $85.5\%$. NIF-Ising therefore does not dominate every dataset, but it remains competitive across both text and image tasks despite its restriction to classically simulable free-fermion dynamics.

The classical LSTM and CNN models remain the strongest in text and image
datasets, respectively. The purpose of this comparison is therefore not to
claim an advantage over specialized classical architectures. Rather, it shows
that restricting the trainable quantum model to matchgate dynamics does not produce a systematic loss of accuracy relative to the more general quantum models considered in~\cite{tiblias2025efficient}. 

The complete NIF-Ising hyperparameter grids and the settings selected on each dataset are
reported in the SM.
\subsection{Regression}
\label{subsec:regression-evaluation}

Figure~\ref{fig:regression_all_data} summarizes the regression results on Friedman~\#1, California Housing, and magnetization datasets. For each dataset, the models are compared separately with ClassicalGP and ClassicalSVR. All comparisons are paired across the same $20$ train-validation-test splits.

\paragraph{Friedman~\#1~:}
The results are strongly dependent on the fermionic architecture and the classical baseline chosen. AngleMpf+ is the only fermionic model that significantly outperforms both ClassicalGP and ClassicalSVR, winning all $20$ splits in both comparisons. VanillaNIFGP and VanillaMpf+ outperform ClassicalSVR but not ClassicalGP. By contrast, VanillaSVR and AngleNIFGP perform worse than both classical baselines. These results also indicate that ClassicalGP provides the stronger classical baseline for this dataset.

\paragraph{California Housing~:}
None of the fermionic models significantly outperforms either classical baseline. VanillaNIFGP is statistically indistinguishable from ClassicalSVR ($p=0.076$), but its comparison with ClassicalGP favors the classical model. The remaining comparisons also favor the corresponding classical baseline. Several of these differences are small in magnitude, particularly for the Pfaffian models, but remain statistically significant because their direction is consistent across most of the splits.

\paragraph{Magnetization~:}
The magnetization dataset produces the clearest advantages for the fermionic models. AngleNIFGP and AngleMpf+ significantly outperform both ClassicalGP and ClassicalSVR, winning all $20$ splits in each comparison. VanillaMpf+ is statistically comparable to ClassicalGP ($p=0.571$), although it performs slightly worse than ClassicalSVR. VanillaSVR and VanillaNIFGP perform worse than both classical baselines. Since the magnetization target is constructed from a known sector of the fermionic feature space, the performance of AngleNIFGP and AngleMpf+ illustrates the benefit of matching the model representation to the structure of the target.

Taken together, the results do not show a uniform advantage for all fermionic regression models. Instead, performance depends on the encoding, regression method, and dataset. The strongest fermionic models can match or outperform the classical baselines when their feature representation is well aligned with the target, while the classical models remain stronger on datasets without the same structural correspondence.
\section{Discussion}
\label{sec:discussion_section}
The performance of a QML model depends on three related but distinct properties. \emph{Expressivity} describes the range of quantum states, feature representations, or input-output functions that can be realized by a chosen circuit architecture and readout \cite{sim2019expressibility,abbas2021power}. \emph{Learnability} concerns whether a useful predictor can be inferred from a finite training set and whether its performance generalizes to previously unseen data (test data) \cite{huang2021powerdata,caro2022generalization}. \emph{Trainability} concerns whether suitable model parameters can be found with computational resources that scale favorably with system size. For parameterized quantum circuits, this is closely related to the structure of the loss landscape and, in particular, to whether the gradients required by an optimizer remain resolvable as the circuit grows \cite{mcclean2018barren,cerezo2021cost,holmes2022connecting}. 

Greater expressivity does not necessarily improve learning. A highly expressive circuit may be difficult to train, may include features that are irrelevant to the target or may require too much data to generalize reliably. Conversely, a restricted model can perform well when its inductive bias is aligned with the data. Increasing circuit expressivity can also move the model closer to random-circuit behavior and suppress variations in the loss landscape, producing a trade-off between expressivity and trainability~\cite{holmes2022connecting}.

\subsection{Expressivity and learning capacity}
\label{subsec:expressivity_learning_capacity}

Matchgate circuits are restricted to fermionic Gaussian transformations and can be efficiently simulated using covariance matrices~\cite{terhal2002classical,josza2008matchgates,bravyi2005lagrangian}. Therefore, despite accessing a smaller family of states than unrestricted quantum circuits, this restriction does not determine their learning performance.

The expressivity of a FermiML model depends on both its circuit architecture and readout. The architecture determines the family of data-dependent Gaussian states that can be prepared. The readout determines which properties of those states are made available to the classical estimator. The fidelity kernel compares complete encoded states through their squared overlap, whereas the Pfaffian covariance kernels retain selected orders of the Majorana correlation structure. Higher-order sectors enlarge this feature representation while remaining within the fermionic setting. Models based on the same matchgate dynamics can therefore define different hypothesis classes. 

We compare unrestricted quantum circuits with matchgate counterparts while keeping preprocessing, readout, classical estimator, and hyperparameter budget as similar as possible. This isolates the effect of restricting the circuit to matchgates. Across benchmarks, matchgate models remain competitive and even outperform their unrestricted QML counterparts, extending our earlier proof-of-concept results~\cite{gince2024fermionic}. Additional expressivity of unrestricted circuits, therefore, provides no systematic predictive advantage on the tasks considered here.

This conclusion does not imply that FermiML models have equivalent expressivity. It simply shows that the observed performance of the unrestricted models cannot be attributed solely to non-Gaussian dynamics, classical intractability, or quantum-resource theoretic arguments involving entanglement and non-stabilizerness. Demonstrating a contribution from these resources requires identifying tasks for which performance decreases when these resources are removed.
\subsection{Trainability}
\label{subsec:trainability}

Classical simulability has different consequences for the kernel and Hamiltonian models. The covariance and fidelity kernels do not optimize circuit parameters, so barren plateaus in circuit-parameter space are not relevant. Their main scaling concern is instead the concentration of kernel values over different input pairs.

As discussed earlier, matchgate circuits explore only a restricted subset of the full unitary group. Their limited expressivity avoids the concentration mechanism associated with embeddings that approach full-unitary designs. It does not, however, rule out concentration from other sources. Fidelity kernels can concentrate because they compare encoded states through a global overlap, even when the embedding has low expressivity and produces little entanglement~\cite{Thanasilp2024_EC}. 

Classical evaluation removes the quantum measurement cost of resolving these small kernel values, but it does not automatically remove data-distinguishability problems. Nevertheless, we did not observe a general collapse in predictive performance over the system sizes and datasets studied. Several fermionic models (both kernel-based or Hamiltonian) remain competitive or outperform their classical and quantum baselines. 

The NIF-Ising model differs from fixed-kernel models because its encoder and single-particle transition matrices are optimized via gradient descent. Both the forward pass and the gradients are evaluated classically with the gradients obtained via backpropagation. The relevant dynamical Lie algebra of matchgate circuits is
\(\mathfrak{so}(2n)\), whose dimension is \(n(2n-1)\) and hence grows polynomially with $n$, the number of qubits~\cite{Ragone2024_LieAlgebraicBP,diaz2023showcasingbarrenplateautheory}. Lie-algebraic results show that a polynomial-dimensional algebra can avoid barren plateaus caused by exploration of an exponentially large unitary group~\cite{Ragone2024_LieAlgebraicBP}. However, this fact alone does not prove trainability. Training also depends on the input state, the measured observable, and the sector of operator space explored by the loss~\cite{cerezo2021cost}. For parameterized
matchgate circuits, Ref.~\cite{diaz2023showcasingbarrenplateautheory} shows that the loss variance is governed by the dimension of the corresponding Lie-group module. This is particularly relevant to NIF-Ising. Under the Jordan-Wigner transformation,
\begin{equation}
    Z_q=-i\,c_{2q-1}c_{2q},
\end{equation}
and hence, for $i\neq j$,
\begin{equation}
    Z_iZ_j=-c_{2i-1}c_{2i}c_{2j-1}c_{2j}.
    \label{eq:zz-four-majoranas}
\end{equation}
Each Ising interaction is therefore an order-four Majorana operator. The order-four sector has dimension
$\binom{2n}{4}$,which is polynomial in $n$. In contrast, Majorana sectors whose order grows with $n$ can have exponentially large dimension and may lead to barren plateaus even if the circuit contains only matchgates~\cite{diaz2023showcasingbarrenplateautheory}.

Keeping the Majorana order fixed is therefore conducive to trainability, but it does not provide a general guarantee. The loss landscape and gradient scaling may also depend on the encoder, circuit parameterization, initialization, and data distribution. Moreover, we do not directly measure the scaling of the gradient variance. Our results show that NIF-Ising can be optimized successfully over the system sizes studied (up to 92 qubits; see SM for details) and produce meaningful results.

\section{Conclusion}
\label{sec:conclusion}

We have developed FermiML as a scalable and versatile benchmarking framework for QML. Like unresticted QML models, FermiML models use superposition, interference, and entanglement as resources. We show that FermiML encompasses essentially all common building blocks of QML, such as quantum-state encodings, state-overlap kernels, and Hamiltonian-based models, and allows for models with nontrivial inductive bias, while the underlying free-fermionic dynamics leads to runtime and memory scaling polynomially with the number of qubits. This scalability enables us to study system and dataset sizes that are impossible to access with most QML models and thus deploy QML at real-world scales.

Across benchmarks, we show that restricting QML models to within the fermionic circuit manifold does not produce a systematic loss of predictive performance. The fermionic surrogates remain competitive and, for some datasets and encodings, outperform the corresponding quantum or classical baselines. These empirical results do not imply that fermionic circuits have the same expressive power as unrestricted quantum circuits. Rather, they show that FermiML provides a controlled and scalable reference for determining whether non-Gaussian quantum resources provide measurable learning benefits beyond classical simulation or not. Simply put: any performance gain or ``advantage'' in QML is only possible if it cannot be reproduced by a FermiML model.

Our work leaves certain questions open to further study. For fermionic kernel models, a direct scaling analysis of kernel variance, spectrum, and alignment can determine whether concentration appears at larger numbers of qubits. For the trainable fermionic models like NIF-Ising, the corresponding question concerns the scaling of the loss landscape and gradient variance with the number of qubits, circuit depth, initialization, and readout. The covariance-matrix representation makes it possible to investigate both questions at sizes outside the reach of direct state-vector simulation.

\section*{Data and code availability}
A tool for automatically constructing fermionic surrogates of QML models is available at~\href{https://github.com/MatchCake/MajoranaKaraoke}{MajoranaKaraoke: A fully functional translator}. The code used to produce the results of this work is publicly available at: \href{https://github.com/MatchCake/Fermionic-Supervised-Learning}{github.com/MatchCake/Fermionic-Supervised-Learning}. All simulations were performed using the \href{https://github.com/MatchCake/MatchCake}{MatchCake} and the \href{https://github.com/MatchCake/TorchPfaffian}{TorchPfaffian} libraries. Data will be made available upon reasonable request.

\section*{Acknowledgments}
This research was financially supported by the Natural Sciences and Engineering Research Council of Canada (NSERC) through a Discovery grant. AS acknowledges support from the Canada First Research Excellence Fund through an Institut quantique Postdoctoral Fellowship. This work made use of compute resources by Calcul Qu\'{e}bec and the Digital Research Alliance of Canada. This work was supported by the QSciTech CREATE-NSERC program. JG gratefully acknowledge the financial support provided by this program.

\bibliographystyle{unsrt}
\bibliography{references}
\appendix 
\onecolumn

\section{Analytical details of the fermionic regression models}
\label{supp:fermionic-regression-details}

This section collects details underlying the
fermionic regression models introduced in Sec.~\ref{subsec:regression-models}. These are not required to follow the main comparison, but clarify the construction, normalization, and
implementation of the kernels.
\subsection{Pfaffian feature representation}
\label{supp:pfaffian-features}

Let $\Gamma(\bm{x})\in\mathbb{R}^{2n\times 2n}$ be the Majorana covariance
matrix associated with the encoded state. For an even Majorana order $m$ and
an index set
$S=\{s_1<\cdots<s_m\}\subseteq\{1,\ldots,2n\}$, we define
\begin{equation}
    \bigl[v_m(\bm{x})\bigr]_S
    =
    \operatorname{Pf}\!\left[
        \Gamma(\bm{x})\big|_S
    \right],
    \label{eq:supp-pfaffian-features}
\end{equation}
where $\Gamma(\bm{x})|_S$ denotes the principal sub-matrix indexed by $S$.
The complete order-$m$ feature vector therefore has dimension
\begin{equation}
    D_m=\binom{2n}{m}.
    \label{eq:supp-pfaffian-dimension}
\end{equation} The corresponding kernel is
\begin{equation}
    \kappa_m(\bm{x},\bm{x}')
    =
    \frac{1}{2^n}
    v_m(\bm{x})^T v_m(\bm{x}').
    \label{eq:supp-pfaffian-kernel}
\end{equation}
Since this is an ordinary Euclidean inner product, $\kappa_m$ is positive
semi-definite, although individual kernel entries may be negative. For $m=2$, the Pfaffian of each $2\times2$ principal sub-matrix is simply the
corresponding covariance-matrix entry. Hence,
\begin{align}
    \kappa_2(\bm{x},\bm{x}')
    &=
    \frac{1}{2^n}
    \sum_{1\leq j<k\leq 2n}
    \Gamma_{jk}(\bm{x})\Gamma_{jk}(\bm{x}')
    \nonumber=
    -\frac{1}{2^{n+1}}
    \Tr\!\left[
        \Gamma(\bm{x})\Gamma(\bm{x}')
    \right].
    \label{eq:supp-second-order-kernel}
\end{align}
The minus sign follows from the antisymmetry of the covariance matrices. For a pure Gaussian state and the complete even-order sector $m=2p$, the
kernel diagonal is
\begin{equation}
    \kappa_{2p}(\bm{x},\bm{x})
    =
    2^{-n}\binom{n}{p}.
    \label{eq:supp-pfaffian-diagonal}
\end{equation}
Thus, within a fixed pure-state sector, the diagonal is independent of the input but depends on the number of modes and the retained Majorana order.
\subsection{Multi-sector covariance kernels}
\label{supp:multisector-kernels}

For a collection of Majorana orders $M_s$, we concatenate the
corresponding feature vectors:
\begin{equation}
    v_{M_{s}}(\bm{x})
    =
    \bigoplus_{m\in M_s}v_m(\bm{x}).
    \label{eq:supp-multisector-features}
\end{equation}
Its dimension is
\begin{equation}
    D_{M_s}
    =
    \sum_{m\in M_s}\binom{2n}{m}.
    \label{eq:supp-multisector-dimension}
\end{equation}
The associated kernel is therefore
\begin{align}
    \kappa_{M_s}(\bm{x},\bm{x}')
    &=
    \frac{1}{2^n}
    v_{M_s}(\bm{x})^T
    v_{M_s}(\bm{x}')
    \nonumber =
    \sum_{m\in M_s}
    \kappa_m(\bm{x},\bm{x}').
    \label{eq:supp-multisector-kernel}
\end{align}
The parity-preserving covariance models use a single even Majorana order,
$m\in\{2,4\}$. For the parity-breaking plus (+) models, we instead concatenate
the features from several orders, using
$M_s\in\{\{1,2,3\},\{1,2,3,4\}\}$.
\subsection{Odd-order features and parity-indefinite reference states}
\label{supp:odd-order-features}

For a state with definite fermionic parity, all odd-order Majorana moments
vanish. The plus(+) models instead use the parity-indefinite reference state
$\ket{+}^{\otimes n}$ together with the generalized product-state
representation implemented in MatchCake. Let $d_j(\bm{x})=\Tr\!\left[\rho(\bm{x})c_j\right]$ denote the first Majorana moments. These moments are incorporated through the
augmented antisymmetric matrix~\cite{bravyi2005lagrangian}
\begin{equation}
    \widetilde{\Gamma}(\bm{x})
    =
    \begin{pmatrix}
        \Gamma(\bm{x}) & \bm d(\bm{x})\\
        -\bm d(\bm{x})^T & 0
    \end{pmatrix}.
    \label{eq:supp-augmented-covariance}
\end{equation}
For an odd-order index set $S$, the corresponding feature is evaluated as
\begin{equation}
    \bigl[v_m(\bm{x})\bigr]_S
    =
    \operatorname{Pf}\!\left[
        \widetilde{\Gamma}(\bm{x})
        \big|_{S\cup\{\star\}}
    \right],
    \label{eq:supp-odd-pfaffian-feature}
\end{equation}
where $\star$ labels the auxiliary row and column. Because
$|S\cup\{\star\}|=m+1$ is even, the Pfaffian is well defined. For a parity
eigenstate, $\bm d(\bm{x})=\bm 0$, and the odd-order features vanish.
\subsection{Gaussian-process posterior inference}
\label{supp:gp-inference}

Fermionic covariance-kernel models use GPR~\cite{rasmussen2006gp}. Let
$\mathcal{D}=\{(\bm{x}_i,y_i)\}_{i=1}^{N_s}$ denote full dataset while \(\mathcal D_{\mathrm{tr}}\subset\mathcal D\) denote the training fold. The targets are standardized using the mean
$\bar{y}_{\mathrm{tr}}$ and standard deviation $s_{y,\mathrm{tr}}$ computed
from the training fold:
\begin{equation}
    \widetilde{y}_i
    =
    \frac{y_i-\bar{y}_{\mathrm{tr}}}{s_{y,\mathrm{tr}}}.
    \label{eq:supp-standardized-targets}
\end{equation}
For a chosen fermionic kernel $\kappa$, we assume the zero-mean prior
\begin{equation}
    f(\bm{x})
    \sim
    \mathcal{GP}\!\left(0,\kappa(\bm{x},\bm{x}')\right)
\end{equation}
and the observation model
\begin{equation}
    \widetilde{y}_i
    =
    f(\bm{x}_i)+\epsilon_i,
    \qquad
    \epsilon_i\sim\mathcal{N}(0,\alpha),
    \label{eq:supp-gp-observation-model}
\end{equation}
where $\alpha$ is the assumed homoscedastic noise variance. Equivalently,
$\alpha$ provides diagonal regularization of the training Gram matrix,
\begin{equation}
    K_{\alpha} = K+\alpha I_N,
    \qquad
    K_{ij}=\kappa(\bm{x}_i,\bm{x}_j).
    \label{eq:supp-regularized-gram}
\end{equation} For a test input $\bm{x}_*$, define
\begin{equation}
    \bm{\kappa}_*
    =
    \bigl(
        \kappa(\bm{x}_1,\bm{x}_*),\ldots,
        \kappa(\bm{x}_N,\bm{x}_*)
    \bigr)^T
\end{equation}
and $\kappa_{**}=\kappa(\bm{x}_*,\bm{x}_*)$. Standard Gaussian
conditioning gives the posterior mean and latent-function variance
~\cite{rasmussen2006gp}:
\begin{align}
    \widetilde{\mu}_*
    &=
    \bm{\kappa}_*^T
    K_\alpha^{-1}
    \widetilde{\bm{y}},
    \label{eq:supp-gp-posterior-mean}
    \\
    \widetilde{s}_{f,*}^{\,2}
    &=
    \kappa_{**}
    -
    \bm{\kappa}_*^T
    K_\alpha^{-1}
    \bm{\kappa}_*.
    \label{eq:supp-gp-latent-variance}
\end{align}
If uncertainty in a future noisy observation is required, its variance is
\begin{equation}
    \widetilde{s}_{y,*}^{\,2}
    =
    \widetilde{s}_{f,*}^{\,2}+\alpha.
    \label{eq:supp-gp-observation-variance}
\end{equation}
The posterior quantities are returned to the original target scale using
\begin{equation}
    \mu_*
    =
    \bar{y}_{\mathrm{tr}}
    +
    s_{y,\mathrm{tr}}\widetilde{\mu}_*,
    \qquad
    s_{f,*}
    =
    s_{y,\mathrm{tr}}\widetilde{s}_{f,*}.
    \label{eq:supp-gp-original-scale}
\end{equation}
The inverse of $K_\alpha$ is not formed explicitly. Instead, we compute its Cholesky factorization,
$K_\alpha=LL^T,$ and evaluate the posterior through triangular solves. This is the standard implementation of exact GPR and provides greater numerical stability than
forming the matrix inverse explicitly
~\cite{rasmussen2006gp,higham2009cholesky}. The covariance-kernel models are implemented using
\texttt{GaussianProcessRegressor} from
\texttt{scikit-learn}~\cite{pedregosa2018scikitlearnmachinelearningpython}. We set
\texttt{optimizer=None}; consequently, the fermionic kernel parameters are
not optimized by maximizing the log marginal likelihood. Instead, the
encoder, retained Majorana sectors, input transformation, and $\alpha
    \in
    \left\{
        10^{-6},10^{-4},10^{-3},10^{-2}
    \right\}$ are selected using validation performance. The final predictive mean is
mapped back to the original target scale and clipped to the range of target
values observed in the corresponding training fold.
\subsection{Finite-dimensional kernels and numerical conditioning}
\label{supp:gp-conditioning}

The Pfaffian covariance kernels have explicit finite-dimensional feature
maps. For a training set of size $N_s$, let
$\Phi\in\mathbb{R}^{N_s\times D}$ denote the feature matrix whose $i^{\text{th}}$ row is
$v(\bm{x}_i)^T$. The corresponding Gram matrix can be written as
\begin{equation}
    \mathcal{K} = 2^{-n}\Phi\Phi^T.
    \label{eq:supp-feature-gram-factorization}
\end{equation}
It therefore satisfies
\begin{equation}
    \operatorname{rank}(
    \mathcal{K})
    \leq
    \min\{N_s,D\}.
    \label{eq:supp-kernel-rank-bound}
\end{equation}
Consequently, $\mathcal{K}$ is necessarily rank deficient when $N_s>D$. Rank deficiency or ill-conditioning can also occur for $N_s\leq D$ when the evaluated feature
vectors are linearly dependent or nearly so.

Although the exact Gram matrix is positive semi-definite, finite-precision evaluation may introduce small symmetry or positivity violations. Moreover,
a positive semi-definite matrix need not be sufficiently well conditioned for a stable un-regularized solve. Adding a positive diagonal term,
$K_\alpha=K+\alpha I_N$ is the standard noise regularization used in GPR and also improves the conditioning of the linear system
~\cite{rasmussen2006gp,higham2009cholesky}. For $\alpha>0$, $K_\alpha$ is positive definite in exact arithmetic; nevertheless, very small
values of $\alpha$ may be insufficient to prevent numerical failure when the kernel is evaluated in finite precision.

In our implementation, the fermionic feature vectors and Gram matrices are evaluated in single precision. If Cholesky factorization fails for the
validation-selected value of $\alpha$, we multiply $\alpha$ by ten and repeat the factorization. At most four such retries are performed. The final value
of $\alpha$ used by every successful fit is recorded. If the Cholesky factorization fails, $\alpha$ is increased only to make the
matrix numerically invertible. It is not optimized using the log marginal likelihood.

This distinction is important because the initial value of $\alpha$ is a model-selection hyperparameter chosen using validation data, whereas any subsequent increase is triggered solely by the failure of numerical factorization. Fermionic models are fitted with \texttt{optimizer=None}, so neither $\alpha$ nor the parameters defining the
fermionic kernel are modified by the internal optimizer of \texttt{GaussianProcessRegressor}.

\section{Classification with kernel methods}
\label{sec:supp-classification-hyperparameters}

This section provides the complete hyperparameter search underlying the
classification results reported in the main text. The benchmark contains ten models evaluated on 31 dataset configurations, using twenty stratified
train--validation splits for each feasible hyperparameter point. In total,
the search comprises 116 hyperparameter points and 69\,360 completed model
fits. For each model and dataset configuration, the hyperparameter point with the highest mean validation accuracy is selected. The fixed test set is accessed only after this selection, so the test accuracy does not influence the choice of hyperparameters. Most models supply a precomputed kernel to a support vector classifier, whereas \textsc{QKS} and f\textsc{QKS} supply explicit random-feature vectors to logistic regression. The summary statistics of the
search are reported in
\cref{tab:classification-hyperparameter-summary}. The complete point-level
results are given below.
\begin{longtable}{@{}llrrrrr@{}}
  \caption{\textbf{Complete classification hyperparameter search.}
  For each hyperparameter point, $W$, $T$, and $L$ count the dataset
  configurations on which that point is uniquely best, tied for best, or
  neither, respectively, according to the mean validation accuracy.
  Consequently,
  $N_{\mathrm{total}}=W+T+L$ is the number of dataset configurations on
  which the point was evaluated. In most cases
  $N_{\mathrm{total}}=31$. Smaller values occur when a requested PQC width
  is unavailable for low-dimensional inputs or when the order-four
  Majorana sector is not defined. Within each model, the points are ordered
  by their mean validation performance. The final column gives the
  corresponding mean test accuracy. Here $r$ is the number of IQP
  repetitions, $E$ is the number of QKS episodes, $qf$ is the QKS
  qubit-allocation rule, $n$ is the number or allocation of qubits, $m$ is
  the Majorana-sector order, and $C$ is the regularization parameter. For
  \textsc{VanillaMpf+}, $m=\{2,3\}$ denotes the joint use of the
  order-two and order-three Majorana sectors.}
  \label{tab:hyperparameters_full}\\
  \toprule
  Model & Hyperparameter point & $W$ & $T$ & $L$ & $N$ & Test acc. \\
  \midrule
  \endfirsthead
  \caption[]{(continued)}\\
  \toprule
  Model & Hyperparameter point & $W$ & $T$ & $L$ & $N$ & Test acc. \\
  \midrule
  \endhead
  \midrule
  \multicolumn{7}{r}{\footnotesize continued on the next page}\\
  \endfoot
  \bottomrule
  \endlastfoot
  IQP & $r=1$, $C=10$ & 1 & 13 & 17 & 31 & 0.858 \\*
   & $r=1$, $C=1$ & 16 & 1 & 14 & 31 & 0.852 \\*
   & $r=1$, $C=100$ & 0 & 13 & 18 & 31 & 0.853 \\*
   & $r=5$, $C=10$ & 0 & 0 & 31 & 31 & 0.710 \\*
   & $r=5$, $C=100$ & 0 & 0 & 31 & 31 & 0.708 \\*
   & $r=5$, $C=1$ & 0 & 0 & 31 & 31 & 0.704 \\*
   & $r=1$, $C=0.1$ & 1 & 0 & 30 & 31 & 0.630 \\*
   & $r=10$, $C=10$ & 0 & 0 & 31 & 31 & 0.634 \\*
   & $r=10$, $C=100$ & 0 & 0 & 31 & 31 & 0.633 \\*
   & $r=10$, $C=1$ & 0 & 0 & 31 & 31 & 0.618 \\*
   & $r=5$, $C=0.1$ & 0 & 0 & 31 & 31 & 0.534 \\*
   & $r=10$, $C=0.1$ & 0 & 0 & 31 & 31 & 0.523 \\
  \midrule
  fIQP & $r=1$, $C=1$ & 7 & 1 & 23 & 31 & 0.719 \\*
   & $r=1$, $C=10$ & 6 & 3 & 22 & 31 & 0.709 \\*
   & $r=1$, $C=100$ & 5 & 3 & 23 & 31 & 0.703 \\*
   & $r=5$, $C=1$ & 1 & 2 & 28 & 31 & 0.667 \\*
   & $r=5$, $C=10$ & 1 & 4 & 26 & 31 & 0.659 \\*
   & $r=5$, $C=100$ & 0 & 3 & 28 & 31 & 0.655 \\*
   & $r=1$, $C=0.1$ & 2 & 2 & 27 & 31 & 0.629 \\*
   & $r=10$, $C=1$ & 0 & 1 & 30 & 31 & 0.625 \\*
   & $r=10$, $C=10$ & 0 & 1 & 30 & 31 & 0.613 \\*
   & $r=10$, $C=100$ & 0 & 1 & 30 & 31 & 0.607 \\*
   & $r=5$, $C=0.1$ & 1 & 2 & 28 & 31 & 0.573 \\*
   & $r=10$, $C=0.1$ & 0 & 2 & 29 & 31 & 0.554 \\
  \midrule
  QKS & $E=2000$, $qf=\mathrm{half}$ & 20 & 1 & 10 & 31 & 0.850 \\*
   & $E=2000$, $qf=\mathrm{full}$ & 10 & 1 & 20 & 31 & 0.834 \\*
   & $E=500$, $qf=\mathrm{half}$ & 0 & 0 & 31 & 31 & 0.819 \\*
   & $E=500$, $qf=\mathrm{full}$ & 0 & 0 & 31 & 31 & 0.806 \\*
   & $E=100$, $qf=\mathrm{half}$ & 0 & 0 & 31 & 31 & 0.747 \\*
   & $E=100$, $qf=\mathrm{full}$ & 0 & 0 & 31 & 31 & 0.733 \\*
   & $E=10$, $qf=\mathrm{half}$ & 0 & 0 & 31 & 31 & 0.613 \\*
   & $E=10$, $qf=\mathrm{full}$ & 0 & 0 & 31 & 31 & 0.591 \\
  \midrule
  fQKS & $E=2000$, $qf=\mathrm{full}$ & 9 & 5 & 17 & 31 & 0.879 \\*
   & $E=2000$, $qf=\mathrm{half}$ & 8 & 4 & 19 & 31 & 0.877 \\*
   & $E=500$, $qf=\mathrm{full}$ & 5 & 1 & 25 & 31 & 0.869 \\*
   & $E=500$, $qf=\mathrm{half}$ & 3 & 0 & 28 & 31 & 0.860 \\*
   & $E=100$, $qf=\mathrm{full}$ & 1 & 0 & 30 & 31 & 0.831 \\*
   & $E=100$, $qf=\mathrm{half}$ & 0 & 0 & 31 & 31 & 0.807 \\*
   & $E=10$, $qf=\mathrm{full}$ & 0 & 0 & 31 & 31 & 0.704 \\*
   & $E=10$, $qf=\mathrm{half}$ & 0 & 0 & 31 & 31 & 0.686 \\
  \midrule
  PQC & $n=\mathrm{full}$, $C=10$ & 6 & 4 & 21 & 31 & 0.894 \\*
   & $n=\mathrm{full}$, $C=1$ & 16 & 1 & 14 & 31 & 0.888 \\*
   & $n=\mathrm{half}$, $C=10$ & 0 & 1 & 28 & 29 & 0.890 \\*
   & $n=\mathrm{full}$, $C=100$ & 1 & 3 & 27 & 31 & 0.886 \\*
   & $n=\mathrm{half}$, $C=1$ & 0 & 2 & 27 & 29 & 0.884 \\*
   & $n=\mathrm{half}$, $C=100$ & 0 & 1 & 28 & 29 & 0.882 \\*
   & $n=\mathrm{quarter}$, $C=10$ & 0 & 1 & 26 & 27 & 0.875 \\*
   & $n=\mathrm{quarter}$, $C=1$ & 0 & 1 & 26 & 27 & 0.867 \\*
   & $n=\mathrm{quarter}$, $C=100$ & 0 & 0 & 27 & 27 & 0.865 \\*
   & $n=\mathrm{full}$, $C=0.1$ & 0 & 2 & 29 & 31 & 0.811 \\*
   & $n=\mathrm{half}$, $C=0.1$ & 0 & 0 & 29 & 29 & 0.809 \\*
   & $n=\mathrm{quarter}$, $C=0.1$ & 0 & 0 & 27 & 27 & 0.806 \\*
   & $n=\mathrm{eighth}$, $C=10$ & 0 & 0 & 22 & 22 & 0.794 \\*
   & $n=\mathrm{eighth}$, $C=100$ & 0 & 0 & 22 & 22 & 0.791 \\*
   & $n=\mathrm{eighth}$, $C=1$ & 0 & 0 & 22 & 22 & 0.795 \\*
   & $n=\mathrm{eighth}$, $C=0.1$ & 0 & 0 & 22 & 22 & 0.756 \\
  \midrule
  fPQC & $n=\mathrm{double}$, $C=1$ & 16 & 0 & 15 & 31 & 0.888 \\*
   & $n=\mathrm{double}$, $C=10$ & 8 & 2 & 21 & 31 & 0.894 \\*
   & $n=\mathrm{double}$, $C=100$ & 1 & 2 & 28 & 31 & 0.886 \\*
   & $n=\mathrm{full}$, $C=10$ & 1 & 0 & 30 & 31 & 0.817 \\*
   & $n=\mathrm{full}$, $C=100$ & 0 & 0 & 31 & 31 & 0.814 \\*
   & $n=\mathrm{full}$, $C=1$ & 0 & 1 & 30 & 31 & 0.801 \\*
   & $n=\mathrm{double}$, $C=0.1$ & 2 & 0 & 29 & 31 & 0.811 \\*
   & $n=\mathrm{half}$, $C=100$ & 0 & 0 & 29 & 29 & 0.785 \\*
   & $n=\mathrm{half}$, $C=10$ & 0 & 0 & 29 & 29 & 0.782 \\*
   & $n=\mathrm{quarter}$, $C=10$ & 0 & 0 & 27 & 27 & 0.770 \\*
   & $n=\mathrm{quarter}$, $C=100$ & 0 & 0 & 27 & 27 & 0.776 \\*
   & $n=\mathrm{full}$, $C=0.1$ & 0 & 0 & 31 & 31 & 0.726 \\*
   & $n=\mathrm{half}$, $C=1$ & 0 & 1 & 28 & 29 & 0.746 \\*
   & $n=\mathrm{quarter}$, $C=1$ & 0 & 0 & 27 & 27 & 0.737 \\*
   & $n=\mathrm{half}$, $C=0.1$ & 0 & 0 & 29 & 29 & 0.658 \\*
   & $n=\mathrm{eighth}$, $C=100$ & 0 & 0 & 22 & 22 & 0.686 \\*
   & $n=\mathrm{quarter}$, $C=0.1$ & 0 & 0 & 27 & 27 & 0.631 \\*
   & $n=\mathrm{eighth}$, $C=10$ & 0 & 0 & 22 & 22 & 0.685 \\*
   & $n=\mathrm{eighth}$, $C=1$ & 0 & 0 & 22 & 22 & 0.675 \\*
   & $n=\mathrm{eighth}$, $C=0.1$ & 0 & 0 & 22 & 22 & 0.611 \\
  \midrule
  Vanilla & $n=d$, $C=10$ & 4 & 4 & 23 & 31 & 0.851 \\*
   & $n=d$, $C=1$ & 17 & 2 & 12 & 31 & 0.848 \\*
   & $n=d$, $C=100$ & 5 & 4 & 22 & 31 & 0.848 \\*
   & $n=32$, $C=10$ & 0 & 1 & 30 & 31 & 0.726 \\*
   & $n=32$, $C=100$ & 0 & 1 & 30 & 31 & 0.725 \\*
   & $n=d$, $C=0.1$ & 1 & 1 & 29 & 31 & 0.688 \\*
   & $n=32$, $C=1$ & 0 & 1 & 30 & 31 & 0.719 \\*
   & $n=32$, $C=0.1$ & 0 & 1 & 30 & 31 & 0.528 \\
  \midrule
  VanillaMpf & $n=d$, $m=4$, $C=0.1$ & 2 & 4 & 23 & 29 & 0.875 \\*
   & $n=d$, $m=2$, $C=0.1$ & 8 & 1 & 22 & 31 & 0.858 \\*
   & $n=32$, $m=2$, $C=0.1$ & 6 & 3 & 22 & 31 & 0.857 \\*
   & $n=d$, $m=4$, $C=1$ & 1 & 3 & 25 & 29 & 0.873 \\*
   & $n=d$, $m=4$, $C=10$ & 1 & 3 & 25 & 29 & 0.870 \\*
   & $n=d$, $m=4$, $C=100$ & 0 & 3 & 26 & 29 & 0.867 \\*
   & $n=d$, $m=2$, $C=1$ & 2 & 2 & 27 & 31 & 0.860 \\*
   & $n=64$, $m=2$, $C=0.1$ & 0 & 1 & 30 & 31 & 0.844 \\*
   & $n=32$, $m=2$, $C=1$ & 0 & 0 & 31 & 31 & 0.848 \\*
   & $n=64$, $m=2$, $C=1$ & 0 & 0 & 31 & 31 & 0.838 \\*
   & $n=64$, $m=2$, $C=10$ & 0 & 0 & 31 & 31 & 0.835 \\*
   & $n=64$, $m=2$, $C=100$ & 0 & 0 & 31 & 31 & 0.833 \\*
   & $n=32$, $m=2$, $C=10$ & 0 & 0 & 31 & 31 & 0.844 \\*
   & $n=d$, $m=2$, $C=10$ & 2 & 1 & 28 & 31 & 0.854 \\*
   & $n=32$, $m=2$, $C=100$ & 0 & 0 & 31 & 31 & 0.840 \\*
   & $n=d$, $m=2$, $C=100$ & 1 & 1 & 29 & 31 & 0.852 \\
  \midrule
  VanillaMpf+ & $n=d$, $m=\{2,3\}$, $C=0.1$ & 5 & 4 & 22 & 31 & 0.869 \\*
   & $n=d$, $m=\{2,3\}$, $C=1$ & 3 & 5 & 23 & 31 & 0.867 \\*
   & $n=32$, $m=2$, $C=0.1$ & 6 & 1 & 24 & 31 & 0.857 \\*
   & $n=d$, $m=2$, $C=0.1$ & 3 & 3 & 25 & 31 & 0.859 \\*
   & $n=d$, $m=2$, $C=1$ & 0 & 0 & 31 & 31 & 0.862 \\*
   & $n=d$, $m=\{2,3\}$, $C=10$ & 0 & 4 & 27 & 31 & 0.863 \\*
   & $n=d$, $m=\{2,3\}$, $C=100$ & 1 & 4 & 26 & 31 & 0.861 \\*
   & $n=64$, $m=2$, $C=0.1$ & 2 & 0 & 29 & 31 & 0.844 \\*
   & $n=64$, $m=2$, $C=1$ & 0 & 0 & 31 & 31 & 0.838 \\*
   & $n=64$, $m=2$, $C=10$ & 0 & 0 & 31 & 31 & 0.836 \\*
   & $n=64$, $m=2$, $C=100$ & 0 & 0 & 31 & 31 & 0.834 \\*
   & $n=d$, $m=2$, $C=10$ & 2 & 1 & 28 & 31 & 0.856 \\*
   & $n=32$, $m=2$, $C=1$ & 0 & 0 & 31 & 31 & 0.848 \\*
   & $n=d$, $m=2$, $C=100$ & 1 & 0 & 30 & 31 & 0.852 \\*
   & $n=32$, $m=2$, $C=10$ & 0 & 0 & 31 & 31 & 0.844 \\*
   & $n=32$, $m=2$, $C=100$ & 0 & 0 & 31 & 31 & 0.840 \\
  \midrule
  AngleMpf & $n=d$, $m=2$, $C=100$ & 15 & 2 & 14 & 31 & 0.796 \\*
   & $n=d$, $m=2$, $C=1$ & 2 & 3 & 26 & 31 & 0.800 \\*
   & $n=d$, $m=2$, $C=10$ & 4 & 2 & 25 & 31 & 0.798 \\*
   & $n=d$, $m=2$, $C=0.1$ & 6 & 1 & 24 & 31 & 0.777 \\
\end{longtable}
\paragraph{Comparison with entangling PQC circuits.}

In the main benchmark, hyperparameters are selected independently for every
model and dataset configuration according to the mean validation accuracy. For
\textsc{PQC}, this procedure selects the full-width, depth-one circuit in most
configurations. At depth one, the data-dependent single-qubit rotations are
followed by a fixed CNOT layer. Since the same CNOT layer appears at the end
of both circuits entering the fidelity overlap, it cancels from the kernel.
The selected depth-one \textsc{PQC} therefore reduces to a product of
single-qubit kernels, and its performance does not measure the contribution
of a multilayer entangling circuit.

\begin{figure*}[h]
    \centering
    \includegraphics[width=1\linewidth]{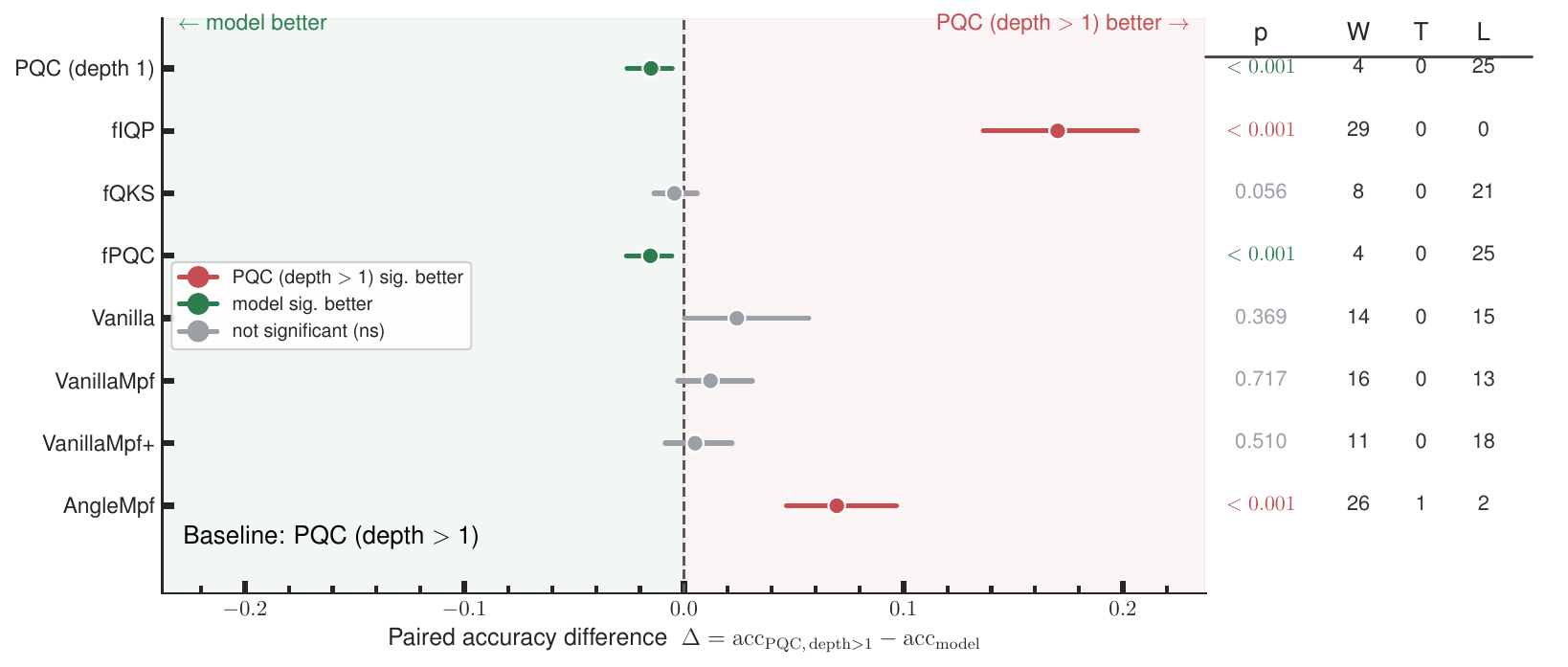}
    \caption{\textbf{Paired comparison with depth-greater-than-one
    \textsc{PQC}.} The \textsc{PQC} baseline is restricted to circuits of depth greater than
    one, and its best hyperparameter point is selected independently for each
    dataset configuration using mean validation accuracy. The comparison
    models retain the hyperparameters selected from their usual full grids.
    The analysis includes the $29$ dataset configurations for which a
    depth-greater-than-one \textsc{PQC} circuit is available. For every
    configuration, test accuracy is averaged over the $20$ data splits and
    the paired difference is defined as
    $\Delta_q=
    \overline{\operatorname{Acc}}_{\mathrm{PQC},\,\mathrm{depth}>1,q}
    -\overline{\operatorname{Acc}}_{\mathrm{model},q}$.
    Points show the mean paired difference, and horizontal bars show the
    corresponding $95\%$ percentile-bootstrap confidence intervals obtained
    from $10^4$ resamples of the configuration-level differences. Negative
    values favor the comparison model, whereas positive values favor the
    depth-greater-than-one \textsc{PQC}. The reported $p$-values are obtained
    from two-sided Wilcoxon signed-rank tests. The columns $W$, $T$, and $L$
    give the numbers of configurations for which the
    depth-greater-than-one \textsc{PQC} performs better than, ties with, or
    performs worse than the comparison model, respectively. Red and green
    markers denote statistically significant differences favoring the
    depth-greater-than-one \textsc{PQC} and the comparison model,
    respectively. Gray markers denote non-significant differences.}
    \label{fig:classification-pqc-depth-forest}
\end{figure*}

To examine whether additional circuit depth changes the comparison, we
repeat the analysis after restricting the \textsc{PQC} baseline to circuits
of depth greater than one. For each dataset configuration, the best circuit
within this restricted grid is selected using mean validation accuracy. All
other models retain the hyperparameters selected from their usual full
grids. A depth-greater-than-one \textsc{PQC} circuit is available for $29$ of
the $31$ dataset configurations; the remaining two configurations are too
low-dimensional to produce more than one encoding layer under the considered
qubit-allocation rules. Figure~\ref{fig:classification-pqc-depth-forest}
compares this restricted \textsc{PQC} baseline with the validation-selected
depth-one \textsc{PQC} and the fermionic models using the same paired
statistical procedure as in the main text.

This comparison provides no evidence that increasing the \textsc{PQC} depth improves classification performance. The validation-selected depth-one \textsc{PQC} outperforms the depth-greater-than-one baseline on $25$ of the $29$ configurations, while the deeper circuit performs better on only four
($p<0.001$). f\textsc{PQC} shows the same win-loss count and also significantly outperforms the deeper \textsc{PQC} baseline
($p<0.001$). Thus, the strong performance of f\textsc{PQC} cannot be
attributed solely to its comparison with a shallow \textsc{PQC}; it remains
strong when the baseline is explicitly restricted to multilayer entangling
circuits.

f\textsc{QKS} performs better than the depth-greater-than-one \textsc{PQC}
on $21$ configurations and worse on eight, although the paired difference
does not reach the conventional significance threshold ($p=0.056$).
\textsc{Vanilla}, \textsc{VanillaMpf}, and \textsc{VanillaMpf+} are also
statistically indistinguishable from the deeper \textsc{PQC}
($p=0.369$, $0.717$, and $0.510$, respectively). In contrast, the deeper
\textsc{PQC} significantly outperforms f\textsc{IQP} on all $29$
configurations and \textsc{AngleMpf} on $26$ configurations, with one tie
($p<0.001$ in both comparisons).

These results show that adding entangling layers \emph{does not} improve
generalization in this benchmark. Indeed, the depth-one product kernel is
more successful than the restricted multilayer \textsc{PQC}, while several
efficiently simulable fermionic models remain competitive with or outperform
the deeper circuit.
\subsection{Dimension-resolved classification performance}
\label{subsec:classification-dimension-resolved}

The violin plots in the main text pool results obtained at different input, ambient, or latent dimensions. Although these aggregated distributions summarize overall performance, they can conceal systematic changes in classification difficulty and model generalization as the dataset dimension varies. Figure~\ref{fig:classification_kernel_vs_dim} therefore reports the mean training and test accuracies separately for each dataset configuration, with shaded regions indicating the corresponding $95\%$ confidence intervals. Bars-and-stripes is omitted because it is evaluated at only one image size, $4\times4$.

The dimension-resolved results include the three matched architectural pairs \textsc{IQP}-f\textsc{IQP}, \textsc{QKS}-f\textsc{QKS}, and \textsc{PQC}-f\textsc{PQC}, together with the native free-fermionic models \textsc{Vanilla}, \textsc{VanillaMpf}, \textsc{VanillaMpf+}, and \textsc{AngleMpf}. To isolate the effect of imposing a matchgate restriction, Fig.~\ref{fig:classification_kernel_diffvs_dim} additionally reports the accuracy differences
$$
    \Delta_{\mathrm{IQP}}
    =
    \operatorname{Acc}_{\mathrm{fIQP}}
    -
    \operatorname{Acc}_{\mathrm{IQP}},
$$
and analogously for the \textsc{QKS} and \textsc{PQC} pairs. Positive values
therefore favor the fermionic counterpart, whereas negative values favor the
unrestricted model.
\subsubsection{Classification Accuracy versus dataset dimension}

\begin{figure*}[tbp!]
    \centering
    \includegraphics[width=1\linewidth]{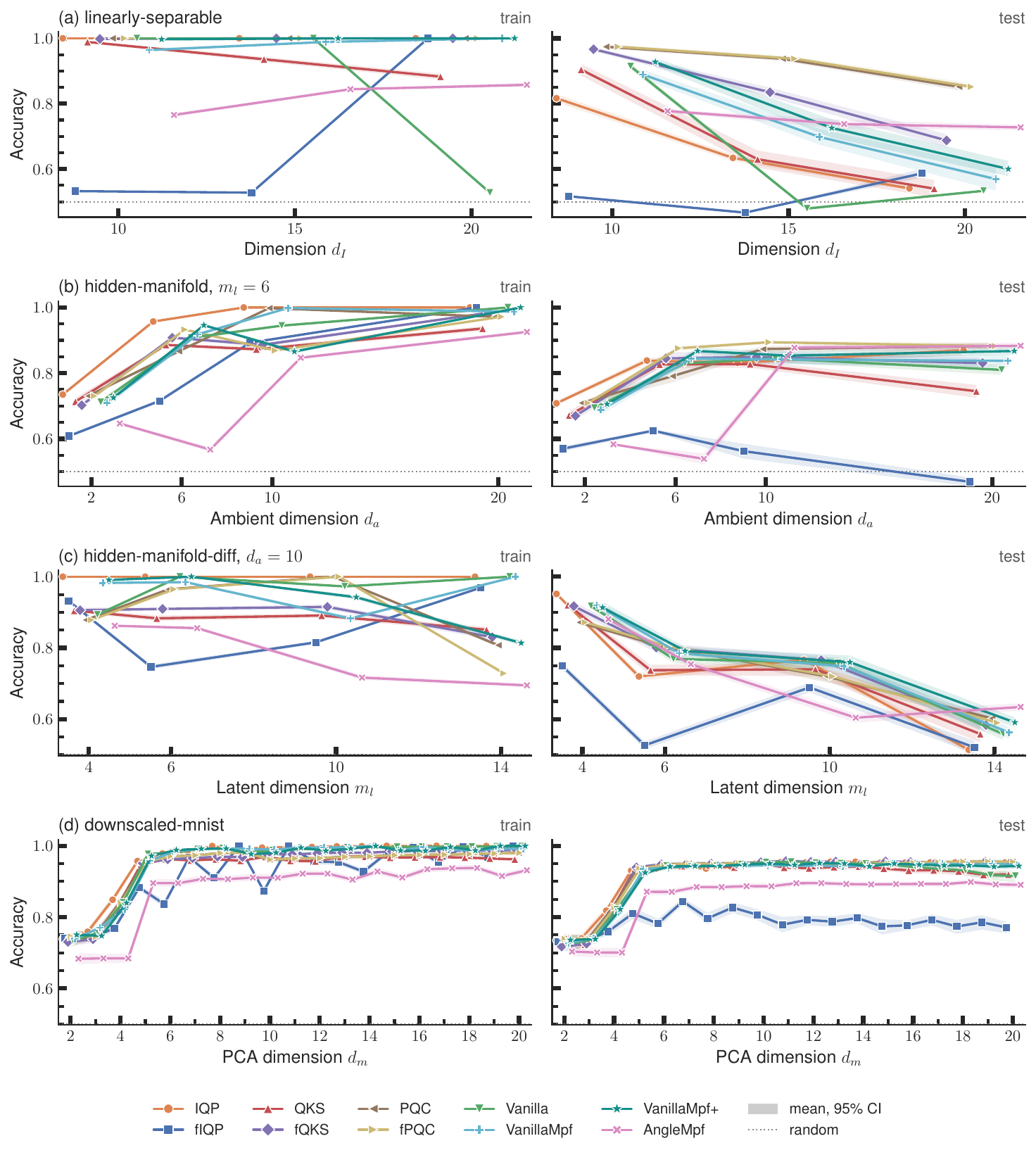}
    \caption{\textbf{Dimension-resolved training and test accuracy.} Points show the mean accuracy over the $20$ data splits, and shaded bands show the corresponding $95\%$ confidence intervals. Results include the three matched architectural pairs \textsc{IQP}-f\textsc{IQP}, \textsc{QKS}-f\textsc{QKS}, and \textsc{PQC}--f\textsc{PQC}, together with the native free-fermionic models \textsc{Vanilla}, \textsc{VanillaMpf}, \textsc{VanillaMpf+}, and \textsc{AngleMpf}. Hyperparameters are selected independently for each model and dataset configuration using the validation data. The panels show (a) the linearly separable datasets as a function of the input dimension $d_I$; (b) the hidden-manifold datasets as a function of the ambient dimension $d_a$, with the latent dimension fixed at $m_l=6$; (c) the hidden-manifold-diff datasets as a function of the latent dimension $m_l$, with the ambient dimension fixed at $d_a=10$; and (d) downscaled MNIST as a function of the number of retained PCA components $d_m$. Training and test accuracies are displayed separately to reveal dimension-dependent generalization gaps. The dotted horizontal line marks the random-classification accuracy of $0.5$.}
    \label{fig:classification_kernel_vs_dim}
\end{figure*}

For the \textbf{linearly separable datasets},
Fig.~\ref{fig:classification_kernel_vs_dim}(a), test accuracy generally decreases as the input dimension $d_I$ increases. This trend is especially clear for the \textsc{QKS} and native free-fermionic models, several of which perform well at $d_I=10$ but approach random classification at the largest dimension. \emph{f\textsc{QKS} is more robust than \textsc{QKS} throughout
the sweep and retains a clear advantage at $d_I=15$ and $20$.}
\textsc{PQC} gives the strongest and most stable test performance, whereas f\textsc{PQC} remains consistently below it. The \textsc{IQP} pair performs poorly on the test data despite attaining much higher training accuracies in several configurations. Among the native fermionic models, \textsc{AngleMpf} varies less strongly with $d_I$ than the Vanilla-based models, although it does not
match the performance of \textsc{PQC}. The separation between the training and test curves shows that fitting the training data does not ensure that the learned representation preserves the linear decision boundary on unseen examples.

For the \textbf{hidden manifold datasets} with fixed latent dimension
$m_l=6$, Fig.~\ref{fig:classification_kernel_vs_dim}(b), most models improve between the smallest ambient dimension and the intermediate values $d_a=6$ and $10$. The principal exception is f\textsc{IQP}, whose test accuracy decreases as $d_a$ increases even though its training accuracy approaches unity. This produces the largest generalization gap in this dataset family. \emph{f\textsc{QKS} performs similarly to \textsc{QKS} at small $d_a$ and becomes more accurate at the largest ambient dimension.} f\textsc{PQC} performs poorly relative to \textsc{PQC} at $d_a=6$, but the two models become comparable at $d_a=10$ and $20$. The native free-fermionic models also perform competitively at the intermediate and
large ambient dimensions. In particular, the initially weaker
\textsc{AngleMpf} improves substantially once the ambient dimension reaches $d_a=10$.

Figure~\ref{fig:classification_kernel_vs_dim}(c) shows the
\textbf{hidden manifold-diff datasets}, for which the ambient dimension is fixed at $d_a=10$ and the latent dimension $m_l$ is varied. Test accuracy decreases for nearly every model as $m_l$ increases, demonstrating that the latent dimension directly controls the difficulty of the classification problem. Several models achieve accuracies near or above $0.9$ at $m_l=4$, whereas most approach $0.5$-$0.6$ by $m_l=14$. The decline in test performance occurs even though many of the corresponding training accuracies remain close to unity, indicating a growing generalization gap. \emph{f\textsc{QKS} is slightly more accurate than \textsc{QKS} over most of the
sweep.} By contrast, f\textsc{IQP} is substantially worse than \textsc{IQP} at the smaller and intermediate latent dimensions. The relative performance of \textsc{PQC} and f\textsc{PQC} is not uniform: f\textsc{PQC} is competitive at $m_l=4$, but falls below \textsc{PQC} at the intermediate latent dimensions. The native free-fermionic models remain among the stronger models at several intermediate configurations, but they also deteriorate as $m_l$ increases.

The \textbf{downscaled-MNIST} results in
Fig.~\ref{fig:classification_kernel_vs_dim}(d) differ substantially from the previous results. Most models improve rapidly as the number of retained PCA components increases from $d_m=2$ to approximately $d_m=5$, after which their accuracies become comparatively stable. \textsc{IQP}, \textsc{QKS}, f\textsc{QKS}, \textsc{PQC}, and several of the native free-fermionic models all maintain high test accuracy over the remainder of the sweep. f\textsc{QKS} has a small but persistent advantage over \textsc{QKS}. \emph{The Vanilla-based Pfaffian models also perform strongly, demonstrating that their broad pooled distributions in the main text should not be interpreted
as uniformly poor performance.} \textsc{AngleMpf} remains below the other native models but still performs well above random classification. f\textsc{PQC} follows the same dimensional trend as \textsc{PQC}, although with consistently lower test accuracy. The clearest degradation caused by the matchgate restriction occurs for f\textsc{IQP}, whose accuracy remains well below that of \textsc{IQP} once more than a few PCA components are
retained.

\subsubsection{Accuracy difference between fermionic and unrestricted counterparts}

\begin{figure*}[tbp!]
    \centering
    \includegraphics[width=1\linewidth]{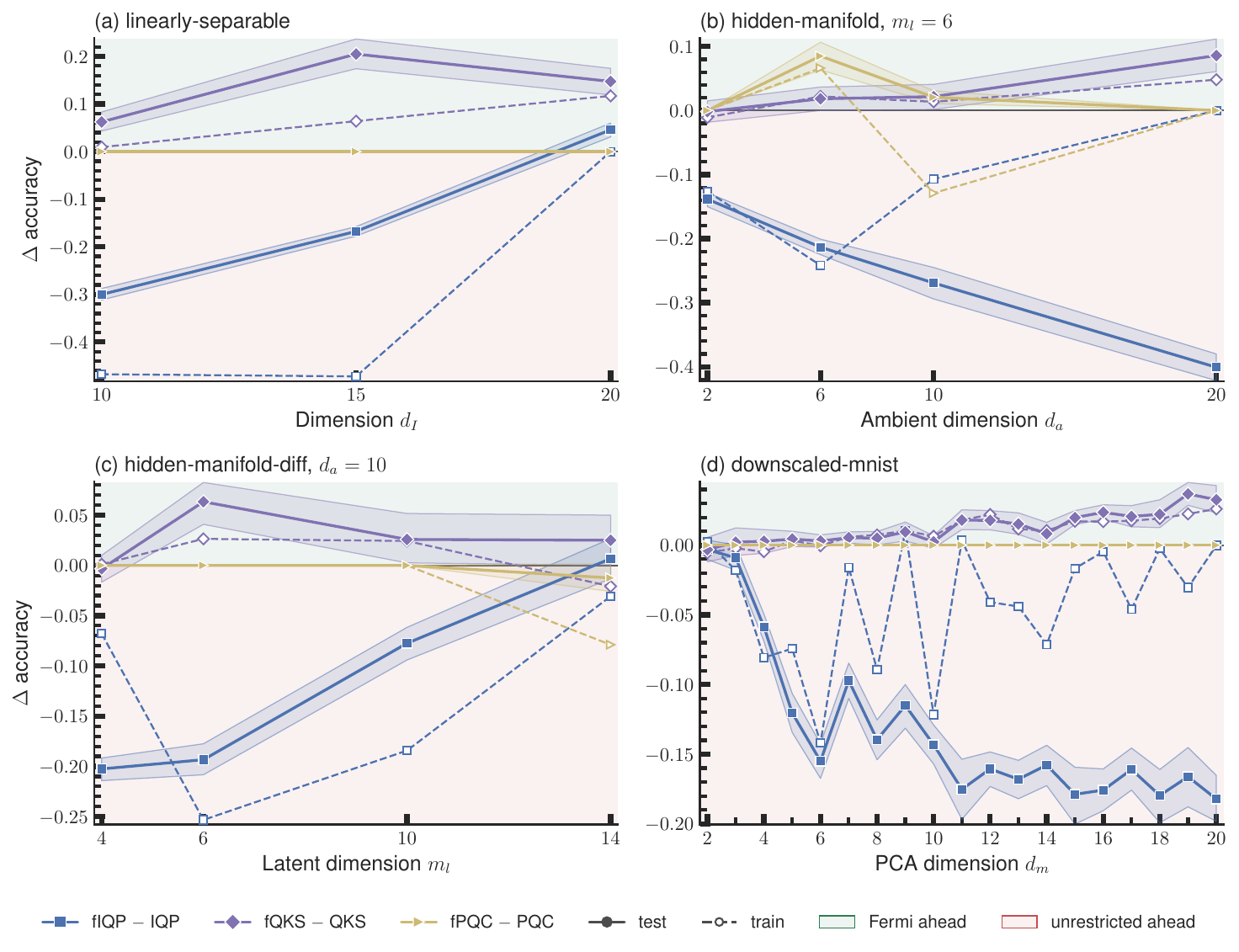}
    \caption{\textbf{Accuracy differences between fermionic models and their unrestricted counterparts.} The plotted differences are $\Delta_{\mathrm{IQP}} =\operatorname{Acc}(\mathrm{fIQP})-\operatorname{Acc}(\mathrm{IQP})$,$\Delta_{\mathrm{QKS}}=\operatorname{Acc}(\mathrm{fQKS})-\operatorname{Acc}(\mathrm{QKS})$, and $\Delta_{\mathrm{PQC}} =\operatorname{Acc}(\mathrm{fPQC})-\operatorname{Acc}(\mathrm{PQC})$. Solid lines with filled markers show the test-accuracy differences, while dashed lines with open markers show the training-accuracy differences. Shaded bands around the test curves denote the corresponding $95\%$ confidence intervals over the $20$ data splits. Positive values, shown against the green background, favor the fermionic model; negative values, shown against the red background, favor its unrestricted counterpart. The horizontal line at zero indicates equal mean accuracy. Panels use the same dataset-configuration sweeps as Fig.~\ref{fig:classification_kernel_vs_dim}.}
\label{fig:classification_kernel_diffvs_dim}
\end{figure*}

Figure~\ref{fig:classification_kernel_diffvs_dim} directly compares each unrestricted model with its fermionic counterpart. The results show that the effect of the matchgate restriction depends on both the underlying architecture and the dataset. It generally improves the \textsc{QKS} construction, reduces the performance of the \textsc{IQP} construction, and has a more configuration-dependent effect on \textsc{PQC}.

For \textbf{linearly separable datasets},
Fig.~\ref{fig:classification_kernel_diffvs_dim}(a), \emph{f\textsc{QKS}
outperforms \textsc{QKS} at every input dimension, with its largest test advantage occurring at $d_I=15$.} In contrast, f\textsc{PQC} remains below \textsc{PQC} throughout the sweep. f\textsc{IQP} is substantially worse than \textsc{IQP} at $d_I=10$ and $15$, but the difference closes and slightly reverses at $d_I=20$, where both models have relatively low test accuracy. The near equality at the largest dimension should therefore not be interpreted as evidence that either IQP construction performs strongly.

For the \textbf{hidden manifold datasets},
Fig.~\ref{fig:classification_kernel_diffvs_dim}(b), the f\textsc{IQP}
disadvantage grows with the ambient dimension and becomes particularly large at $d_a=20$. f\textsc{QKS} is approximately comparable to \textsc{QKS} at the smallest dimension and develops an increasing advantage as $d_a$ increases. The \textsc{PQC} comparison behaves differently: f\textsc{PQC} has a pronounced deficit at $d_a=6$, but the difference becomes small at $d_a=10$ and $20$. Thus, the pooled similarity between the two PQC models conceals a substantial difference at one particular ambient dimension.

For the \textbf{hidden manifold-diff datasets},
Fig.~\ref{fig:classification_kernel_diffvs_dim}(c), \emph{f\textsc{QKS} is
comparable to or slightly better than \textsc{QKS} across the latent-dimension sweep.} f\textsc{IQP} is substantially worse than \textsc{IQP} at $m_l=4$ and $6$, but the difference narrows as both models approach random classification at the largest latent dimension. f\textsc{PQC} is slightly better than \textsc{PQC} at $m_l=4$, but falls below it at $m_l=6$ and $10$ before the difference again becomes small at $m_l=14$. This changing
ordering explains why the pooled comparison between the PQC models is not statistically decisive.

Finally, the \textbf{downscaled-MNIST} comparison in
Fig.~\ref{fig:classification_kernel_diffvs_dim}(d) \emph{shows the most consistent architecture-dependent behavior. Once the number of PCA components exceeds the smallest values, f\textsc{QKS} maintains a modest advantage over \textsc{QKS}, while f\textsc{PQC} remains moderately below \textsc{PQC}.} The f\textsc{IQP} deficit is considerably larger and persists across almost the entire dimensional sweep, reaching its greatest magnitude at the larger
values of $d_m$. The training differences for f\textsc{IQP} fluctuate more strongly than the test differences, but do not alter the overall conclusion that the fIQP architecture generalizes less effectively on this benchmark.

Taken together, these results show that fermionic Gaussian structure is not associated with a universal gain or loss in predictive performance. Its effect depends on how the restriction interacts with the encoding architecture. The matchgate version of \textsc{QKS} is consistently competitive and is often more accurate than its unrestricted counterpart. The same restriction is generally detrimental to \textsc{IQP}, while the difference between \textsc{PQC} and f\textsc{PQC} depends more strongly on the dataset. The native free-fermionic models provide an additional comparison: despite having no unrestricted architectural counterpart, several achieve performance comparable to the strongest
circuit-based baselines over substantial parts of these sweeps.
\section{NIF-Ising Classification Benchmark}
In this section, we provide the details of the hyperparameters per seed/split for the various text (AG News, IMDb) and image (CIFAR2, Fashion-MNIST) datasets classified using the NIF-Ising model.
\begin{table*}[tbp!]
\caption{Hyperparameters per seed/split for the \textbf{AG News} dataset.The table lists all evaluated configurations and their validation results. The selected configuration was determined using validation accuracy within
each split. The held-out test set was evaluated only after selection.}
\label{tab:hparams-agnews}
\resizebox{\textwidth}{!}{\begin{tabular}{rlrrrrrrrl}
\toprule
split\_id & model & batch\_size & opt.automl\_iterations & opt.inner\_max\_epochs & opt.max\_epochs & max\_epochs & learning\_rate & n\_qubits & encoder\_output\_activation \\
\midrule
0 & NIFIsingCls & 256 & 20 & 128 & 1024 & 1024 & 1e-05 & 75 & Tanh \\
1 & NIFIsingCls & 256 & 20 & 128 & 1024 & 1024 & 1e-05 & 42 & Tanh \\
2 & NIFIsingCls & 256 & 20 & 128 & 1024 & 1024 & 1e-05 & 75 & Tanh \\
3 & NIFIsingCls & 256 & 20 & 128 & 1024 & 1024 & 1e-05 & 69 & Tanh \\
4 & NIFIsingCls & 256 & 20 & 128 & 1024 & 1024 & 1e-05 & 29 & Tanh \\
5 & NIFIsingCls & 256 & 20 & 128 & 1024 & 1024 & 1e-05 & 69 & Tanh \\
6 & NIFIsingCls & 256 & 20 & 128 & 1024 & 1024 & 1e-05 & 76 & Tanh \\
7 & NIFIsingCls & 256 & 20 & 128 & 1024 & 1024 & 1e-05 & 48 & Tanh \\
\textbf{8} & \textbf{NIFIsingCls} & \textbf{256} & \textbf{20} & \textbf{128} & \textbf{1024} & \textbf{1024} & \textbf{1e-05} & \textbf{92} & \textbf{Tanh} \\
9 & NIFIsingCls & 256 & 20 & 128 & 1024 & 1024 & 1e-05 & 55 & Tanh \\
10 & NIFIsingCls & 256 & 20 & 128 & 1024 & 1024 & 1e-05 & 36 & Tanh \\
11 & NIFIsingCls & 256 & 20 & 128 & 1024 & 1024 & 1e-05 & 56 & Tanh \\
12 & NIFIsingCls & 256 & 20 & 128 & 1024 & 1024 & 1e-05 & 45 & Tanh \\
\textbf{13} & \textbf{NIFIsingCls} & \textbf{256} & \textbf{20} & \textbf{128} & \textbf{1024} & \textbf{1024} & \textbf{1e-05} & \textbf{92} & \textbf{Tanh} \\
14 & NIFIsingCls & 256 & 20 & 128 & 1024 & 1024 & 1e-05 & 73 & Tanh \\
15 & NIFIsingCls & 256 & 20 & 128 & 1024 & 1024 & 1e-05 & 61 & Tanh \\
\textbf{16} & \textbf{NIFIsingCls} & \textbf{256} & \textbf{20} & \textbf{128} & \textbf{1024} & \textbf{1024} & \textbf{1e-05} & \textbf{92} & \textbf{Tanh} \\
17 & NIFIsingCls & 256 & 20 & 128 & 1024 & 1024 & 1e-05 & 87 & Tanh \\
18 & NIFIsingCls & 256 & 20 & 128 & 1024 & 1024 & 1e-05 & 56 & Tanh \\
19 & NIFIsingCls & 256 & 20 & 128 & 1024 & 1024 & 1e-05 & 30 & Tanh \\
\bottomrule
\end{tabular}}
\end{table*}


\begin{table*}[tbp!]
\caption{Hyperparameters per seed/split for the \textbf{CIFAR2} dataset. The table lists all evaluated configurations and their validation results. The selected configuration was determined using validation accuracy within
each split; the held-out test set was evaluated only after selection.}
\label{tab:hparams-cifar2}
\resizebox{\textwidth}{!}{\begin{tabular}{rlrrrrrrrl}
\toprule
split\_id & model & batch\_size & opt.automl\_iterations & opt.inner\_max\_epochs & opt.max\_epochs & max\_epochs & learning\_rate & n\_qubits & encoder\_output\_activation \\
\midrule
0 & NIFIsingCls & 256 & 20 & 128 & 1024 & 1024 & 0.00855718 & 53 & Tanh \\
1 & NIFIsingCls & 256 & 20 & 128 & 1024 & 1024 & 1e-05 & 4 & Tanh \\
2 & NIFIsingCls & 256 & 20 & 128 & 1024 & 1024 & 0.00612338 & 43 & Tanh \\
3 & NIFIsingCls & 256 & 20 & 128 & 1024 & 1024 & 1e-05 & 35 & Tanh \\
4 & NIFIsingCls & 256 & 20 & 128 & 1024 & 1024 & 0.00710432 & 24 & Tanh \\
5 & NIFIsingCls & 256 & 20 & 128 & 1024 & 1024 & 1e-05 & 18 & Tanh \\
6 & NIFIsingCls & 256 & 20 & 128 & 1024 & 1024 & 0.00593611 & 41 & Tanh \\
7 & NIFIsingCls & 256 & 20 & 128 & 1024 & 1024 & 0.00802518 & 52 & Tanh \\
8 & NIFIsingCls & 256 & 20 & 128 & 1024 & 1024 & 0.0103934 & 62 & Tanh \\
9 & NIFIsingCls & 256 & 20 & 128 & 1024 & 1024 & 0.0108232 & 37 & Tanh \\
\textbf{10} & \textbf{NIFIsingCls} & \textbf{256} & \textbf{20} & \textbf{128} & \textbf{1024} & \textbf{1024} & \textbf{0.0115764} & \textbf{92} & \textbf{Tanh} \\
11 & NIFIsingCls & 256 & 20 & 128 & 1024 & 1024 & 1e-05 & 26 & Tanh \\
\textbf{12} & \textbf{NIFIsingCls} & \textbf{256} & \textbf{20} & \textbf{128} & \textbf{1024} & \textbf{1024} & \textbf{0.008502} & \textbf{92} & \textbf{Tanh} \\
13 & NIFIsingCls & 256 & 20 & 128 & 1024 & 1024 & 0.0109247 & 67 & Tanh \\
14 & NIFIsingCls & 256 & 20 & 128 & 1024 & 1024 & 1e-05 & 92 & Tanh \\
15 & NIFIsingCls & 256 & 20 & 128 & 1024 & 1024 & 1e-05 & 25 & Tanh \\
16 & NIFIsingCls & 256 & 20 & 128 & 1024 & 1024 & 0.00614892 & 52 & Tanh \\
17 & NIFIsingCls & 256 & 20 & 128 & 1024 & 1024 & 1e-05 & 45 & Tanh \\
18 & NIFIsingCls & 256 & 20 & 128 & 1024 & 1024 & 0.00906249 & 57 & Tanh \\
\textbf{19} & \textbf{NIFIsingCls} & \textbf{256} & \textbf{20} & \textbf{128} & \textbf{1024} & \textbf{1024} & \textbf{0.00716284} & \textbf{92} & \textbf{Tanh} \\
\bottomrule
\end{tabular}}
\end{table*}


\begin{table*}[tbp!]
\caption{Hyperparameters per seed/split for the \textbf{FashionMNIST} dataset. The table lists all evaluated configurations and their validation results. The selected configuration was determined using validation accuracy within
each split; the held-out test set was evaluated only after selection.}
\label{tab:hparams-fashion}
\resizebox{\textwidth}{!}{\begin{tabular}{rlrrrrrrrl}
\toprule
split\_id & model & batch\_size & opt.automl\_iterations & opt.inner\_max\_epochs & opt.max\_epochs & max\_epochs & learning\_rate & n\_qubits & encoder\_output\_activation \\
\midrule
0 & NIFIsingCls & 256 & 20 & 128 & 1024 & 1024 & 1e-05 & 61 & Tanh \\
1 & NIFIsingCls & 256 & 20 & 128 & 1024 & 1024 & 1e-05 & 71 & Tanh \\
2 & NIFIsingCls & 256 & 20 & 128 & 1024 & 1024 & 1e-05 & 30 & Tanh \\
3 & NIFIsingCls & 256 & 20 & 128 & 1024 & 1024 & 1e-05 & 71 & Tanh \\
4 & NIFIsingCls & 256 & 20 & 128 & 1024 & 1024 & 1e-05 & 48 & Tanh \\
5 & NIFIsingCls & 256 & 20 & 128 & 1024 & 1024 & 1e-05 & 37 & Tanh \\
6 & NIFIsingCls & 256 & 20 & 128 & 1024 & 1024 & 1e-05 & 50 & Tanh \\
7 & NIFIsingCls & 256 & 20 & 128 & 1024 & 1024 & 1e-05 & 74 & Tanh \\
8 & NIFIsingCls & 256 & 20 & 128 & 1024 & 1024 & 1e-05 & 71 & Tanh \\
9 & NIFIsingCls & 256 & 20 & 128 & 1024 & 1024 & 1e-05 & 57 & Tanh \\
10 & NIFIsingCls & 256 & 20 & 128 & 1024 & 1024 & 1e-05 & 37 & Identity \\
11 & NIFIsingCls & 256 & 20 & 128 & 1024 & 1024 & 1e-05 & 64 & Tanh \\
12 & NIFIsingCls & 256 & 20 & 128 & 1024 & 1024 & 1e-05 & 37 & Tanh \\
13 & NIFIsingCls & 256 & 20 & 128 & 1024 & 1024 & 1e-05 & 51 & Tanh \\
14 & NIFIsingCls & 256 & 20 & 128 & 1024 & 1024 & 1e-05 & 64 & Tanh \\
\textbf{15} & \textbf{NIFIsingCls} & \textbf{256} & \textbf{20} & \textbf{128} & \textbf{1024} & \textbf{1024} & \textbf{1e-05} & \textbf{87} & \textbf{Tanh} \\
16 & NIFIsingCls & 256 & 20 & 128 & 1024 & 1024 & 1e-05 & 45 & Tanh \\
17 & NIFIsingCls & 256 & 20 & 128 & 1024 & 1024 & 1e-05 & 31 & Tanh \\
18 & NIFIsingCls & 256 & 20 & 128 & 1024 & 1024 & 1e-05 & 60 & Tanh \\
19 & NIFIsingCls & 256 & 20 & 128 & 1024 & 1024 & 1e-05 & 79 & Tanh \\
\bottomrule
\end{tabular}}
\end{table*}
\begin{table}[tbp!]
\caption{Hyperparameters per seed/split for the \textbf{IMDb} dataset. The table lists all evaluated configurations and their validation results. The selected configuration was determined using validation accuracy within
each split; the held-out test set was evaluated only after selection.}
\label{tab:hparams-imdb}
\resizebox{\textwidth}{!}{\begin{tabular}{rlrrrrrrrl}
\toprule
split\_id & model & batch\_size & opt.automl\_iterations & opt.inner\_max\_epochs & opt.max\_epochs & max\_epochs & learning\_rate & n\_qubits & encoder\_output\_activation \\
\midrule
0 & NIFIsingCls & 64 & 20 & 128 & 1024 & 1024 & 1e-05 & 92 & Tanh \\
1 & NIFIsingCls & 64 & 20 & 128 & 1024 & 1024 & 1e-05 & 51 & Tanh \\
2 & NIFIsingCls & 64 & 20 & 128 & 1024 & 1024 & 1e-05 & 60 & Tanh \\
3 & NIFIsingCls & 64 & 20 & 128 & 1024 & 1024 & 1e-05 & 66 & Tanh \\
4 & NIFIsingCls & 64 & 20 & 128 & 1024 & 1024 & 0.00531045 & 4 & Tanh \\
\textbf{5} & \textbf{NIFIsingCls} & \textbf{64} & \textbf{20} & \textbf{128} & \textbf{1024} & \textbf{1024} & \textbf{1e-05} & \textbf{92} & \textbf{Tanh} \\
6 & NIFIsingCls & 64 & 20 & 128 & 1024 & 1024 & 1e-05 & 68 & Tanh \\
7 & NIFIsingCls & 64 & 20 & 128 & 1024 & 1024 & 1e-05 & 61 & Tanh \\
8 & NIFIsingCls & 64 & 20 & 128 & 1024 & 1024 & 1e-05 & 73 & Tanh \\
9 & NIFIsingCls & 64 & 20 & 128 & 1024 & 1024 & 1e-05 & 77 & Tanh \\
10 & NIFIsingCls & 64 & 20 & 128 & 1024 & 1024 & 1e-05 & 45 & Tanh \\
11 & NIFIsingCls & 64 & 20 & 128 & 1024 & 1024 & 1e-05 & 68 & Tanh \\
12 & NIFIsingCls & 64 & 20 & 128 & 1024 & 1024 & 1e-05 & 53 & Tanh \\
13 & NIFIsingCls & 64 & 20 & 128 & 1024 & 1024 & 1e-05 & 38 & Tanh \\
14 & NIFIsingCls & 64 & 20 & 128 & 1024 & 1024 & 1e-05 & 54 & Tanh \\
15 & NIFIsingCls & 64 & 20 & 128 & 1024 & 1024 & 1e-05 & 71 & Tanh \\
16 & NIFIsingCls & 64 & 20 & 128 & 1024 & 1024 & 1e-05 & 4 & Tanh \\
17 & NIFIsingCls & 64 & 20 & 128 & 1024 & 1024 & 1e-05 & 73 & Tanh \\
18 & NIFIsingCls & 64 & 20 & 128 & 1024 & 1024 & 1e-05 & 46 & Tanh \\
19 & NIFIsingCls & 64 & 20 & 128 & 1024 & 1024 & 1e-05 & 65 & Tanh \\
\bottomrule
\end{tabular}}
\end{table}

\end{document}